\documentclass[prx,twocolumn,showpacs,amsfonts,amsmath,floatfix]{revtex4-1}

\usepackage[svgnames]{xcolor}

 \usepackage[
            colorlinks=true,        
            allcolors = black,  
            citecolor=DarkBlue, 
            linkcolor=DarkBlue,
            urlcolor=DarkBlue
        ]{hyperref}  
\usepackage{amsmath, amsfonts, amssymb, amsthm, bbm}
\usepackage{dsfont}
\usepackage{enumitem}
\usepackage{bm} 
\usepackage{mathtools}
\usepackage{graphicx}
\usepackage{epstopdf} 
\usepackage{epsfig}
\usepackage{dcolumn}
\usepackage{bm}
\usepackage{times}
\usepackage{physics}
\usepackage{graphicx}
\usepackage{caption}
\usepackage{subcaption}
\usepackage{epsfig}
\usepackage{listings}
\usepackage{multirow}
\usepackage{colortbl}
\usepackage{float}
\usepackage{autobreak}
\usepackage{amsmath}
\usepackage{algorithm}
\usepackage[noend]{algpseudocode}
\usepackage{color}
\begin{document}

\title{Gaussian conversion protocols for cubic phase state generation}

\author{Yu Zheng$^{1}$, Oliver Hahn$^{1}$, Pascal Stadler$^{1}$, Patric Holmvall$^{1}$, Fernando Quijandr\'{\i}a$^{1}$,  Alessandro Ferraro$^2$ and Giulia Ferrini$^{1}$}%

\affiliation{%
 $^{1}$ Wallenberg Centre for Quantum Technology, Department of Microtechnology and Nanoscience, Chalmers University of Technology, Sweden  \\
 $^2$ Centre for Theoretical Atomic, Molecular and Optical Physics, Queen's University Belfast, Belfast BT7 1NN, United Kingdom
}
 \email{hahno@chalmers.se} 

\date{\today}

\begin{abstract}
Universal quantum computing with continuous variables requires non-Gaussian resources, in addition to a Gaussian set of operations. A known resource enabling universal quantum computation is the cubic phase state, a non-Gaussian state whose experimental implementation has so far remained elusive. In this paper, we introduce two Gaussian conversion protocols that allow for the conversion of a non-Gaussian state that has been achieved experimentally, namely the trisqueezed state [Sandbo Chang \textit{et al.},  Phys. Rev. X \textbf{10}, 011011 (2020)], to a cubic phase state. The first protocol is deterministic and it involves active (in-line) squeezing, achieving large fidelities that saturate the bound for deterministic Gaussian protocols. The second protocol is probabilistic and it involves an auxiliary squeezed state, thus removing the necessity of in-line squeezing but still maintaining significant success probabilities and fidelities even larger than for the deterministic case. The success of these protocols provides strong evidence for using trisqueezed states as resources for universal quantum computation.
\end{abstract}

\maketitle

\section{Introduction}
\label{sec:level1}

Continuous-Variable (CV) systems~\cite{serafini2017quantum} are promising candidates to implement quantum computation in a variety of physical settings where quantum systems cannot be described within a finite dimensional Hilbert space, including optical~\cite{pfister2019continuous} and microwave radiation~\cite{blais2020circuit, grimsmo2017squeezing, hillmannUniversalGateSet2020}, trapped ions~\cite{serafini2009manipulating, fluhmann2019encoding}, opto-mechanical systems~\cite{schmidt2012optomechanical, houhou2015generation, nielsen2017multimode}, atomic ensembles~\cite{stasinska2009manipulating, milne2012composite, ikeda2013deterministic, motes2017encoding}, and hybrid systems~\cite{aolita2011gapped}. A major feature of CV systems is their potential in terms of scalability and noise resilience. In the optical domain, dual-rail cluster states composed of up to one-million modes have been implemented~\cite{yoshikawa2016}, as well as large bi-dimensional cluster states~\cite{larsen2019deterministic, asavanant2019generation}, with the potential of on-chip integrability \cite{lenzini2018integrated}.
In the microwave regime, the use of bosonic codes in superconducting cavities has allowed for extending the life-time of quantum information with respect to the constituents of the system~\cite{ofek2016}, and recent architectures allow for a lifetime of photons in 3D-cavities of up to two seconds~\cite{romanenko2020three}. Furthermore, bosonic codes that render CV quantum computation fault-tolerant against arbitrary errors, namely Gottesman-Kitaev-Preskill (GKP) codes~\cite{gottesman2001,menicucci2014fault}, have also been recently experimentally achieved~\cite{campagne-ibarcq2019}. 

In CV quantum computation, Gaussian operations play a central role~\cite{ferraro2005, weedbrook2012, adesso2014continuous}, given that in general they are relatively easy to implement regardless the chosen experimental platform. However, Gaussian operations alone cannot achieve computational universality~\cite{lloyd1999quantum, gottesman2001}, and genuine quantum non-Gaussianity is required as a resource~\cite{albarelli2018resource, takagi2018}. In particular, two main routes have been identified in order to promote Gaussian operations to universality by means of resourceful states. The first one relies upon the states that embody the specific codewords of the already mentioned GKP code~\cite{gottesman2001,  baragiola2019, yamasaki2020cost}. The second route is instead based on the so called cubic phase state~\cite{gottesman2001} which, by enabling the implementation of a non-linear gate~\cite{gu2009quantum}, can in principle unlock universality regardless the use of a specific encoding~\cite{lloyd1999quantum} --- including, for example, the generation of GKP states via the probabilistic protocol introduced in Ref.~\cite{douce2019}.

Whereas GKP codeword states have recently been produced experimentally~\cite{fluhmann2018, campagne-ibarcq2019}, the generation of a cubic phase state has proven elusive thus far, despite the considerable theoretical~\cite{gottesman2001, ghose2007, miyata2016, arzani2017, sabapathy2019production, yanagimoto2019engineering} as well as experimental~\cite{yukawa2013} effort. In this work, we provide viable solutions for the generation of a cubic phase state, exploiting a family of non-Gaussian Wigner-negative states that have been recently generated experimentally.

A number of experiments have demonstrated the generation of non-Gaussian states, both in the optical domain --- typically using photon subtraction and addition operations~\cite{dakna1997generating, wenger2004non, parigi2007probing} --- and in the microwave domain --- using controlled qubit-mediated operations~\cite{hofheinz2008generation, Heeres2015, campagne-ibarcq2019} or other form of non-linearities~\cite{Kirchmair2013,   svensson_multiplication_2018, Touzard2018, Grimm2019, chang_observation_2019,  Lescanne2020}.
However, it is not known currently which non-Gaussian states can be converted into resource states for quantum computation, such as the cubic phase state, by means of resourceless (Gaussian) protocols. In Ref.~\cite{albarelli2018resource}, a bound on the number of copies needed for the conversion, based on the ratio of the negativities of the Wigner function of the input and target state, has been derived. However, this bound is non-constructive, in the sense that even if the bound is satisfied, it is not guaranteed that a conversion protocol saturating the bound exists. In general, conversion protocols that yield as an output state a resource state starting from experimentally accessible states have not been studied thoroughly yet.

In this work we focus on the cubic phase state as a resource state, and we provide explicit protocols to convert a non-Gaussian state that has been recently generated within microwave circuits --- namely the trisqueezed state~\cite{chang_observation_2019} --- into a cubic phase state, with simple Gaussian operations that are readily available in the laboratory, in both the optical and the microwave regimes.

More specifically, we introduce two conversion protocols. The first one is based on squeezing and displacement transformations, and it belongs to the family of deterministic Gaussian maps. We first provide a bound on the fidelity to the target cubic phase state that can be achieved with the most general deterministic Gaussian map, and we subsequently show that our protocol saturates this bound. We then introduce a second protocol that includes conditional measurements on ancillary states. Belonging to the larger set of non-deterministic Gaussian maps, we are able to show that this second probabilistic protocol achieves higher fidelities with respect to the deterministic bound --- yet retaining success probabilities that are high compared to existing protocols~\cite{sabapathy2019production}.
For both protocols, we rely on numerical optimization in order to determine the best parameters to be used.

With our work, we therefore establish that it is possible to convert a trisqueezed state onto a cubic phase state by means of Gaussian operations alone. As recalled above, since the availability of cubic phase states and Gaussian operations enables universal quantum computation over CV systems, this corroborates the use of trisqueezed states as a resource for CV quantum computation. In fact, once that our protocols have established the Gaussian equivalence of the trisqueezed state and the cubic phase state, one can make direct use of the former to implement universal non-Gaussian gates. Specifically, in Appendix~\ref{appendix:Gate_teleportation} we derive the teleportation gadget needed to implement a cubic phase gate directly from a trisqueezed state.

The paper is structured as follows. In Sec.\ref{def-states} we define the input and target states for our conversion protocols, and we motivate their study. In Sec.\ref{unitary-Gaussian-maps} we calculate the upper bound on the fidelity of the state conversion --- i.e., the fidelity to the desired target state --- obtainable with deterministic Gaussian maps. In Sec.\ref{protocol1} we define our deterministic Gaussian conversion protocol, and we show that it corresponds to a simple squeezing and displacement operation on the input mode, achieving high fidelity of conversion --- for example, a fidelity of $0.971$ for a target cubic non-linearity of approximately $0.156$. In Sec.\ref{protocol2}, we introduce our probabilistic Gaussian conversion protocol, analyse thoroughly its properties, and  show that it yields higher fidelities as compared to the deterministic protocol (for example, up to $0.997$ for the same target), for success probabilities as high as $5\%$. In Sec.\ref{kerr} we discuss the experimental implementability of our protocols in both microwave and optical systems, before presenting conclusive remarks in Sec.\ref{conclusions}. In Appendix~\ref{appendix:numerics} we provide an extensive discussion of the numerical methods used for our optimizations, while other technical details are provided in the remaining Appendixes.

\section{Purpose of our conversion protocols}
\label{def-states}

Before starting, it is useful to recall some standard definitions and notations for CV systems that we are going to use extensively in this paper, as well as the definition of the input and target states. We are going to indicate the vector of quadrature operators for $N$ bosonic modes as $\hat{\Vec{r}}=\qty(\hat{q}_1,\hat{p}_1... \hat{q}_N,\hat{p}_N)^T$, and for each mode we use the following convention for the relation between the quadrature operators and the creation and annihilation operators: $\hat{q} = (\hat{a} +\hat{a}^\dagger)/2$ and $\hat{p} =  ( \hat{a} - \hat{a}^\dagger)/(2i)$, corresponding to setting $\hbar = 1/2$. 
The squeezing $\hat{S}(\xi)$, displacement $\hat{D}(\beta)$ and phase rotation $\hat{U}_p(\gamma)$ operators are defined respectively as
\begin{eqnarray}\label{eq:squeezing}
    \hat{S}(\xi)=e^{\frac{\xi^*}{2}\hat{a}^{2} -\frac{\xi}{2}\hat{a}^{\dagger 2}},
\end{eqnarray}
\begin{eqnarray}
    \hat{D}(\beta)=e^{{\beta\hat{a}^{\dagger }}-\beta^{\ast}\hat{a}},
\end{eqnarray}
\begin{eqnarray}\label{eq:phase-rot}
    \hat{U}_p(\gamma)=e^{-i\gamma\hat{n}},
\end{eqnarray}
with $\hat{n} = \hat a^{\dagger} \hat a$ the number operator, $\gamma \in \mathbb{R},\; \beta \in \mathbb{C},\; \xi \in \mathbb{C}$, and $\xi = |\xi| \; e^{i \phi}$ with $\phi \in [0,2\pi]$. The successive application of a squeezing and displacement operator onto the vacuum state yields a displaced squeezed state 
\begin{equation}\label{eq:input-squeezed-state}
    |\Psi_{\xi,\beta} \rangle =  \hat{D}(\beta) \hat{S}(\xi) |0 \rangle .
\end{equation}

\begin{figure}[h]
\begin{subfigure}{.235\textwidth}
  \centering
  \includegraphics[width=1\linewidth]{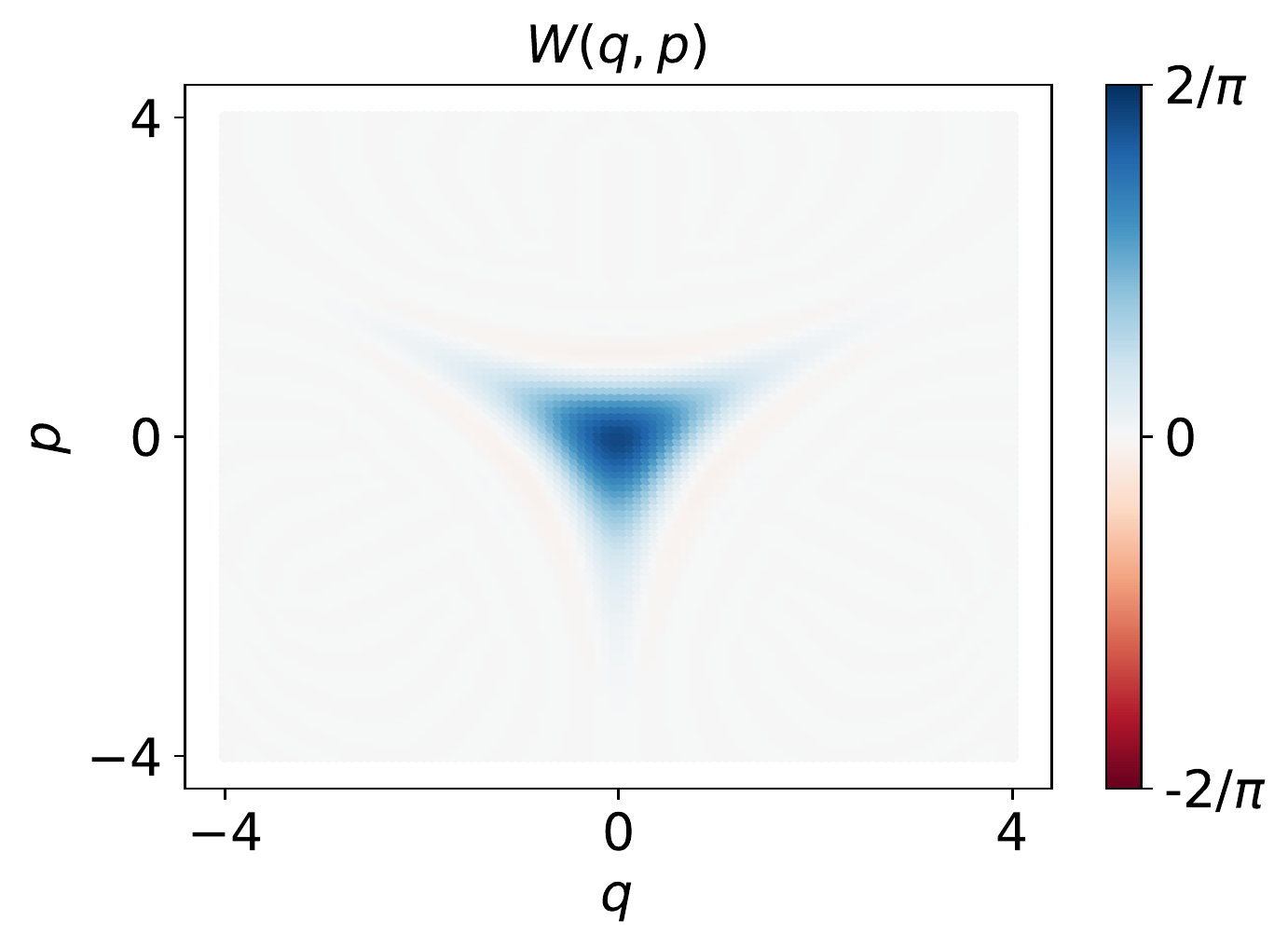}  
  \caption{Trisqueezed state $|\Psi_{\textrm{in}}\rangle$. }
  \label{fig:triple}
\end{subfigure}
\begin{subfigure}{.235\textwidth}
  \centering
  \includegraphics[width=1\linewidth]{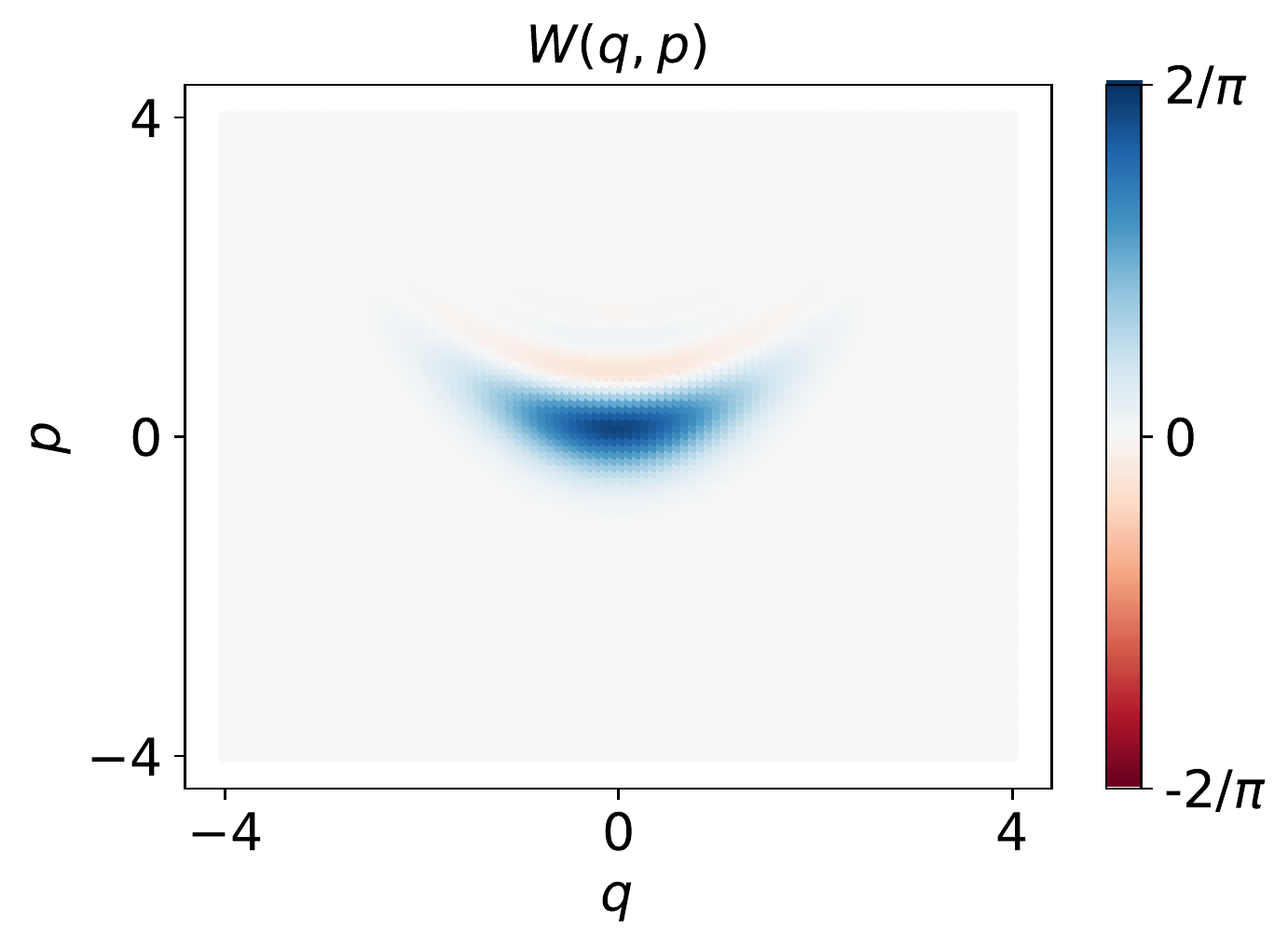}  
  \caption{Cubic phase state $|\Psi_{\textrm{target}}\rangle$.}
  \label{fig:target}
\end{subfigure}
\caption{Wigner functions of the input state (trisqueezed state) and of the target state (cubic phase state) of our protocols. 
The parameter defining the strength of the triple-photon interaction, i.e. the triplicity, is set to $t =0.1 $, while the corresponding parameter characterizing the cubic phase state, the cubicity, is set to $r = 0.1558$, with 5dB squeezing.}
\label{fig:input_target}
\end{figure}

In the following, we are going to address conversion protocols from an experimental available state to a state that, as said, is known to be pivotal for quantum computation.
The input state discussed in this paper is the trisqueezed state defined as~\cite{braunstein1986generalized,banaszek1997quantum}:
\begin{align}
    \label{eq:input_state}
  \ket{\Psi_{\textrm{in}}} =  e^{i (t^* \hat{a}^3 + t \hat{a}^{\dagger 3})}  \ket{0}.
\end{align}
In what follows, we are going to refer to the complex parameter $t$ that characterizes the strength of the tri-photon interaction in Eq.(\ref{eq:input_state}) as the \emph{triplicity}.
Fig.\ref{fig:triple} shows the Wigner function of the trisqueezed state with triplicity $t =0.1$ as an example. As it can be seen, this state is symmetric with respect to the momentum axis $q = 0$, and it also possess a $2\pi/3$-rotational symmetry. The rotational symmetry is a direct consequence of the  Hamiltonian generating the state in Eq.(\ref{eq:input_state}), and is equivalently also reflected in the Fock expansion of the trisqueezed state, where only Fock states with photon numbers that are multiple of three are present~\cite{albert2014symmetries}.

Our target state, the cubic phase state, is defined as~\cite{gu2009quantum}
\begin{equation}
\label{eq:target_state}
    \ket{\Psi_{\textrm{target}}}  
   = e^{i r \hat{q}^3} \hat S(\xi_{\rm target}) \ket{0},
\end{equation}
where the subscript ``target" is used in order to distinguish this squeezing parameter from those of other squeezed states that will be introduced later.
In what follows, we are going to refer to the parameter that characterizes the strength $r$ of the cubic interaction in Eq.(\ref{eq:target_state}) as the \emph{cubicity}.
Fig.\ref{fig:target} shows the Wigner function of the cubic phase state with cubicity $ r = 0.0551$~\footnote{Note that the value of the cubicity depends on the convention used. In our case, the values refer to $\hbar = 1/2$.}. This state, too, is symmetric with respect to the momentum axis $q = 0$.
For convenience, we will fix the squeezing strength of the target state Eq.~(\ref{eq:target_state}) as 5dB, which implies that $\xi_{\textrm{ target }}=-\textrm{log}10^{\frac{{\rm 5dB}}{20}}$\footnote{The relation between the value of a quantity with respect to a reference value and its counterpart in decibel (dB) is $\Delta_{\text{dB}}= -10 \textrm{log}_{10}(\frac{\Delta_0^2}{\Delta^2})$, where here for instance $\Delta_0^2$ and $\Delta^2$ are the variances of the vacuum and a squeezed state respectively. 
Here, we have $\hbar=1/2$, $\Delta_0^2=1/4$ and after applying the squeezing operator Eq.(\ref{eq:squeezing}) to the vacuum the variance of the $\hat q$ quadrature is given by $\Delta^2 =\exp[-2 \abs{\xi} ]/4$.}. The properties of the Wigner function for trisqueezed states and cubic phase states have been considered before in references~\cite{bencheikh2007triple, banaszek1997quantum} and~\cite{ghose2007,arzani2017, brunelli2019linear} respectively.

In order to fix, for a given input state triplicity, the target state cubicity, we use considerations from quantum resource theory. 
As said, it has been proved that operations or initial states characterised by negative Wigner functions are necessary for quantum speed-up~\cite{mari2012positive}. Wigner negativity is thereby regarded as a resource for computational advantage. A convenient measure of the 
negativity content of the Wigner function is the Wigner logarithmic negativity or \emph{mana}  $M(\rho) = \textrm{log}(\int d\Vec{r} |W_\rho(\Vec{r})|)$, where $W_\rho(\Vec{r})$ is the Wigner function of the state $\rho$ and where the integral runs over the whole phase-space. The concept of mana was originally introduced for discrete-variable Wigner functions~\cite{veitch2014}, and later extended to continuous variables~\cite{albarelli2018resource, takagi2018}.
The main features of the CV mana is that it is invariant under Gaussian unitary operations (namely, unitary operations that are generated by Hamiltonian at most quadratic in the canonical bosonic operators), and more generally under deterministic Gaussian protocols. In addition, it does not increase on average under probabilistic Gaussian protocols~\cite{albarelli2018resource}; namely, one has that 
\begin{equation}
M(\rho_{\textrm{in}})\geq p M(\rho_{\textrm{target}}),
\label{eq:mono}
\end{equation}
where $M(\rho_{\textrm{in}})$ and  $M(\rho_{\textrm{target}})$ are the mana of the input and target states, respectively, and $p$ is the probability of success of the conversion protocol.
Therefore, given that we want to assess both deterministic and probabilistic Gaussian conversion protocols for a given pair of input and target states, it is reasonable to choose the latter states in such a way that they possess the same amount of mana. Given a certain input triplicity, a corresponding cubicity can be estimated numerically by the requirement of keeping the mana invariant.
Limited only by numerical accuracy, we choose to start from trisqueezed states with triplicities 0.1, 0.125 and 0.15, while targeting corresponding cubic phase states with the same mana. 
As a side remark, note that the trisqueezed state Eq.(\ref{eq:input_state}) has higher mana than the cubic phase state state Eq.(\ref{eq:target_state}) for the same average photon number.

Notice that the choice of the target cubicity is not crucial in terms of computational universality, and it is taken here only to ease the quantitative comparison of different protocols. In fact, a cubic phase state of any given cubicity $r$ can be used to generate a unitary operation of the form $\exp[i r\hat{q}^3]$, via Gaussian deterministic gate teleportation~\cite{gu2009quantum}. The latter is usually denoted as cubic phase gate and, equipped with arbitrary Gaussian unitaries, unlocks universality for any value of $r$~\cite{weedbrook2012}. In fact, it is easy to show that 
$\hat{S}(s)^\dagger \exp[i r\hat{q}^3]\hat{S}(s)= \exp[i r e^{3s}\hat{q}^3]$, where $s$ is the strength of a squeezing gate. In other words, under the assumption of having at disposal arbitrary squeezing, the non-linearity can be enhanced or reduced by changing the strength of a supplementary squeezing gate.

In order to characterise conversion protocols that map the trisqueezed state onto the cubic phase state or aim at approximating the latter as well as possible, we need to define a measure of the distance between the target state and the transformed input state. For this we choose the fidelity~\cite{jozsa1994fidelity}
\begin{align}
\label{eq:fidelity}
    \mathcal{F}(\rho_1,\rho_2) = \qty( \Tr{\sqrt{ \sqrt{\rho_1}\rho_2 \sqrt{\rho_1}  }})^2.
\end{align} 
As our target state is a pure state, this expression can be simplified to
\begin{align}
\label{eq:fidelity-pure}
    \mathcal{F}(\rho,\Psi_{\textrm{target}}) = \langle \Psi_{\textrm{target}} |\rho|\Psi_{\textrm{target}} \rangle.
    \end{align}

In what follows, we detail two Gaussian protocols enabling the approximate conversion of a trisqueezed state onto a cubic phase state, and we characterise their performances.

\section{Deterministic Gaussian Conversion Protocol}
\label{section-all-maps}

In this section we introduce our deterministic Gaussian conversion protocol. Before doing so, we provide numerically an upper-bound to the fidelity of conversion that can be achieved by the class of trace-preserving Gaussian completely-positive (CP) maps, and we show that our deterministic protocol saturates this bound. In other words, the optimal maps are symplectic maps with displacement, and we show that the dominant contribution consists of squeezing. 

\subsection{Fidelity bound with general Gaussian maps}
\label{unitary-Gaussian-maps}

Completely-positive trace-preserving (CPTP) maps are called Gaussian if they map Gaussian states into Gaussian states. Note that the target Gaussian state does not have to be necessarily a pure state. These maps are characterized by their action onto the symmetrically ordered characteristic function~\cite{ferraro2005}:
\begin{align}
\label{eq:generating-function}
    \chi_{\rho}(\Vec{r}) = \Tr{ \hat{D}(-\Vec{r}) \rho} 
\end{align}
with the arbitrary displacement operator being
\begin{align}
    \hat{D}(-\Vec{r}) = e^{-i( \Vec{r}^T \Omega \hat{\Vec{r}})}, 
\end{align}
with $ \Vec{r} \in \mathbb{R}^{2N}$ and
\begin{align*}
    \Omega =\bigoplus^{N}_{j=1} \begin{pmatrix}
    0& 1\\
    -1 &0 
    \end{pmatrix}
\end{align*}
being the symplectic form for N modes.
Beyond unitary deterministic processes, these Gaussian maps may also include non-unitary maps representing noise or processes where ancillary modes are measured. In the latter case, however, feed-forward is then assumed to take place, to restore determinism.

The action of any Gaussian CPTP-map $\Phi$  on the characteristic function can then be written as~\cite{depalmaNormalFormDecomposition2015}:
\begin{align}
\label{eq:gen_Gausssian_map}
    \chi_\rho(\Vec{r})\rightarrow& \chi_{\Phi(\rho)}(\Vec{r}) = e^{-\frac{1}{4}\Vec{r}^T\Omega^T Y \Omega \Vec{r} + i \Vec{l}^T \Omega \Vec{r}}\chi_\rho (\Omega^T X^T\Omega \Vec{r})
\end{align}
with $X$,$Y$ being $2N\times 2N$ real matrices, $\Vec{l}$ being a $2N$ real vector, $Y$ being symmetric, and fulfilling the following positive semi-definite matrix constraint:
\begin{align}
\label{eq:gen_Gausssian_map_ineq}
    Y & \pm i(\Omega - X\Omega X^T)\geq 0 \;.
\end{align}
Notice that Eq.~(\ref{eq:gen_Gausssian_map_ineq}) in turn implies that $Y$ has to be a positive semi-definite matrix. The requirement for positive semi-definiteness needs to hold for both signs, since transposition does not affect the positive (semi-) definiteness of a matrix. 
It has to be noted that Eq.~(\ref{eq:gen_Gausssian_map}) characterises general trace-preserving open Gaussian dynamics, where $\Vec{l}$ is the displacement on the state and $Y$ denotes additive Gaussian noise. 

Since the conversion protocol we are investigating has one mode only at  both the input and the output, we set $N=1$ in the following paragraphs.
In order to determine numerically the matrices $X$, $Y$ and the vector $\Vec{l}$ that map the trisqueezed state onto the cubic phase state (for a given pair of cubicity and triplicity parameters), or approximate it as good as possible, we re-express the fidelity defined in Eq.~(\ref{eq:fidelity-pure}) in terms of the characteristic functions of the input and target states:
\begin{align}
    \label{eq:fidelity_det_protocol}
    \mathcal{F}(\rho,\rho_{\textrm{target}}) = 
    \bra{\Psi_\textrm{target}} \hat \rho \ket{\Psi_\textrm{target}} 
    =\frac{1}{4 \pi} \int d\Vec{r} \;\chi_{\rho}(\Vec{r})\; \chi_{\rho_{\text{target}}}(-\Vec{r}),
\end{align}
where $\rho_{\textrm{target}} = \ket{\Psi_\textrm{target}} \bra{\Psi_\textrm{target}}$. 
We calculated numerically the characteristic functions for both the input and target states given respectively by Eqs.~(\ref{eq:input_state}) and (\ref{eq:target_state}) and then transformed the input characteristic function given the Gaussian CPTP-map in Eq.~(\ref{eq:gen_Gausssian_map}). We then maximized the fidelity between the transformed state and the target state by optimizing $X$, $Y$ and $\Vec{l}$, while still fulfilling Eq.~(\ref{eq:gen_Gausssian_map_ineq}). Since this optimization involves a number of potential evaluations of the characteristic functions, using the  analytical expressions for the matrix elements of the displacement operator [see Eqs.(\ref{eq:exp1}) and (\ref{eq:exp2})] significantly speeds-up the computation, compared to direct matrix exponentiation. More details regarding the numerical calculations are provided in Appendix~\ref{appendix:numerics}.

The results of this fidelity optimization are shown in Table~\ref{tab:det_maps}, for various values of the triplicity of the input trisqueezed state. For the value of triplicity  $t=0.1$, we obtain a conversion fidelity of 0.9708. The fidelity of conversion decreases at increasing triplicity. The parameters in terms of the matrices $X$, $Y$ and $\Vec{l}$ that optimize the conversion are given in Appendix \ref{appendix: Opt_res_det}, in Table~\ref{tab:gauss_cp}. 

By looking at the optimized parameters, a few considerations can be made. The matrix $Y$ is essentially a null matrix, implying that the conversion can be done unitarily. Furthermore, $X$ is nearly diagonal with reciprocal entries, i.e., it corresponds to squeezing. The vector $\Vec{l} = \begin{pmatrix}  l_q\\l_p    \end{pmatrix}$  has a non-trivial contribution for the displacement $l_p$, the displacement along the momentum axis $q = 0$, so we expect that a purely symplectic conversion will not saturate the bound given by the trace-preserving  Gaussian  completely-positive  maps. This result can be understood at an intuitive level: $Y$ should be close to zero, so we do not add any noise to the conversion; the displacement along the position axis $p = 0$ should be zero to conserve the symmetry between both Wigner functions, and positive along the momentum axis $q = 0$, to match the maxima of their respective Wigner functions. 
Finally, as we will show in the next paragraph, an exact conversion can be done asymptotically with only squeezing.


\subsection{Unitary protocols}
\label{protocol1}
We now specialize to the case of symplectic transformations, which allows us to design a specific deterministic conversion protocol. Symplectic transformations are special cases of the protocols introduced in Sec.\ref{unitary-Gaussian-maps} and correspond to a class of unitary operations for which the noise matrix $Y$ and the displacement vector $\Vec{l}$ are set to zero, whereas $X \in Sp_{2,\mathbb{R}}$ is a symplectic matrix~\cite{serafini2017quantum}.


We denote the unitary operation associated to $X$ as $\hat{U}_X$, acting as follows on the trisqueezed state:
\begin{align}
    \hat{U}_X e^{it (\hat{a}^3 +\hat{a}^{\dagger 3})}  \ket{0} =   \hat{U}_X e^{it (\hat{a}^3 +\hat{a}^{\dagger 3})}  \hat{U}_X^\dagger  \hat{U}_X\ket{0}.
\end{align}
For exact state conversion, the following relation needs to hold:
\begin{align}
    \hat{U}_X e^{it (\hat{a}^3 +\hat{a}^{\dagger 3})} \hat{U}_X^\dagger \rightarrow e^{i r \hat{q}^3}.
\end{align}
This can be achieved asymptotically in the infinite squeezing limit.
Squeezing implements a Bogoliubov transformation
\begin{align}
   \hat{a}\rightarrow \hat{S}(\xi) \hat{a} \hat{S}^\dagger(\xi) =  u\hat{a} + v\hat{a}^\dagger\\
 \hat{a}^\dagger\rightarrow \hat{S}(\xi) \hat{a}^\dagger \hat{S}^\dagger(\xi) =  u^* \hat{a}^\dagger + v^* \hat{a}
\end{align}
with $u = \cosh{(|\xi|)}$ and $v = \sinh{(|\xi|)} \, e^{-i\phi}$ (see Eq. \eqref{eq:squeezing})
In the case of $u=v$ and $u^*=v^*$, this transformation gives us the required form, because 
$\hat{q} \propto \hat{a}+\hat{a}^\dagger$. This means that a conversion with asymptotically unit fidelity is possible for  $|\xi| \rightarrow \infty $ and $\phi = 0$. In other words, in the high squeezing limit negligible contributions of displacements are expected, since first, exact conversion is asymptotically possible  and second, displacements on the input state add lower orders of $\hat{p}/\hat{q}$ to the exponent in Eq.~(\ref{eq:input_state}) while the target is only cubic in $\hat{q}$ in the exponent. 

Since the squeezing parameter $\xi_\text{target}$ associated with the target cubic phase state Eq.(\ref{eq:target_state}) is finite, one expects that the optimal squeezing operation will be a trade-off between matching the target state squeezing and transforming the trisqueezed state.
Moreover, in the regime of finite squeezing we expect that the contributions to the Gaussian map coming from the displacements cannot be neglected. Note that in view of the above discussion, we expect the fidelity of our protocol to increase at increasing squeezing in the target state.
\begin{table*}
\centering
\setlength{\tabcolsep}{2.5mm}
\begin{tabular}{|c c c c|}
\hline
Triplicity  & Cubicity(5dB) & Gaussian CPTP-map & Symplectic  \\
\hline
0.1& 0.1558& 0.9708& $0.9335$    \\
\hline
0.125 & 0.2757 &0.9273& $0.8810$   \\
\hline
0.15 &0.4946& 0.8557 & $ 0.8113$     \\
\hline
\end{tabular}
\caption{Maximized fidelities for Gaussian CPTP-maps (third column) and purely symplectic maps (fourth column), for different input-target pairs (first and second column). The squeezing of the target cubic phase state is fixed to be 5 dB. The optimized parameters are in Appendix \ref{appendix: Opt_res_det}.}
\label{tab:det_maps}
\end{table*}

\begin{table*}
\centering
\setlength{\tabcolsep}{2.5mm}
\begin{tabular}{|ccccccc|}
\hline
Triplicity & Cubicity & Fidelity & Squeezing & Displacement $l_p$ & Displacement $l_q$ & $\text{Mana}_{\rm out}$  \\
\hline
0.1& 0.1558&$0.9708 $  & 0.6741 (3.4 dB)& $0.1547$ &  $2\cdot10^{-9}$&0.1658 \\
\hline
0.125 &0.2757 &$0.9273$& 0.7816 (2.1 dB)& $0.2268$ & $-10^{-8}$&0.3338\\
\hline
0.15 &0.4946& $0.8557$& 0.9463 (0.5 dB) & $ 0.3029$& $ -5\cdot 10^{-8}$ &0.5450  \\
\hline
\end{tabular}
\caption{Maximized fidelities and optimized parameters for the deterministic Gaussian conversion protocol including squeezing and displacement, as depicted in Fig. \ref{fig:scheme}, for different input-targets pairs. The squeezing of the target cubic phase state is fixed to be 5 dB. Notice that the squeezing values that are given in absolute numbers correspond to the first diagonal element of the symplectic matrix $X$, while the values in parenthesis correspond to the squeezing parameter of Eq.(\ref{eq:squeezing}). The two values are related by $\abs{\xi} = - \log(X_{00})$. The last column is the mana of the transformed state. }
\label{tab:squeezing+disp}
\end{table*}

Similarly to Sec. \ref{unitary-Gaussian-maps}, we determine the maximum fidelity between the input and target states that is achievable with symplectic transformations by transforming the input characteristic function. The advantage with respect to the general Gaussian maps of section~\ref{unitary-Gaussian-maps} is that here, since we know that $X \in Sp_{2,\mathbb{R}}$, we can parametrize the transformation as~\cite{dopicoParametrizationMatrixSymplectic2009}:
\begin{align}
    X = \begin{pmatrix}
    g & g e\\
    c g &\; g^{-1} + c g e
    \end{pmatrix},
    \label{eq:symplectic-X}
    \end{align}
for $g,e,c \in \mathbb{R}$ and non-zero $g$. In other words, we are using  $3$ real parameters to parametrize a real symplectic transformation, which is precisely the dimension of the real symplectic group $Sp_{2,\mathbb{R}}$. 

\begin{figure}[h]
\begin{subfigure}{.235\textwidth}
  \centering
  \includegraphics[width=1\linewidth]{input_deter.pdf}  
  \caption{Trisqueezed state $|\Psi_{\textrm{in}}\rangle$. }
  \label{fig:triple_deter}
\end{subfigure}
\begin{subfigure}{.235\textwidth}
  \centering
  \includegraphics[width=1\linewidth]{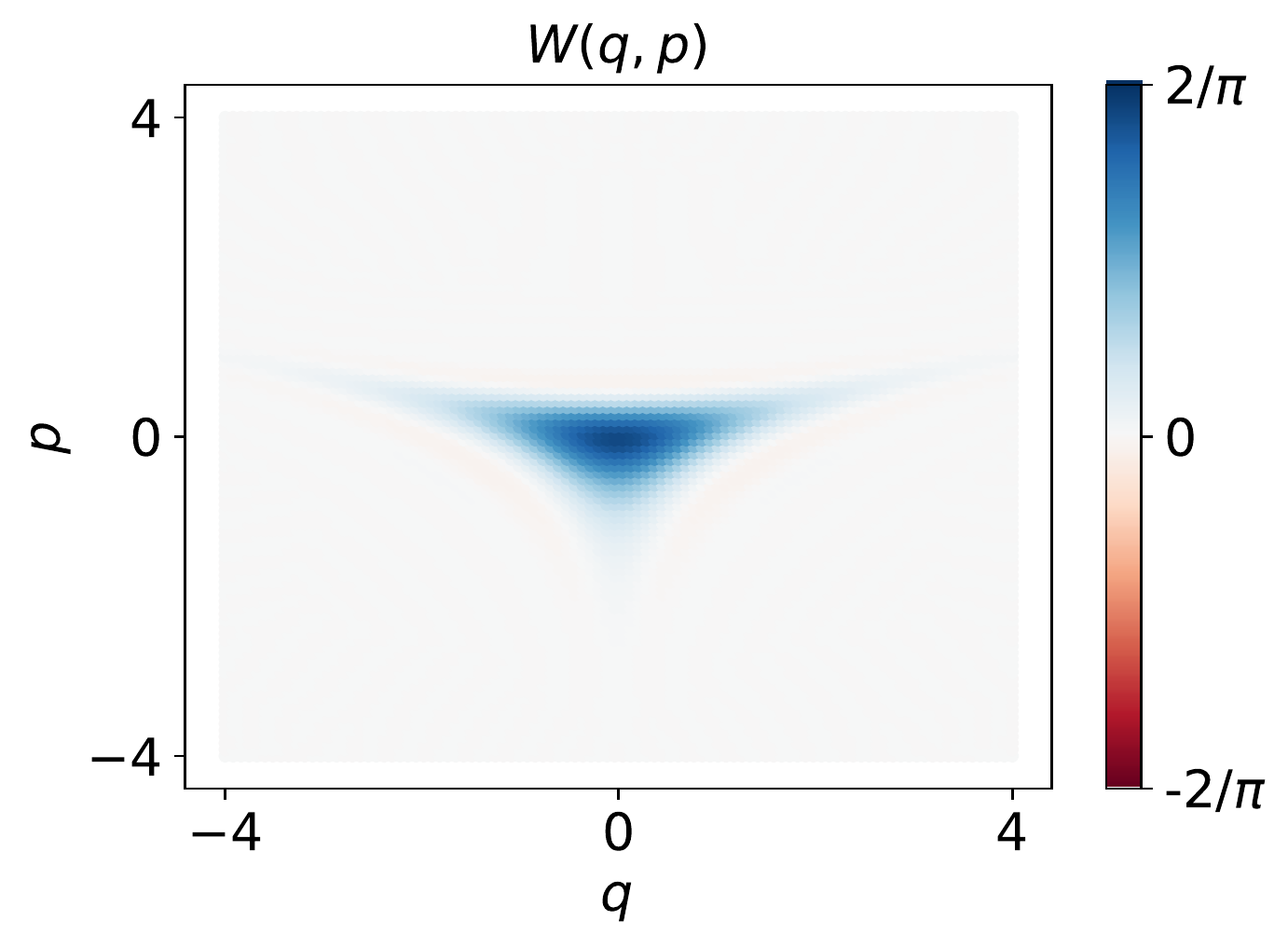}  
  \caption{Output state $|\Psi_{\textrm{out}}\rangle$.}
  \label{fig:out_deter}
\end{subfigure}
\caption{
Right panel (b): Wigner function of the output state obtained via our deterministic symplectic Gaussian conversion protocol $|\Psi_{\textrm{out}}\rangle$, i.e. excluding the final displacement. In order to ease the comparison, we reproduce in panel (a) the Wigner function of the input trisqueezed state with triplicity $t=0.1$, already plotted in Fig.\ref{fig:triple}. The full list of parameters characterising the optimisation protocol is given in  Table~\ref{tab:squeezing+disp}.}
\label{fig:output_1}
\end{figure}

Fig.~\ref{fig:output_1}b shows the Wigner function of the output state corresponding to the optimization of the symplectic transformation in Eq. (\ref{eq:symplectic-X}). It can be seen that  the Wigner function of the output state and that of the target state in Fig.\ref{fig:target} are qualitatively similar, which is expected as the fidelity is equivalent to the Wigner overlap~\cite{cahill1969density}.
The results of the optimizations are shown in Table~\ref{tab:det_maps}. Squeezing is the dominant contribution of the symplectic transformation, with the off-diagonal elements in the symplectic matrix being negligible small, as can be seen in Appendix~\ref{appendix: Opt_res_det}. 

Interestingly, from Table~\ref{tab:det_maps} we see that the obtained values for the fidelity of conversion are, for all values of triplicitly, below the ones from the Gaussian maps. This is to be expected, since the optimal Gaussian maps had a non-vanishing contribution of displacement.
The effectiveness of squeezing on the input state towards reaching a cubic phase state can also be intuitively understood from Fig.\ref{fig:output_1}.

As we have seen, given only symplectic single-mode transformations, squeezing is the relevant contribution.
An intuitive extension to the investigated purely symplectic transformations and motivated by the results obtained from the trace-preserving  Gaussian  completely-positive  maps  are displacements. 
The characteristic function transforms under displacements as:
\begin{align}
    \chi_\rho(\Vec{r})\rightarrow \chi_{\hat{D}(\Vec{l})\rho \hat{D}(\Vec{l})^\dagger  }(\Vec{r})= e^{i \Vec{l}^T \Omega \Vec{r}} \chi(\Vec{r}).
\end{align}

\begin{figure}[h]
\includegraphics[scale=0.85]{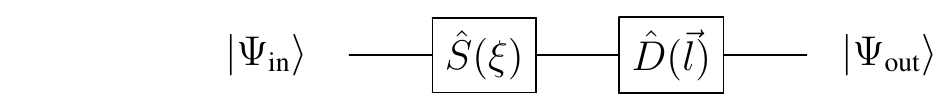}  
\caption{Sketch of our  deterministic Gaussian conversion protocol. We apply a squeezing operator and a displacement on a trisqueezed state.   $\vert \Psi_{\textrm{out}} \rangle$ is the output state after optimizing the parameters in the circuit. }
\label{fig:scheme}
\end{figure}

Since the off-diagonal terms in the symplectic case were trivial, we focus on squeezing and displacement only. 
The resulting conversion scheme is depicted in Fig.\ref{fig:scheme}.

The optimized fidelities for this approach are shown in Table~\ref{tab:squeezing+disp} as well as the parameters for the optimized protocol. The maximum achieved fidelity is improved further with displacements along the momentum axis $q=0$,  saturating the bounds given by the Gaussian CPTP-maps and achieving for example the value of $0.971$ for triplicity $t=0.1$.   

Note that in the present case, the achieved output state has by construction the same mana as the input state, because unitary Gaussian operations conserve the Wigner negativity~\cite{albarelli2018resource}. 

\section{Probabilistic Gaussian conversion protocol}
\label{protocol2}

\label{results}

The deterministic conversion protocol introduced in Sec.~\ref{section-all-maps} belongs to the class of trace-preserving Gaussian maps~\cite{depalmaNormalFormDecomposition2015}. In this section, we relax the requirement for a deterministic protocol, and we introduce instead a probabilistic conversion protocol which belongs to the larger class of CP Gaussian maps \cite{albarelli2018resource}. While becoming associated to a success probability, the conversion fidelity that can be reached with probabilistic protocols can in principle be higher. As we will study, the mana of the output state can be larger than the one of the input state, since the protocol is solely constrained by Eq.~(\ref{eq:mono}). 

Let us mention in advance that the probabilistic protocol will achieve squeezing by means of an ancillary squeezed state, rather than of a squeezing operator acting directly on the input state. The former is usually referred to as offline squeezing, in contrast to the latter that is known as inline squeezing. The possibility to substitute inline with offline squeezing is a well known result \cite{filip2005measurement}, which has proven to be of practical relevance especially in quantum optical set-ups \cite{miwa2014exploring}. However, as we will show below, our probabilistic protocol is not merely  an offline-squeezing version of the deterministic protocol. Rather, as said, it belongs to a larger class of protocols and can therefore achieve better conversion performances.

\begin{figure}[h]
\includegraphics[scale=0.85]{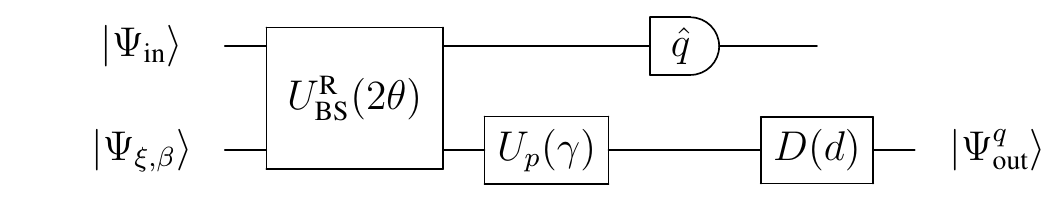}  
\caption{Sketch of our probabilistic Gaussian conversion protocol. We apply a beam-splitter $U_{\textrm{BS}}^\textrm{R}(2\theta)$ on a trisqueezed state and an ancillary displaced squeezed state. After a rotation $U_p(\gamma)$, we post-select the homodyne measurement on the first mode to value $q=0$ and displace the second mode with $D(d)$. The output state is denoted as $\ket{\Psi^q_{\text{out}}}$.}
\label{fig:circuit2}
\end{figure}

The conversion protocol that we consider is sketched in Fig.~\ref{fig:circuit2}. It takes as inputs a trisqueezed state, Eq.\eqref{eq:input_state}, in the upper rail and a displaced squeezed state, Eq.\eqref{eq:input-squeezed-state}, in the lower rail. These are fed into a beam-splitter corresponding to the symplectic transformation  $U_{\textrm{BS}}^\textrm{R}(2\theta)$, where 
\begin{equation}
\label{eq:beam-splitter}
    U_{\textrm{BS}}^\textrm{R}(2\theta)= \begin{pmatrix}
   \cos{\theta}& \sin{\theta} \\
  -\sin{\theta} & \cos{\theta} 
  \end{pmatrix}.
\end{equation}
Notice that the $2\times2$ matrix in Eq.(\ref{eq:beam-splitter}) refers to two modes, i.e. is meant to act on the annihilation operators $\hat a_1, \hat a_2$, in contrast to the $2\times2$ matrices of the previous sections and in particular of Eq.(\ref{eq:symplectic-X}), acting on the annihilation and creation operators of a single bosonic mode.
Next, a phase rotation, Eq.\eqref{eq:phase-rot}, is performed on the lower rail and a homodyne measurement is performed on the upper rail. Upon post-selection on the measurement result $q = 0$ on the upper rail, a displacement is performed on the state on the lower rail. 

Notice that the deterministic protocol analyzed in Sec.~\ref{protocol1},  which makes use of active in-line squeezing, could be converted to a (deterministic) protocol that uses only off-line squeezing. Ideally, this could be accomplished via a gate-teleportation gadget~\cite{nielsen1997programmable, gottesman1999b, bartlett2003quantum, weedbrook2012}  composed of a control-phase gate whose two input modes are fed by the trisqueezed input state and an auxiliary infinitely squeezed state; an additional phase-space rotation and a final displacement of the latter mode using feed-forward (i.e., depending on the outcome of the homodyne measurement on the first mode) would implement the required transformation. Such measurement-based squeezers have been proposed theoretically in Ref.~\cite{filip2005measurement} and implemented experimentally in Ref.~\cite{yoshikawa2007demonstration}. Here we generalise this strategy, by using a displaced squeezed input state, a beam-splitter operation with variable amplitude, a non-conditional displacement, and we post-select on the measurement outcome.
As already said, this has the advantage that the map implemented does not belong to the set of CPTP Gaussian maps analysed in Sec.~\ref{section-all-maps}, and hence it allows in principle for achieving higher fidelities, at the price of introducing a success probability. We also note that, in comparison to the protocol of Ref.~\cite{filip2005measurement} where the output state is a mixed state due to the finite squeezing in the ancillary squeezed state, in our probabilistic protocol the purity of the output state is preserved~\cite{fuwa2014noiseless}, the only source of impurity stemming - as we will see - from the finitely-resolved homodyne detector.

As an additional remark, it is interesting to compare our conversion protocol to the probabilistic synthesis protocols in Ref.~\cite{sabapathy2019production}, aiming at generating a cubic phase state starting by means of tunable optical circuits with optimised parameters, and the deterministic protocol in Ref.~\cite{yanagimoto2019engineering}. In these protocols, the non-Gaussian element is provided, respectively, by the measurement (photon-number resolving detector) and by the nonlinear medium (self Kerr effect). In our scheme, instead, both evolution and measurement are described by Gaussian processes and are hence resourceless, but the input state is non-Gaussian.

\begin{figure}
  \centering
  \begin{subfigure}{.235\textwidth}
  \centering
  \includegraphics[width=1\linewidth]{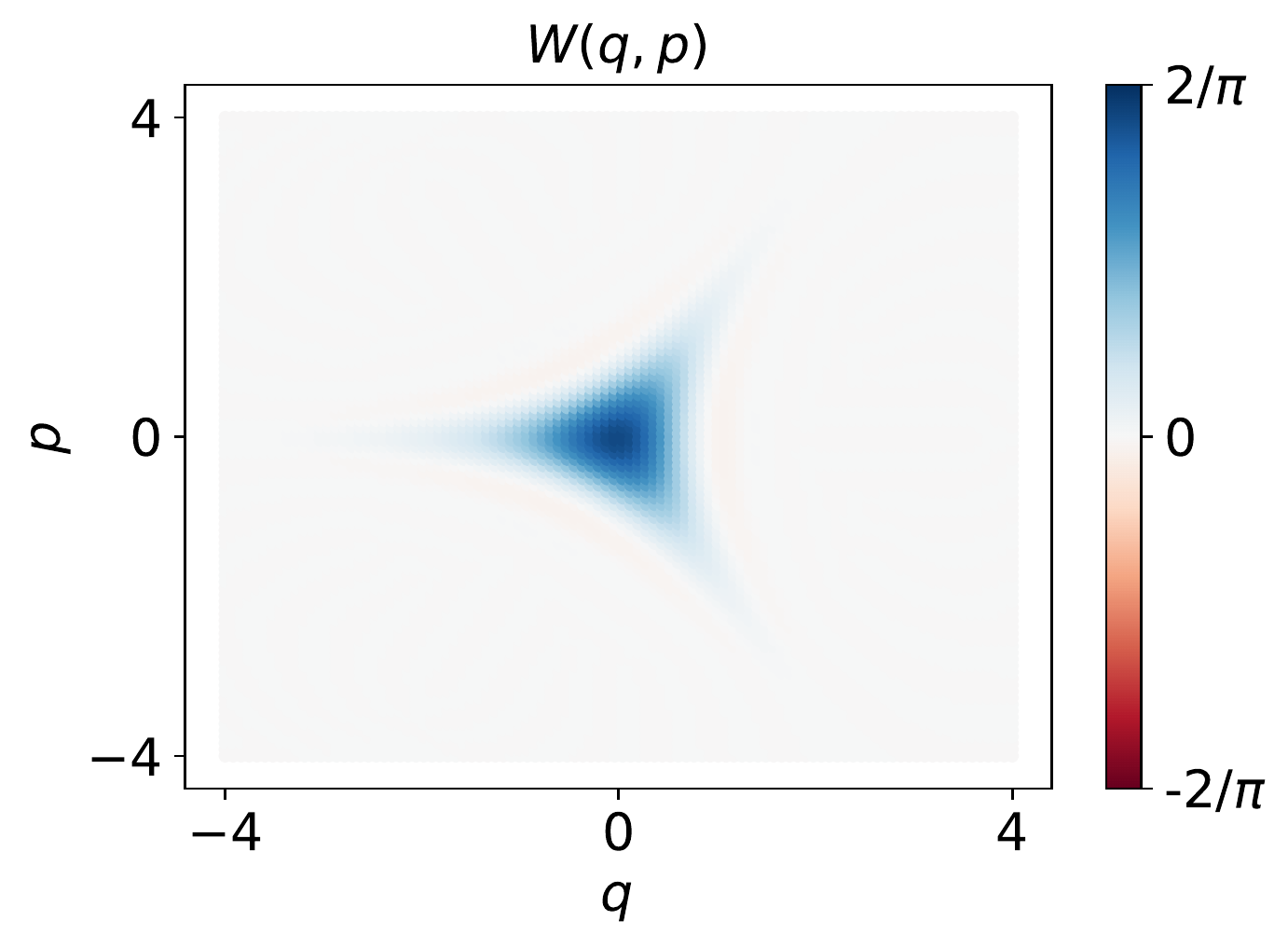}  
  \caption{Rotated trisqueezed state. }
  \label{fig:triple_c}
\end{subfigure}
\begin{subfigure}{.235\textwidth}
  \centering
  \includegraphics[width=1\linewidth]{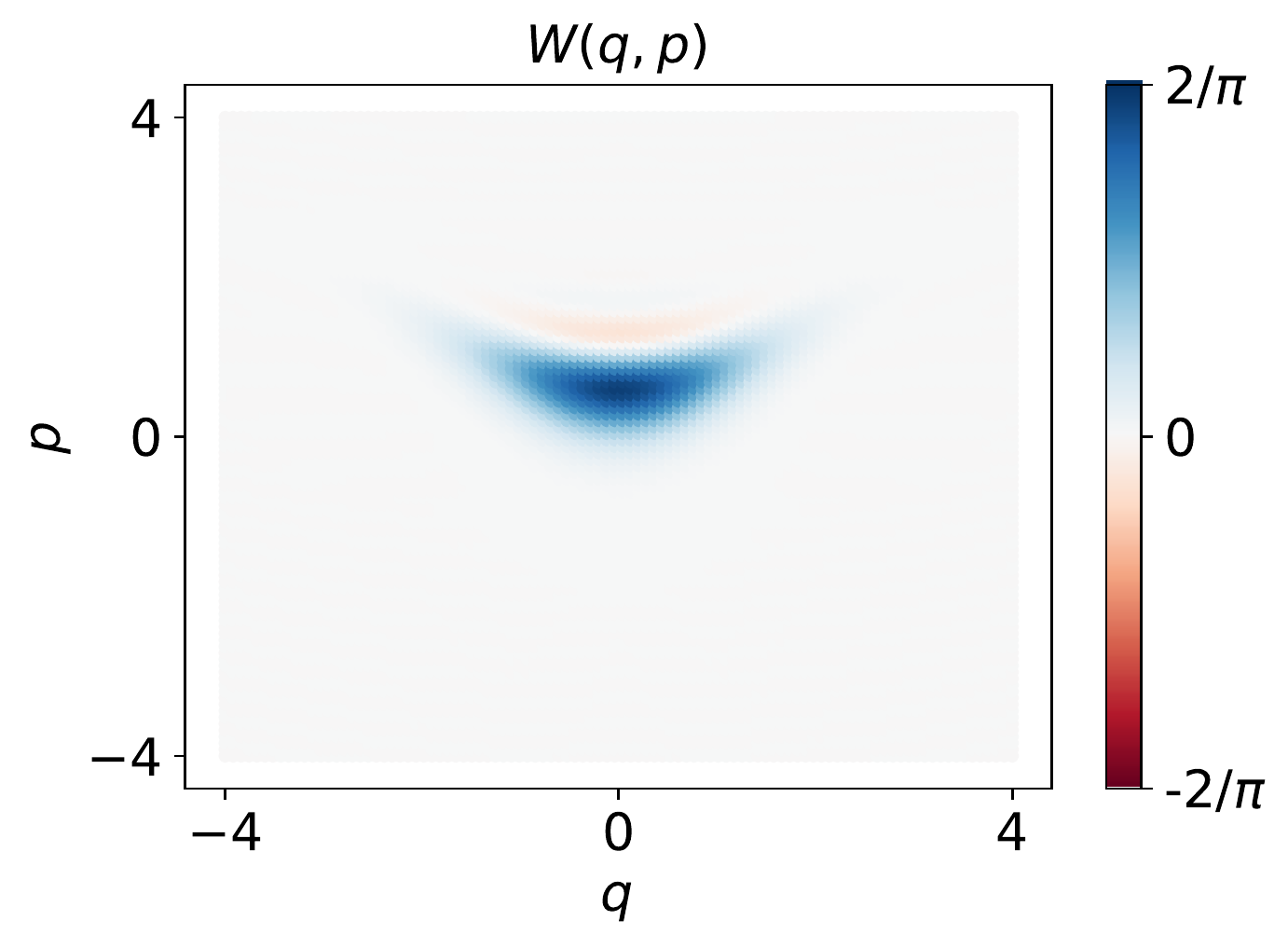}  
  \caption{Output state.}
  \label{fig:output_2}
\end{subfigure}
  
\caption{Wigner functions of the rotated input state (corresponding to a  trisqueezed state with triplicity of $t =0.1 \exp(i \pi/2)$) and  of the output state $|\Psi^q_{\textrm{out}}\rangle$ of the probabilistic Gaussian conversion protocol. The corresponding parameters of the circuit are shown in the first row of Table~\ref{table:probabilistic}.}

\end{figure}

As will come clear later, in this section we will consider an input trisqueezed state with triplicity $\arg(t)=\pi/2$ [see Fig.\ref{fig:triple_c}]. This is done in order to exploit the symmetries of input and target states and ease the numerical optimization of the circuit parameters.

We now calculate explicitly the output state of the circuit sketched in Fig.~\ref{fig:circuit2}.
To simplify the calculation, we consider the output state before the displacement on the lower mode. 
For the purpose of calculating the fidelity or the overlap between output and target states, this is equivalent to displacing the target in the opposite direction.
In other words, the fidelity is the same whether we displace the output state, or the target state by the opposite displacement.  

Note that the quadrature operator $\hat q$, associated with ideal homodyne detection, has eigenvalues $q$ in the real axis. Then, the probability to measure a particular eigenvalue is infinitely small.
In order to overcome this limitation, and properly model a finite-resolution homodyne detector, the real axis can be discretized into bins of width $2 \delta$ around the real values $q_n$, where $n \in \mathbb{Z}$ is the integer labelling the $n$-th bin. 
The probability associated to $q_n$ corresponds to the probability that a measurement outcome belongs to the $n$-th bin.
In particular, the probability of obtaining $q_n=0$ defines the success probability of our protocol.

As shown in Appendix~\ref{appendix:output}, the output state that corresponds to a general measurement outcome $q$ associated to an infinitely resolved homodyne detector can be written as: 
\begin{equation}
\begin{split}
&|\Psi^{q}_{\textrm{out}}(\xi,\beta,\theta,\gamma)\rangle 
 \\
 =&
\frac{1}{\pi}
\int\, \,dq_2\,\,d\alpha
 \Psi_{\textrm{in}}(q\cos{\theta}+q_2\sin{\theta}) \\
 &
 \Psi_{{\xi,\beta}}(-q\sin{\theta}+q_2\cos{\theta}) \langle\alpha|q_2\rangle |\alpha e^{-i\gamma}\rangle,
\label{eq:output2}
\end{split}
\end{equation}
where the displaced squeezed state parameters $\xi$ and $\beta \equiv q_\beta+i p_\beta$ are complex numbers, while the angle $\theta$ parameterizing the beam-splitter and the phase rotation $\gamma$ are real numbers. 
A consequence of binning is that the conditional output state of our protocol is a mixed state,
\begin{equation}
\begin{split}
\hat{\rho}_{n,\textrm{cond}} = \frac{1}{\textrm{Prob}[q_n]}\int^{q_n+\delta}_{q_n-\delta} dq  |\Psi^{q}_{\textrm{out}}\rangle\langle\Psi^{q}_{\textrm{out}}|,
 \label{eq:cond-out}
 \end{split}
\end{equation}
where $|\Psi^{q}_{\textrm{out}}\rangle$  is given in Eq.~\eqref{eq:output2} and $\textrm{Prob}[q_n]$ is the probability of obtaining the outcome $q_n$, explicitly calculated in Appendix~\ref{appendix:output} (see in particular Eq.~\eqref{prob}).

 The fidelity defined in Eq.(\ref{eq:fidelity}) between the target state and the output conditional density matrix in Eq.(\ref{eq:cond-out}) is then expressed as
\begin{equation}
\begin{split}
F_{q_n}= &\langle\Psi_{\textrm{target}}|\hat{\rho}_{n,\textrm{cond}}|\Psi_{\textrm{target}}\rangle
\\
= & \frac{1}{\textrm{Prob}[q_n]}
\int^{q_n+\delta}_{q_n-\delta}\, \,dq \,|\langle\Psi_{\textrm{target}}|\Psi^q_{\textrm{out}}\rangle|^2.
 \label{eq:fidelity3}
 \end{split}
\end{equation}
As already stated, we focus in particular on $q_n =0$.

There are seven parameters that can be optimized for maximizing the fidelity in Eq.(\ref{eq:fidelity3}), including the displacement parameter $d$ after the measurement.
The numerical optimization of the fidelity is hence a challenging task. It involves three computationally expensive numerical integrations, and the total necessary time grows exponentially with the dimension of the space to be explored. However, we empirically find that, as a consequence of the symmetries of the input and  target states, some of the parameters  can be fixed. As shown in  Fig.~\ref{fig:triple_c} 
we consider a trisqueezed state with Wigner function symmetric with respect to the position axis  $p = 0$, while the target state is symmetric with respect to the momentum axis $q = 0$ (Fig.~\ref{fig:target}). 
We fix the phase of the ancillary displaced squeezed state so as to yield a position-squeezed state, i.e., $\xi$ real and positive,  
and we consider a real displacement for the ancillary squeezed state, hence $p_\beta =0$. With these choices, the full two-mode input state is symmetric with respect to the position axis $p = 0$. 
Then, we set the phase rotation to $\gamma=-\pi/2$ so that our output state upon post-selection over $q_n = 0$ 
has the same symmetry of the target state, i.e., it is symmetric with respect to the momentum axis $q = 0$. Hence, we are left with tuning the magnitudes of the 
squeezing and displacement parameters of the ancillary state, 
the real beam-splitter parameter and the final momentum displacement, in order to achieve the maximal fidelity to our target state. 
In Appendix~\ref{parameters-control} we provide an analysis of how the various tunable parameters in our protocol affect the properties of the output states. 

We carry out the numerical optimizations by running three independent codes, namely a Python code running on a personal computer, a C++ code running in serial on central processing units (CPUs) in a cluster environment, and finally a CUDA~~\cite{CUDA2019} code running in parallel on graphics processing units (GPUs)~~\cite{Matthews2018} in a cluster environment. We provide the relevant details on these approaches in Appendix~\ref{appendix:numerics}.

\begin{figure}[h]
\centering
\includegraphics[width=0.99\linewidth]{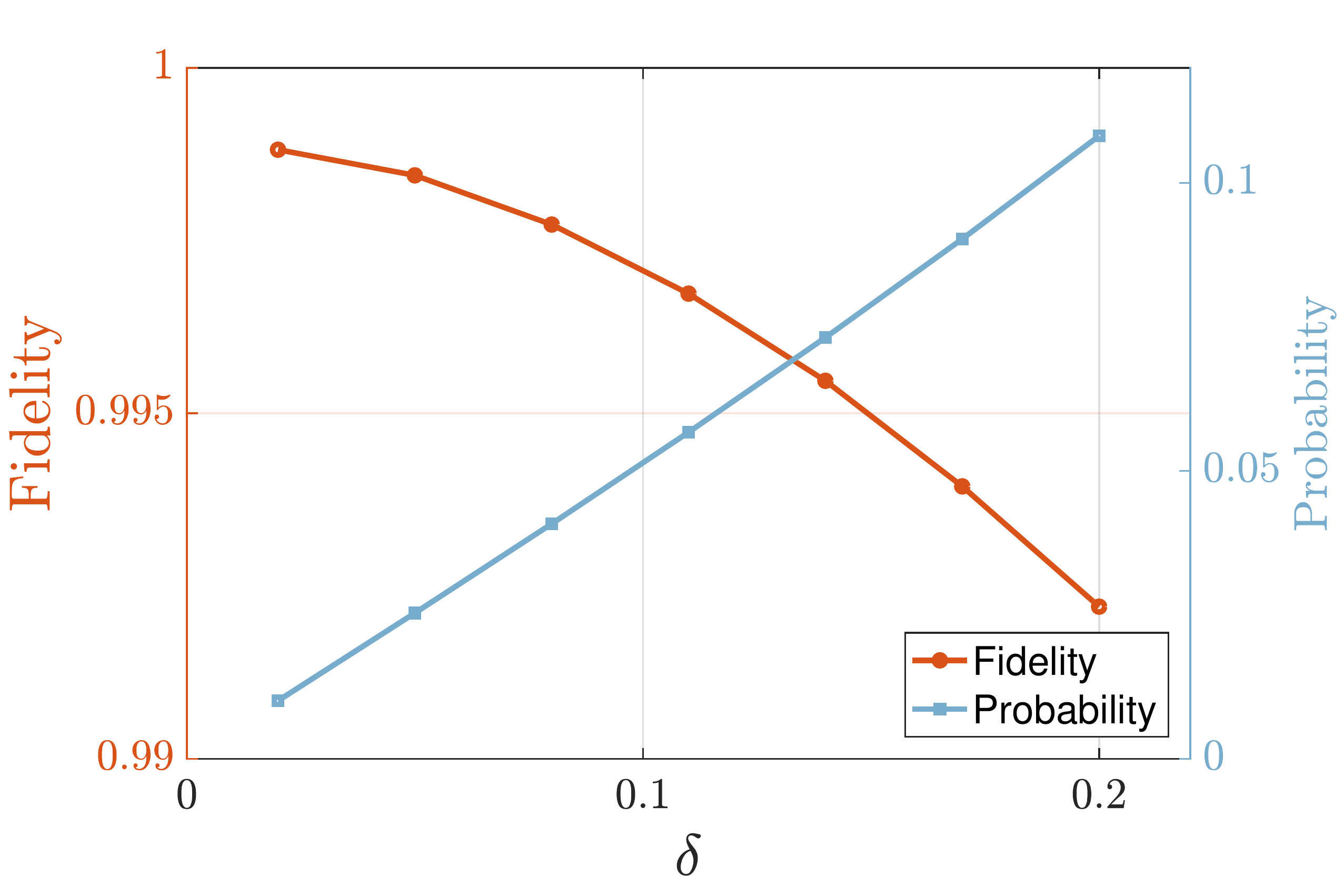}  
\caption{Fidelity and success probability as a function of the width of the acceptance region $\delta$. The parameters in the circuit, including input and target mana, correspond to the first row of Table.~\ref{table:probabilistic}.}
\label{fig:eta}
\end{figure}

\begin{table*}[htbp]
\begin{tabular}{|c|c|c|c|c|c|c|c|c|}
\hline
Triplicity&Mana$_{\textrm{in}}$&Fidelity&Probability&$\theta$&$q_\beta$&$\xi$& $d$ &Mana$_{\textrm{out}}$\\
\hline

0.1&0.1576 &  0.9971&0.0513 &1.0133 & 0.8304& 0.3257 (2.83 dB)&-0.9525&0.1103\\
\hline
0.125&0.3350 &  0.9866& 0.0434&0.7992&1.2153& 0.001 (0.01dB) & -1.1104&0.1945\\
\hline
0.15&0.5737 & 0.9284&0.0508&0.6378&0.001&1.4184 (12.3dB)& -1.3639&0.2197\\
\hline
\end{tabular}
\caption{Fidelity, success probability, and optimal circuit parameters when targeting 
an output state with the same mana than that of the corresponding input state within the probabilistic Gaussian conversion protocol. The squeezing of the target cubic phase state is fixed to be 5 dB.
Here $\theta$ is the beam-splitter parameter, $\xi$ is the squeezing strength and $q_\beta$ is the position displacement of the ancillary displaced squeezed state, and $d$ is the displacement of the output state along the $p$ direction.}
\label{table:probabilistic}
\end{table*}
Finally, in Fig.~\ref{fig:eta}, we analyse the effect of the width $\delta$ of the acceptance region on the success probability, as well as on the fidelity of our protocol. As expected, for larger values of $\delta$, the fidelity decreases due to the lower purity of the output state Eq.(\ref{eq:cond-out}), while the success probability increases. 
In Appendix~\ref{app:ineff} we have studied the case of inefficient homodyne detection. For realistic efficiency values, our probabilistic conversion protocol still achieves high fidelities.

\subsection{Conversion protocol performance at fixed input mana} 
In order to compare our results with those obtained in Sec.\ref{section-all-maps}, we first consider the case where the target state has the same mana as the input trisqueezed state, which is always the case for the deterministic maps.
The mana in the target state is determined by the parameters $r$ and $\xi_{\rm target}$. However, we will only come 
 as close as possible to this state with our probabilistic protocol. Therefore, we expect that the output mana will not be the same as that of the input state. We anticipate that the deviations may be significant  because the mana is an extremely sensitive quantity that can vary significatively even if the overlap (quantified by the fidelity) of two states is very high.

\begin{figure}[h]
  \centering
  \centering
  \includegraphics[width=1\linewidth]{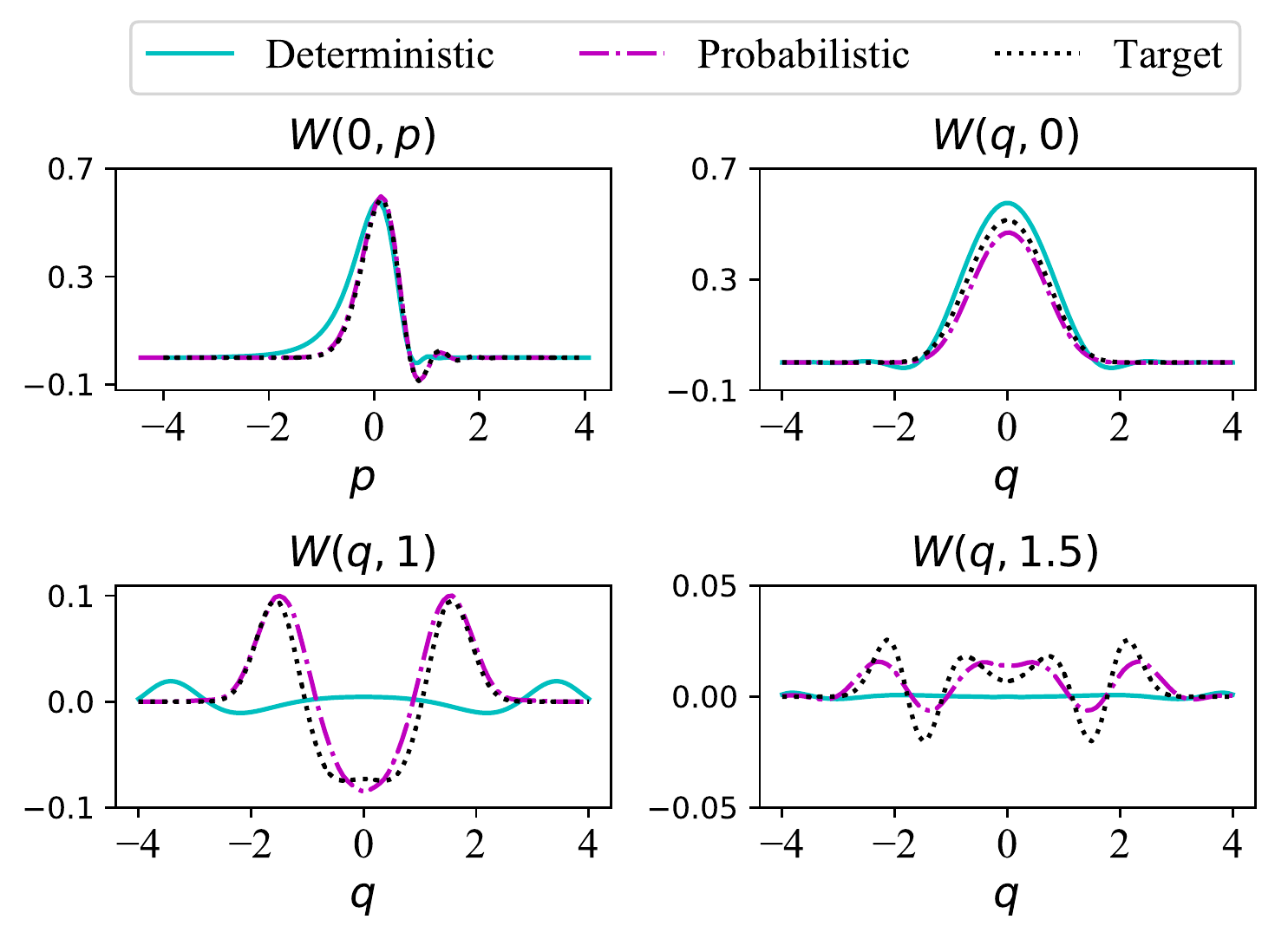}  
\caption{Slices of Wigner functions to compare the output states of the deterministic and probabilistic protocols with the target cubic phase states for cubicity $r=0.1558$. The value of the target squeezing is fixed at 5\,dB. }
 \label{fig:cut}
\end{figure}

As can be seen in Fig.~\ref{fig:output_2}, the Wigner function of the output state is qualitatively similar to the one of the target state (Fig.~\ref{fig:target}). 
This can be observed more precisely in Fig.~\ref{fig:cut}
where we plot the cross sections of the Wigner functions of the target cubic phase state and the output states generated by the probabilistic and the deterministic protocols corresponding to 
$q=0$, $p=0$, $p=1$, and $p=1.5$. 
In general, the probabilistic protocol gives a very good approximation to the target state. This in contrast to the
deterministic protocol which fails to reproduce some of the features of the target state Wigner function.
Table~\ref{table:probabilistic} shows the results for the achieved fidelity of conversion after optimizing the tunable parameters in our probabilistic protocol. We can see that a fidelity as high as 0.997 can be obtained with  success probability of 5$\%$ when the triplicity of the input state is moderate. As can be observed in Table \ref{table:probabilistic}, and as expected, the mana obtained in the output state of our protocol can be sensibly different with respect to the one of the target state, regardless of the high fidelity.

\subsection{Conversion protocol performance at  varying input mana} 
\begin{figure}[h]
\centering
\includegraphics[width=0.99\linewidth]{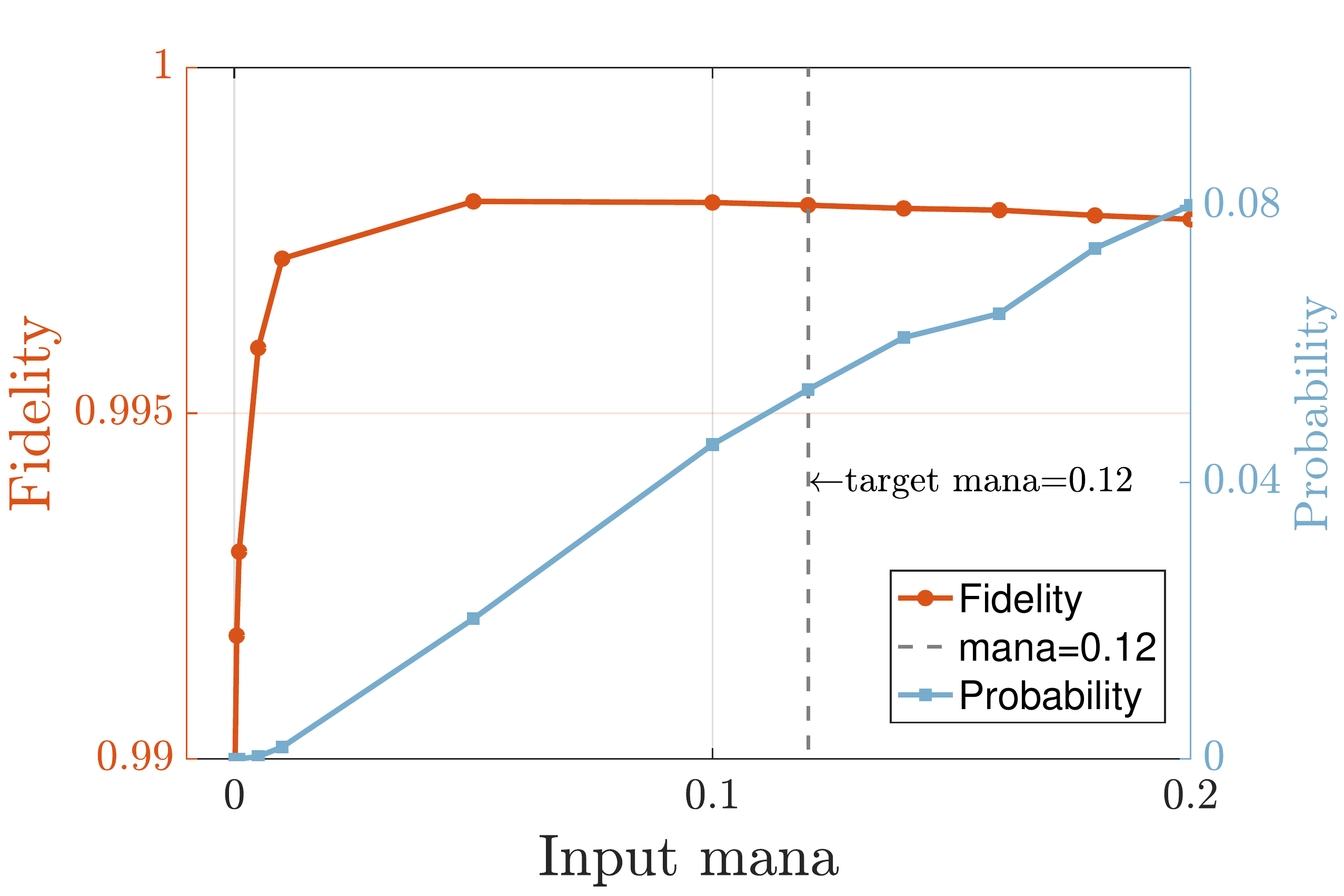}  
\caption{Fidelity and success probability of the probabilistic conversion protocol as a function of the mana of the input state, considered to be a perfect trisqueezed state. The mana of the target state here is fixed at 0.12, corresponding to a cubicity of $r=0.133$. We set $\delta=0.1$. 
}
\label{fig:plot_mana}
\end{figure}

We now relax the requirement that the target state must possess the same mana as the input state. Eq.~(\ref{eq:mono}) implies that the mana will still be conserved on average, even when we are targeting a state with higher mana with respect to the one at the input, which will succeed with a certain probability. So,  what is the performance of our protocol when we start from an input state with lower, or higher, mana with respect to the one of the target state? Fig.~\ref{fig:plot_mana} shows the fidelity and success probability of our protocol, as a function of the mana in the input state, for fixed target state mana.
The dashed line in Fig.~\ref{fig:plot_mana} corresponds to the (fixed) mana in the target state. On the left of this line, i.e., when the mana of the input is smaller than that of the target state, the obtained high fidelity corresponds hence to a probabilistic concentration protocol.
We observe that this is possible to achieve with success probabilities up to roughly 5$\%$.
On the right of the dashed line we
observe that the fidelity does not increase when having higher mana in the input state. However, there is a positive correlation between the success probability and the input mana. A possible explanation of this fact is provided by Eq.~(\ref{eq:mono}). The latter sets an upper bound on the success probability given as the ratio of the input and target states mana.
For a fixed target state mana, by increasing that of the input state we also increase the upper bound on the success probability.
Moreover, the fidelity is robust against a decrease in the input mana
up to a value of roughly 0.01 where it drops very quickly.
At each point of the figure, the bound of Eq.~(\ref{eq:mono}) is satisfied, as can be verified by multiplying the mana of the output state with the success probability.

\section{Experimental implementability of the protocols}
\label{links-experiments}

In this Section, we discuss the feasibility of our protocols in state-of the art experiments, based either on microwave circuits, or on optical systems. 

\subsection{Microwave circuits}
\label{kerr}

In superconducting microwave circuits, the field commonly referred to as circuit QED,
nonlinear interactions between microwave photons are mediated by Josephson junctions or Josephson junction-based devices such as the Superconducting QUantum Interference Device (SQUID)~\cite{Devoret2004, gu2017}. 
Arranging Josephson junctions in loops allows for magnetic flux biasing. 
This, in turn, gives the possibility to drive different parametric processes when these devices interact with superconducting resonators or propagating microwaves. 
For our purposes, here we are going to restrict to the case of interacting resonant modes.

It is well-established  
that in the linear regime of the SQUID, parametrically-mediated interactions between microwave modes permit the engineering of Gaussian operations such as the squeezing and beam-splitting required for this proposal~\cite{vitaly-parametric-2013, vitaly-parametric-2017, gao2018programmable}.
The full set of linear operations follows trivially with the addition of monochromatic microwave tones which implement linear displacements.
Finally, homodyne detection can be implemented via phase-sensitive parametric amplification~\cite{eichler-linear-2012}. The highest quantum efficiency reported today for 
microwave
homodyne detection is about 0.7~\cite{walter2017rapid}.
For the particular case of our deterministic protocol, the resonator field would need to be released into a waveguide
(in a controlled fashion) in order to subject it to the homodyne detection.
This can be done, for instance, following~\cite{pfaff2017controlled}.

In a similar way, higher-order processes can be exploited from the SQUID non-linearity. Recently, the three-photon drive Hamiltonian giving rise to the trisqueezed state studied in this work, Eq.(\ref{eq:input_state}), has been realized experimentally~\cite{svensson_multiplication_2018,chang_observation_2019}. 
The experiments by Chang et al. have established the possibility to engineer strong degenerate as well as non-degenerate three-photon interactions in the microwave regime.
Whereas their results correspond to the stationary state emerging in the continuous driving regime, their Hamiltonian engineering is by no means restricted to the latter. Therefore, it is possible to operate these devices in a gate-based regime, in which the degenerate three-photon interactions will give rise to the trisqueezed state as defined by Banaszek and Knight~\cite{banaszek1997quantum}.
Evolving an initial vacuum state for a time $\tau$ with the Hamiltonian $g_3 (\hat a^3 + \hat a^{\dagger 3})$ results in a trisqueezed state with triplicity $t = g_3 \times \tau$.
Considering typical parameters corresponding to planar microwave architectures, we have estimated a three-photon strength $g_3$ of a few MHz. The corresponding resonator lifetimes imply operation times $\tau$ of a few hundred nanoseconds in order to avoid dissipation effects. 
Following this analysis, a figure of merit for the triplicity corresponds to $t \sim 0.1$ which we use along this manuscript.

Finally, a residual Kerr interaction which might be detrimental for the protocols presented here
is unavoidable in these implementations.
In what follows, we analyse the effect of the residual Kerr term on our probabilistic conversion protocol, namely, how the fidelity between the output and the target state decreases with an increasing residual Kerr non-linearity strength.
Later, we interpret these fidelities  in operational terms. For this, the output state is used 
in a gate teleportation gadget in order to exert a non-Gaussian gate on an input state. We will assess whether the presence of non-idealities has a detrimental effect on the gate fidelity.

\subsubsection{Fidelity in the presence of residual Kerr interactions}
\label{se:residual-kerr-sub}

\begin{figure}[h]
\includegraphics[width=0.99\linewidth]{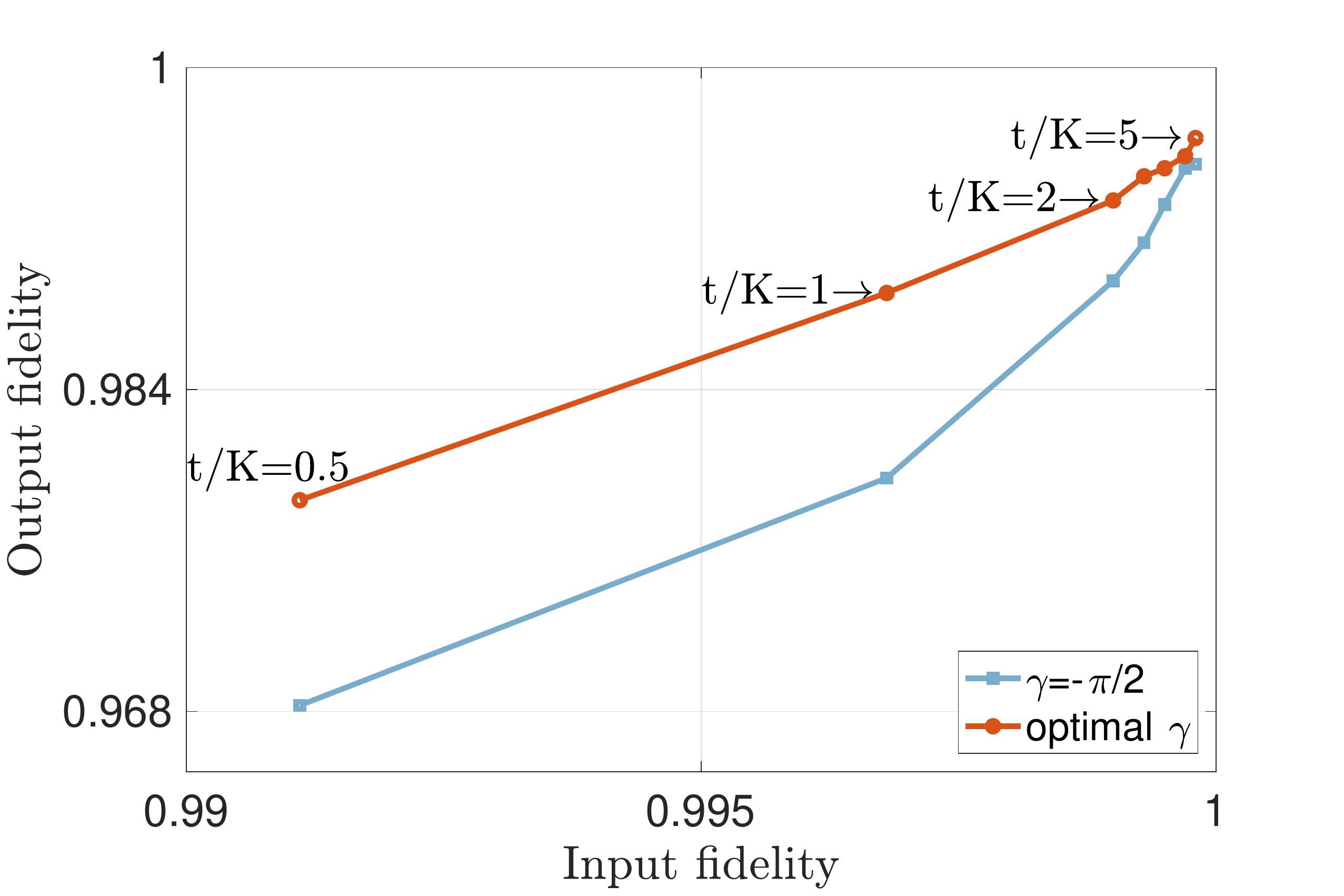}  
\caption{Output fidelity as a function of the input fidelity when the input state is a trisqueezed state generated in the presence of residual Kerr interaction. After relaxing the condition of fixed $\gamma$, 
higher output fidelities can be achieved. This implies that the 
main effect of the Kerr effect is an undesired rotation.
The plot also shows that for a relevant ratio of the triplicity and Kerr strengths  which is accessible experimentally, namely $t/K = 2$, the fidelity of conversion is still very high, up to 0.993. Success probabilities vary between $5\%$ and $13\%$.}
\label{fig:input_vs_output}
\end{figure}

In order to assess the robustness of our probabilistic conversion protocol (outlined in Sec.\ref{protocol2}) against imperfections in the input state due to the residual Kerr interaction $K \hat a^{\dagger 2} \hat a^2$, we repeat the fidelity optimization over the circuit parameters when we introduce a Kerr deformation in the input state for different Kerr strengths $K$.
We perform this analysis for the probabilistic protocol because it is the one yielding the highest conversion fidelity.    
The results are shown in Fig.~\ref{fig:input_vs_output}. We plot the output fidelities as a function of the input fidelity when the input state is a trisqueezed state generated in the presence of residual Kerr non-linearities. The input fidelity refers to the fidelity of a perfect trisqueezed state with respect to a Kerr-deformed-trisqueezed state. In order to compare with the result in the first row of Table.~\ref{table:probabilistic}, we fix the triplicity $t$ at 0.1 here.
Since the Kerr term introduces not only a deformation, but also a rotation of the trisqueezed state, for this analysis we also optimise on the parameter $\gamma$ entering the probabilistic protocol, which we had previously fixed due to symmetry considerations.

From Fig.~\ref{fig:input_vs_output} we see that high fidelities can still be obtained for $t/K > 1$, i.e., when the Kerr interaction is weak as compared to the triplicity, and in particular, when the ratio $t/K \simeq 2$ which is a relevant value for the experiments in Ref.~\cite{chang_observation_2019}. 
We conclude that our results still hold in the case where an unwanted Kerr non-linearity introduced by the SQUID affects the generation of the trisqueezed state~\cite{chang_observation_2019}.

\subsubsection{Operational interpretation of the achieved fidelities}
\label{se:fidelity}

One natural question stemming at this point is: how does the non-unity fidelity of the generated cubic phase state affects a quantum computation? In this Section, we analyse how the output fidelity translates into a gate error when the cubic phase state generated with our protocol is used to implement a cubic phase gate onto an arbitrary state via gate-teleportation.

\begin{figure}[h]
\centering
\includegraphics[scale=0.9]{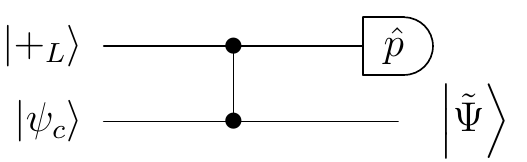}  
\caption{Sketch of the simplified gate-teleportation gadget to implement a cubic gate on an input GKP state $\ket{+_L}$ using an ancillary state  $\ket{\psi_c}$. The ancillary state  $\ket{\psi_c}$ can be either the cubic phase state that is the target of our conversion protocols, $\ket{\psi_c} = e^{i r \hat{q}^3} \hat S(\xi_{\rm target}) \ket{0}$, or the actual output mixed  state of our Gaussian conversion protocol $\rho_\textrm{cond}$, given in  Eq.(\ref{eq:cond-out}) for $n = 0$. In the former case, the corresponding output state is denoted $\vert \tilde{\Psi}\rangle$ while in the latter,
the output state is a mixed state $\tilde{\rho}$. We consider post-selection onto $p=0$.
}
\label{fig:gadget-simp}
\end{figure}

Consider the gate-teleportation gadget in Fig.~\ref{fig:gadget-simp}. We define the gate error $\epsilon$ as the infidelity $1 - F$ between (i) the output state of the gate-teleportation gadget when we use as ancillary state the target (perfect) cubic phase state of our protocol, or (ii) the output state of the gate-teleportation gadget when we use as ancillary state the actual output state of our conversion protocol (see Appendix~\ref{app:gate-error} for a formal definition and calculation). In this way, the  infidelity between our target and output  states is interpreted in an operational way. Since a similar analysis has shown to yield deceivingly small gate errors when the state onto which the gate is applied is a coherent state or a displaced squeezed state, following \cite{yanagimoto2019engineering} we use instead a hard instance of an arbitrary state, namely a GKP state in the encoded $|+\rangle$ state, which is expressed in the momentum representation as
\begin{align}
\label{eq:gkp-plus}
  \ket{+_L} = \frac{1}{N}  \int dp \sum_{s=-\infty}^\infty e^{-\frac{\Delta^2}{2} (2s)^2 \pi}  e^{-\frac{1}{2\Delta^2} (p-2s \sqrt{\pi})^2 } \ket{p},
\end{align} 
with $\Delta$ the variance of the individual peaks in the the GKP code~\cite{gottesman2001}, and $N$ a normalisation constant.

\begin{figure}[h]
\includegraphics[width=0.99\linewidth]{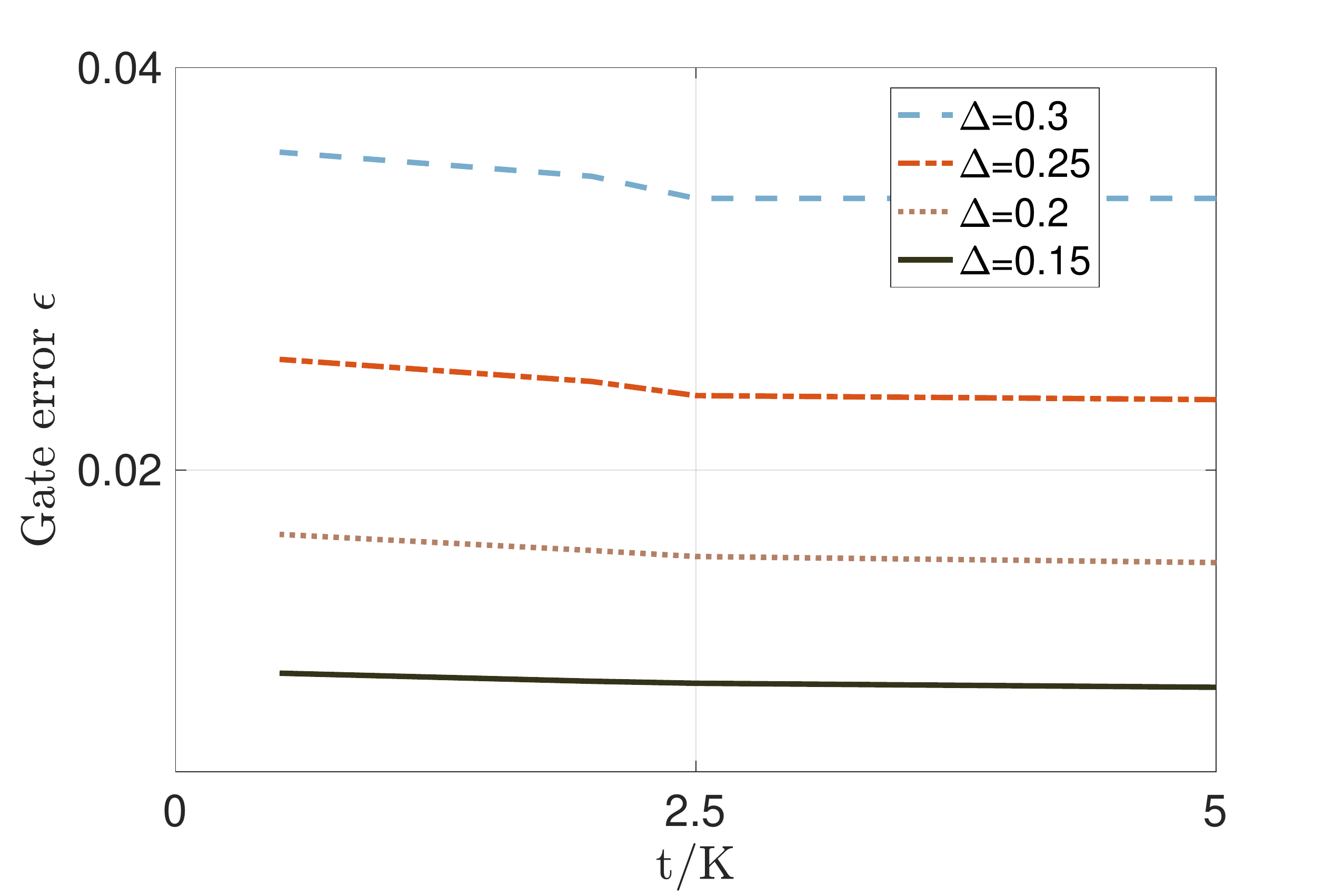}  
\caption{Gate error as a function of $t/K$. As the Kerr effect is stronger, namely the ratio of the triplicity and Kerr strength $t/K$ decreases, the error associated with using the cubic phase state at the output of our probabilistic conversion protocol in a teleportation gadget such as the one of Fig.\ref{fig:gadget-simp}, instead than a perfect cubic phase state,  increases.}
\label{fig:gateerror}
\end{figure}

A numerical plot of the gate error as a function of the infidelity of the ancillary cubic phase state is provided in Fig.~\ref{fig:gateerror}. The latter infidelity is accounted for by the residual Kerr interactions studied in Sec.~\ref{se:residual-kerr-sub}. From the plot, we  observe that the gate error does not increase significatively 
with a decreasing ratio $t/K$, i.e., with an increasing infidelity due to the Kerr effect in the initial trisqueezed state.

 We also observe that the gate error decreases when the squeezing of the GKP state on which we applied the gate upon increases or, equivalently, the variance $\Delta$ decreases.

Note that a recent preprint shows that implementing a T-gate on encoded GKP qubits via the use of a cubic phase state is unsuitable for reasonably squeezed GKP states~\cite{hastrup2020}. However, the purpose of our analysis here is to show that using  the state that is generated from our protocol for gate teleportation does  not introduce a significant discrepancy with respect to the use of a cubic phase state. Cubic phase gates through gate teleportation are still relevant in the context of non-encoded QC with continuous variables, e.g. within a CV-NISQ approach~\cite{hillmannUniversalGateSet2020}, to implement quantum gates beyond the regime of simulatable quantum computation.

\subsection{Optical systems}
\label{optics}

In quantum optics, availability of ancillary squeezed states characterized by squeezing parameters ranging from a few to 15dB~\cite{vahlbruch2016detection,cai2017}, beam-splitters and homodyne detection, i.e. the elements composing our probabilistic conversion protocol introduced in Sec.\ref{protocol2}, is well established. Hence, our probabilistic conversion protocol can be implemented with technology that is readily available in quantum optics labs. In-line squeezing, i.e., the application of a squeezing operation on a state different from the vacuum, as it is required by our first protocol, has been demonstrated~\cite{miwa2014exploring}, but is generally regarded as more challenging. In this sense, the probabilistic protocol presented in this paper appears to be easier to implement than the deterministic protocol when it comes to optical setups. 
Also note that implementability of the present protocols in optical devices  holds in contrast to the proposal of Ref.~\cite{hillmannUniversalGateSet2020}, which discusses the deterministic (gate-based) generation of a cubic phase state in the context of a specific microwave architecture.

Entanglement properties of triple-photon states - a three-mode version of our trisqueezed states, corresponding to the Hamiltonian $a^{\dagger}_1 a^{\dagger}_2 a^{\dagger}_3 +$ h.c. - have been studied theoretically in Ref.~\cite{gonzalez2018continuous}, while preliminary experimental results on the optical trisqueezed state have been reported in Ref.~\cite{bencheikh2007triple}. Third and higher-order processes in spontaneous parametric down conversion and other non-linear parametric interactions have also been analysed theoretically in Ref.~\cite{okoth2019seeded, zhang2020non}.

\section{Conclusions and perspective views}
\label{conclusions}

In this paper, we have studied two Gaussian conversion protocols that allow for the conversion of an experimentally available non-Gaussian state, namely the trisqueezed state, into a known resource state for universal quantum computation over continuous variables, the cubic phase state. 

Depending on the experimental set-up and on the needs, one or the other conversion method might be preferable. Our first protocol presents the advantage of being deterministic, while requiring in-line squeezing --- possible in micro-wave set-ups, while challenging in quantum optical ones. On the other hand, our second protocol is probabilistic, but achieves higher fidelities and could be implemented using off-line squeezing --- therefore feasible in various platforms, in particular both within optical and microwave set-ups.
The squeezing required in the two  protocols, relative to the conversion with highest fidelity, is of the order of 3.4 dB and 2.8 dB respectively, both achievable in either microwave or optical devices.

In Sec.\ref{se:fidelity} we have seen that it is possible to interpret operationally the infidelity of the generated state in terms of a gate error induced when using the generated state as an ancillary state to implement a non-Gaussian gate. However, in order to conclude unequivocally that the trisqueezed state is a universal resource, one needs to address the question as to whether in turn the resulting non-Gaussian gate, combined with Gaussian operations, allows one to implement fault-tolerant, universal quantum computation~\cite{gottesman2001}. This can be assessed using the framework of Ref.~\cite{douce2019}, where the cubic phase gate is used in combination to Gaussian resources in order to implement approximate GKP states in a self-consistent way. In this way, the use of qubit error-correction codes concatenated to the GKP code allows one to determine a target fidelity for the generated GKP states. This target fidelity, in turn, is translated into a requirement for the fidelity of the required cubic phase gates.
We leave this analysis for future work.

Finally, note that the approach that we have developed for the study of our probabilistic protocol, namely the calculation of the output fidelity provided in Appendix~\ref{appendix:output}, Eq.(\ref{eq:fidelity_result}), combined with the numerical optimization tools detailed in Appendix~\ref{appendix:numerics}, are valid for arbitrary input, ancillary state and target state. Therefore, our approach can be readily employed, upon replacement of the input and target wave functions, for the study of further arbitrary conversion protocols. 
The study of these extensions is an interesting perspective stemming from our work. Ultimately, Gaussian conversion protocols can shed light on the resourcefulness of generic non-Gaussian states for universal quantum computation. 

\section{Acknowledgements}

O. H. and Y. Z. contributed equally to this work.
We thank Chris Wilson, Timo Hillmann, Laura Garc\'{\i}a-\'{A}lvarez and Simone Gasparinetti for fruitful discussions. G.F. acknowledges support from the Swedish Research Council (Vetenskapsrådet) through the project grant QuACVA. F. Q., Y. Z., G. F. and O. H. acknowledge support from the  Knut and Alice Wallenberg Foundation through the Wallenberg  Center for Quantum Technology (WACQT).
The numerical computations were performed on resources at Chalmers Centre for Computational Science and Engineering provided by the Swedish National Infrastructure for Computing.

\appendix

\section{Gate Teleportation}
\label{appendix:Gate_teleportation}

In this Appendix we derive how the Gaussian feed-forward operations, necessary in order to implement a deterministic cubic phase gate via gate teleportation, are effected by the use of an ancillary trisqueezed state, instead than a cubic phase state. In order to derive these corrections, we use our deterministic conversion protocol developed in Sec.\ref{protocol1}.

\begin{figure}[h]
\includegraphics[scale=0.9]{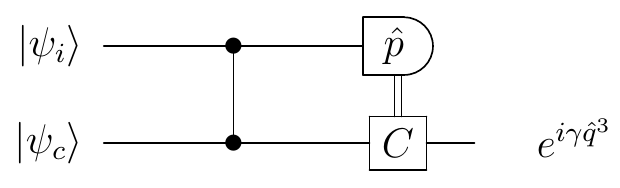}  
\caption{Sketch of the standard gate-teleportation gadget to implement the cubic phase gate on an input state $\ket{\psi_i}$ with the cubic-phase state $\ket{\psi_c} = e^{i r \hat{q}^3} \hat S(\xi_{\rm target}) \ket{0}$.}
\label{fig:gadget}
\end{figure}

\begin{figure}[h]
\includegraphics[scale=0.9]{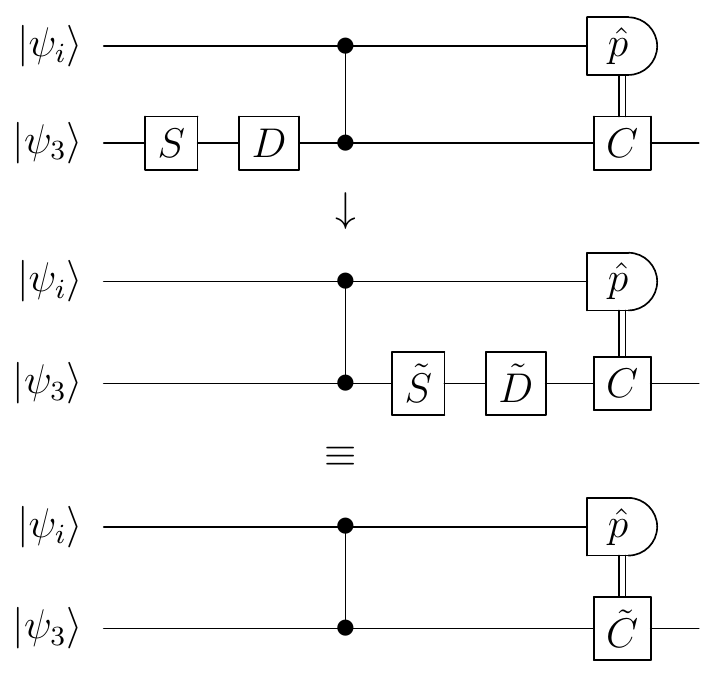}  
\caption{Sketch of the modified teleportation gadget to implement the cubic phase gate with using the trisqueezed state $\ket{\psi_3}= e^{i (t^* \hat{a}^3 + t \hat{a}^{\dagger 3})}  \ket{0}$. The top circuit  shows the gates needed to approximate the cubic-phase state with a triple squeezed state. The last two depict the changes needed in the feed-forward, with respect to the teleportation gadget of Fig.~\ref{fig:gadget}, if one wants to use the trisqueezed state directly. }
\label{fig:mod_gadget}
\end{figure}

The gadget used for gate teleportation can be seen in Fig.~\ref{fig:gadget}. When one wants to use the trisqueezed state as a resource for implementing a cubic phase gate, according to our symplectic conversion protocol a gadget such as the one shown in Fig.~\ref{fig:mod_gadget} can be used. However, it is also possible to commute the squeezing and displacement operators appearing in the top circuit  through the $\hat{C}_Z$ gate. Since these operators do not commute, this yields to modified Gaussian corrections, that can be obtained by computing the commutators between $\hat{C}_Z$ and $\hat{S}$, and $\hat{C}_Z$ and $\hat{D}$.

The operations are defined as:
\begin{align*}
    \hat{C}_Z &= e^{i \hat{q}\otimes \hat{q}}\\
    \hat{S} &= \mathds{1} \otimes e^{-i\frac{s}{2}( \hat{q}\hat{p} +  \hat{p}\hat{q} )}\\
    \hat{D} &= \mathds{1} \otimes e^{-i( q \hat{p} - p  \hat{q} )}
\end{align*}
By representing the operators as a series, it is easy to see that we only need to calculate the action of 
\begin{align*}
     e^{-i \hat{q}\otimes \hat{q}} (\mathds{1}\otimes \hat{p})  e^{i \hat{q}\otimes \hat{q}} &=\mathds{1} \otimes\hat{p} + [- i \hat{q}\otimes \hat{q}, \mathds{1}\otimes \hat{p} ]\\ 
     &= \mathds{1} \otimes\hat{p} + \mathds{1} \otimes \mathds{1}.
\end{align*}
Then we obtain
\begin{align*}
     \tilde{D} = \hat{C}_Z^\dagger  \hat{D} \hat{C}_Z &= \mathds{1} \otimes e^{-i (q ( (\hat{p} + \mathds{1}))- p  \hat{q}  )} \\
     & =  \mathds{1} \otimes e^{-i( q \hat{p} - p \hat{q} )}  e^{-i q}\\
\end{align*}
and
\begin{align*}
   \tilde{S} =    \hat{C}_Z^\dagger \hat{S} \hat{C}_Z  &= \mathds{1} \otimes e^{-i\frac{s}{2}(  \hat{q}(\hat{p} + \mathds{1}) + (\hat{p} + \mathds{1})\hat{q} )}\\
     &= \mathds{1} \otimes e^{-i\frac{s}{2}( \hat{q}\hat{p} +  \hat{p}\hat{q}+2 \hat{q} )}\\
     &=  \qty(\mathds{1} \otimes e^{-i\frac{s}{2}(  \hat{q}\hat{p} +  \hat{p}\hat{q} )}) \qty(\mathds{1 } \otimes e^{-is \hat{q}}) \qty(\mathds{1 } \otimes e^{-\frac{1}{2} [...,...]})\\
     &=  \qty(\mathds{1} \otimes e^{-i\frac{s}{2}(   \hat{q}\hat{p} +  \hat{p}\hat{q} )})  \qty(\mathds{1} \otimes e^{-i(s-\frac{s^2}{2} )  \hat{q}}).
\end{align*}
These operations can be merged onto the Gaussian feed-forward operations needed in the standard gate teleportation gadget, resulting in a modified Gaussian feed-forward operation.

\section{Details on the numerical optimization}
\label{appendix:numerics}

In both the probabilistic and deterministic protocols, the fidelity is a relatively expensive function to evaluate numerically, making the numeric optimization challenging. To tackle this challenge, we turned to high-performance computing, and tried different numerical optimization strategies. Furthermore, for each protocol, we developed three independent implementations that were benchmarked against each other, and against analytic calculations, to ensure numeric consistency. The first two codes were implemented in Python and C++ to run in serial on central processing units (CPUs), and the third code in CUDA~~\cite{CUDA2019,Nickolls2008} for high-performance computing and massive parallelization on NVIDIA Tesla V100 graphics processing units (GPUs). The Python code was run on personal computers (both laptop and desktop machines), while the C++ and CUDA codes were run both on personal computers and on a computer cluster.

For the probabilistic protocol, the Python code used the library GPyOpt~~\cite{gpyopt2016} and Bayesian optimization (BO)~~\cite{Kushner:1964,Mockus:1989}, while the C++ and CUDA codes used the library Thrust~~\cite{thrust1,thrust2,thrust3} and particle swarm optimization (PSO)~~\cite{kennedy:1995,yang:2011}. For the deterministic protocol, the Python code used the library QuTiP~~\cite{qutip1,qutip2} with BFGS optimization~~\cite{Broyden1970,Fletcher1970,Shanno1970,Goldfarb1970}, while the C++ and CUDA codes used the libraries Thrust, Armadillo~~\cite{Sanderson2016,Sanderson2018} and OptimLib~~\cite{OptimLib} with both particle-swarm optimization and differential evolution optimization (DE)~~\cite{Storn1997}. Both particle-swarm optimization and differential evolution are inspired by natural evolution, and were chosen because they are versatile methods with good performance in higher dimensions, and are easy to parallelize efficiently. In contrast to many quasi-Newton methods, they do not rely on the gradient of the objective function (the function to be minimized, i.e. one minus the fidelity), and can get out of local minima. Similarly, Bayesian optimization does not rely on the gradient, and was chosen as it is a powerful and popular method for global optimization. BFGS, which is a quasi-Newton method, was chosen for comparison. Some of these methods are described in greater detail further below.

In the end, all implementations gave the same results for the same choice of input parameters, and all the optimization methods eventually found the same maxima in the fidelity. The CUDA implementation managed to properly harness the performance of the GPUs~~\cite{Matthews2018}, and was therefore more than an order of magnitude faster than the C++ implementation (measured in number of fidelities evaluated per second), which in turn was more than an order of magnitude faster than the Python implementation.

Since the integrands in Eqs.~\eqref{eq:fidelity_det_protocol} and \eqref{eq:fidelity3} behave smoothly as a function of the integration parameters, the triple integrals in the deterministic and probabilistic Gaussian conversion protocol can be calculated using standard numeric integration. In both optimization algorithms, we limited the range of the optimization parameters according to Tab.~\ref{table:bounds}. The bounds for the displacement are limited by the probability, which decreases exponentially with the displacement of the input state.

In both optimization algorithms, we limited the range of the optimization parameters according to the following arguments. We adjust $\theta$ in between $[0, \pi/2]$ as the sign of $\cos{\theta}$ and $\sin{\theta}$ does not play a role in the fidelity and probability. Considering that the position range of our target states is around [-2,2], we choose [0, 1.5] as the displacement range of the ancillary displaced squeezed state $q_\beta$ and [0, 3] as the range for the displacement operator $d$. Since the value of the current record for squeezing is 15dB in quantum optics, we set $1.5$ as the bound of the squeezing strength in the ancillary squeezed state $\xi$, which corresponds to $13\,\text{dB}$.

\begin{table}[!htbp]
\begin{tabular}{|c|c|c|c|}
\hline
$\theta$& $q_\beta$&$\xi$&$d$\\
\hline
[0,$\pi/2$] &  [0,1.5]& [0,1.5] & [-3,0]\\
\hline
\end{tabular}
\caption{The bounds for optimizing parameters.}
\label{table:bounds}
\end{table}

The Bayesian and particle-swarm optimization strategies will now be explained in greater detail.

\subsection{Bayesian optimization}
Bayesian optimization~~\cite{Kushner:1964,Mockus:1989} (BO) is a global optimization algorithm which is applied for the search of optimal parameters in computationally expensive functions. The general BO algorithms iterates between function evaluations and predictions about optimal parameters, and terminates when a certain number of iterations has been executed. Writing the optimization parameters at iteration step $i$ in a vector $\boldsymbol{x}_i$, BO tries to minimize the number of function evaluations by carefully selecting the next point $\boldsymbol{x}_{i+1}$ where to compute the objective function. In each iteration step, BO considers the complete history of so far collected points $\boldsymbol{x}_{i}$ and function evaluations. 

The two main components of BO are (i) a prior probabilistic belief of an objective function and (ii) an acquisition functions~~\cite{ekstrom:2019}. The prior probabilistic belief of the objective function is in general sampled from a Gaussian process. The obtained value of the objective function is then used in the acquisition function, which determines the optimization parameters for the next position $\boldsymbol{x}_{i+1}$.  In our approach, we applied the square exponential kernel as a model of similarity in a Gaussian process, and the expected improvement criterion as acquisition function. The maximal number of iterations was obtained empirically by running the algorithm several times and benchmarking with the optimal values predicted by PSO. Our implementation uses the library GPyOpt~~\cite{gpyopt2016} for BO. 

\subsection{Particle swarm optimization}
Particle-swarm optimization (PSO) attempts to find the global maximum to an objective function by adjusting the trajectories of $N_{\textrm{PSO}}$ individual particles. Each particle is described by a position vector $\boldsymbol{x}_i$ whose components correspond to each of the optimization parameters $\theta,q_\beta,\xi$ and $d$. The particles are either distributed randomly or initialized on a grid in the landscape of optimization parameters. Additionally to the position vector, each particle is attributed with a velocity vector $\boldsymbol{v}_i$ that iteratively updates the particle's position. 

In standard PSO, the movement of particles depends on a stochastic and a deterministic component reflecting the tradeoff between exploration and exploitation. To move from a current position at iteration step $t$ to a next position at iteration step $t+1$, each particle is attracted to the global best particle $\boldsymbol{g}^*$ and its own best location $\boldsymbol{x}_i^*$ in its past trajectory, while the full update also contains random numbers $\varepsilon_1\in (0,1)$ and $\varepsilon_2\in (0,1)$. Introducing the learning parameters $\alpha$ and $\beta$, the velocity and position at iteration step $t+1$ follow from the equations~~\cite{yang:2011}
\begin{eqnarray}
  \boldsymbol{v}_i^{t+1} &=& \Omega_t \boldsymbol{v}_i^{t} + \alpha  \varepsilon_1 (\boldsymbol{g}^*-\boldsymbol{x}_i^t) + \beta  \varepsilon_2 (\boldsymbol{x}_i^*-\boldsymbol{x}_i^t)
\\
\boldsymbol{x}_i^{t+1}  &=& \boldsymbol{x}_i^{t} +\boldsymbol{v}_i^{t}  \, . \end{eqnarray}
where  $\Omega_t \in (0,1)$ is called the inertia function. 
In each iteration step, we update the global best particle  $\boldsymbol{g}^*$ and the best location in the history of each particle $\boldsymbol{x}_i^*$. Additionally, we ensure that the particle's positions stay within the boundaries of the optimization parameter. The PSO algorithm terminates when a predefined number of iterations $N_{\textrm{iter}}$ has been executed. 
The number $N_{\textrm{iter}}$  was empirically determined by running the simulation a few times for the same values, and observing that the maximum of the objective function converged to the same value with same parameters. After every run, we checked that most of the particles ended in the same position. We set the default value of  $N_{\textrm{iter}}$ to $10^3$. 

GPUs allow for a massively parallel implementation of the PSO algorithm. In our implementation, we addressed each particle $\boldsymbol{x}_i$ to a single thread on the GPU, such that a maximal number of $N_{\textrm{PSO}}=10^8$ particles can search in parallel for the optimal optimization parameters. The optimal fidelities and parameters in Fig.~\ref{fig:plot_mana} and Table~\ref{table:probabilistic} were computed with $N_{\textrm{iter}} = 10^2$, $\alpha = 0.05$ and $\beta=1.05$, and we set the inertia function to $\Omega_t = 0.5$. 
\section{Optimal parameters for the deterministic conversion protocol}
\label{appendix: Opt_res_det}

\begin{table*}[!htbp]
\centering
\begin{tabular}{|ccccccccccc|}
\hline
Triplicity & Cubicity(5dB) & $X_{11}$& $X_{12}$ & $X_{21}$ & $X_{22}$ & $Y_{11}$ & $Y_{12}$ & $Y_{22}$& $l_q$ & $l_p$ \\
\hline
0.1& 0.1558 & 1.4837 & 0.0004 & -0.0004 & 0.67400 & $2 \cdot 10^{-7}$ &  $-3 \cdot 10^{-10}$ & $8 \cdot 10^{-8}$ & $-9 \cdot 10^{-5}$ & 0.15865  \\
\hline
0.125 & 0.2757 &  1.2786& 0.0003 & -0.0001 & 0.7821 & $8 \cdot 10^{-7}$ & $1 \cdot 10^{-7}$ & $3 \cdot 10^{-7}$ & 0.0001 & 0.2275 \\
\hline
0.15 &0.4946&   1.0570 & -0.0005 & -0.0004 & 0.9461 & $3 \cdot 10^{-7}$ & $-7 \cdot 10^{-8}$ & $4 \cdot 10^{-8}$ & -0.0002 & 0.3031  \\
\hline
\end{tabular}
\caption{Optimized parameters for different triplicity, cubicity pairs for the Gaussian CP map.}
\label{tab:gauss_cp}
\end{table*}

In this Appendix we provide the result of the optimizations for the  Gaussian maps corresponding to  our deterministic conversion protocol introduced in Sec.\ref{section-all-maps}, in terms of the optimal parameters, that maximize the fidelity to the target state.

We start by noting that in order for speeding up the numerical calculation of the characteristic function, it is useful to rewrite Eq.(\ref{eq:generating-function}) using the Fock state basis, as 
\begin{align}
\label{eq:generating-function-bis}
    \chi_{\rho}(\Vec{r}) = \Tr{ \hat{D}(-\Vec{r}) \rho} =\sum^{\infty}_{n,n'=0}\rho_{nn'}\bra{n'}\hat{D}(-\Vec{r})\ket{n}.
\end{align}
The matrix elements of the displacement operator appearing at the RHS of Eq.(\ref{eq:generating-function-bis}) can now be written for  $m \geq n $ as~\cite{PhysRev.177.1857}
\begin{align}
\label{eq:exp1}
    \bra{m} \hat{D}(\alpha)\ket{n} =\sqrt{\frac{n!}{m!}}e^{-|\alpha|^2 /2}  \alpha^{m-n} L_n^{m-n}(|\alpha|^2)
\end{align}
and for $m \leq n $
\begin{align}
\label{eq:exp2}
    \bra{m} \hat{D}(\alpha)\ket{n} =\sqrt{\frac{m!}{n!}}e^{-|\alpha|^2 /2}  (-\alpha^*)^{m-n} L_m^{n-m}(|\alpha|^2),
\end{align}
where $L_m^{n-m}(|\alpha|^2)$ are the associated Laguerre polynomials.

The results for the parameters yielding optimization of general Gaussian CP maps are given in Table \ref{tab:gauss_cp}.

\subsection{Symplectic Transformation}

Here we show the symplectic transformations $X$ that stem out of our optimization and yield the maximal values of the fidelity in Table~\ref{tab:det_maps} of the main text. The subindices correspond to the triplicity of the input trisqueezed state:

\begin{align*}
    X_{0.1} = \begin{pmatrix}
    1.2324 & 2\cdot10^{-7}\\
    -4\cdot10^{-6}&  0.8114
    \end{pmatrix},
\end{align*}

\begin{align*}
    X_{0.125} = \begin{pmatrix}
  1.0002 &4\cdot10^{-8}\\
 4\cdot10^{-7} &0.9998
    \end{pmatrix},
\end{align*}

\begin{align*}
    X_{0.15} = \begin{pmatrix}
0.7976 &-1\cdot10^{-6} \\
 -5\cdot10^{-6} & 1.2538
    \end{pmatrix}.
\end{align*}

\section{Analytical derivations of the output state in the probabilistic protocol}
\label{appendix:output}

In this Appendix, we present the analytical derivation of the output state corresponding to 
the probabilistic protocol sketched in Fig.\ref{fig:circuit2}. 

The input state of this protocol is:
\begin{equation}
|\Psi_{\textrm{in}}\rangle|\Psi_{\xi,\beta}\rangle=
\int\, dq_1 \,dq_2\,\Psi_{\textrm{in}}(q_1)\Psi_{\xi,\beta}(q_2)|q_1,q_2\rangle   .
 \label{input}
\end{equation}
The wave function of a general ancillary displaced squeezed state Eq.(\ref{eq:input-squeezed-state}) is given by~\cite{donodov2003}
\begin{eqnarray}
&& \Psi_{\xi,\beta} (q) 
=\langle q \ket{\Psi_{\xi,\beta}}\nonumber\\
=&& \langle q | e^{\beta \hat a^{\dagger}-\beta^* \hat a} e^{\frac{\xi^*}{2} \hat a^2 -\frac{\xi}{2} \hat a^{\dagger 2}} |0 \rangle \nonumber\\
=&& \left( \frac{2}{\pi}\right)^\frac{1}{4} \frac{(1 - |\xi(\xi)|^2)^\frac{1}{4}}{\sqrt{1-\zeta(\xi)}} e^{-\frac{1+ \zeta(\xi)}{1- \zeta(\xi)}(q-q_\beta)^2+ 2 i p_\beta (q - \frac{q_\beta}{2})},
\label{eq:squeeze}
\end{eqnarray}  
where $\vert q \rangle$ is an eigenstate of the quadrature operator $\hat q$ with real eigenvalue $q$,
$\zeta(\xi) = \xi   \tanh |\xi|/|\xi|$,
and we have introduced the notation $\beta = q_\beta + i p_\beta$. 
Note that in the case of a real squeezing parameter, the wave function of a displaced squeezed state reduces to~\cite{leonhardt2010essential}
\begin{align}
& \Psi_{|\xi|,\beta} (q)= \left(\frac{2}{\pi}\right)^{\frac{1}{4}}e^{|\xi|/2} \nonumber \\
& \exp{(-e^{2|\xi|}(q-q_\beta)^2 + i2p_\beta q -i p_\beta q_\beta)}.
\end{align} 

After the real-valued beam-splitter transformation $U_{BS}^R(2\theta)$, we have
\begin{align}
\Psi_{12}(\xi,\beta,\theta)\rangle &= U_{BS}^R(2\theta)|\Psi_{\textrm{in}}\rangle \ket{\Psi_{\xi,\beta}} \nonumber \\
&=  \int\, dq_1 \,dq_2\,\Psi_{\textrm{in}}(q_1)\Psi_{\xi,\beta}(q_2)\,\,|q_1'\rangle|q_2'\rangle,
 \label{output1}
\end{align}
where
\begin{equation}
|q_1'\rangle=|q_1\cos{\theta}-q_2\sin{\theta}\rangle    ,
\label{eq:q1'}
\end{equation}
\begin{equation}
|q_2'\rangle=|q_1\sin{\theta}+q_2\cos{\theta}\rangle    .
\end{equation}
Indicating with $J(q_1',q_2')$ the Jacobian of the transformation, we have
\begin{equation}
\begin{split}
\int dq_1 dq_2 F(q_1,q_2)=\int dq_1'dq_2'| J(q_1',q_2')|F(q_1',q_2'),
\label{eq:J}
\end{split}
\end{equation}
where with a slight abuse of notation $F(q_1',q_2') = F((q_1(q_1',q_2'),(q_2(q_1',q_2'))$. 
Here we have:
$$|J(q_1',q_2')|=\begin{vmatrix} \frac{\partial q_1}{\partial q_1'} & \frac{\partial q_1}{\partial q_2'} \\ \frac{\partial q_2}{\partial q_1'}  & \frac{\partial q_2}{\partial q_2'}  \end{vmatrix}=\begin{vmatrix}
\cos{\theta}& \sin{\theta} \\\ -\sin{\theta}  & \cos{\theta} \end{vmatrix}=1.$$
Hence, we can rewrite the state in Eq.(\ref{output1}) as:
\begin{equation}
\begin{split}
|\Psi_{12}(\xi,\beta,\theta)\rangle=&\int\, dq_1 d q_2 \Psi_{\textrm{in}}(q_1\cos{\theta}+q_2\sin{\theta})
\\&
\Psi_{\xi,\beta}(-q_1\sin{\theta}+q_2\cos{\theta})
|q_1\rangle|q_2\rangle,
\label{outputstate}
\end{split}
\end{equation}
where we have renamed $q_1' \rightarrow q_1$, $q_2'\rightarrow q_2$.

After the phase rotation $U_p(\gamma)$ on state $|\Psi_{12}(\xi,\beta,\theta)\rangle$, using the closure relation $(1/\pi)\int |\alpha\rangle\langle\alpha|d\alpha^2=I$
we obtain the state:
\begin{equation}
\begin{split}
&U_p(\gamma)|\Psi_{12}(\xi,\beta,\theta)\rangle\\=
&\frac{1}{\pi}
\int\, dq_1 \,dq_2\,\,d\alpha\,
 \Psi_{\textrm{in}}(q_1\cos{\theta}+q_2\sin{\theta})
\\&
\Psi_{{\xi,\beta}}(-q_1\sin{\theta}+q_2\cos{\theta})\langle\alpha|q_2\rangle\,\, |q_1\rangle |\alpha e^{-i\gamma}\rangle
\\
\equiv&
|\Psi_{12}(\xi,\beta,\theta,\gamma)  \rangle,
 \label{output2}
 \end{split}
\end{equation}
where
\begin{equation}
    \langle \alpha|q_2\rangle=\left(\frac{2}{\pi}\right)^{1/4}e^{iab}e^{-2ibq_2}e^{-(q_2-a)^2},
\end{equation}
and $\alpha=a+ib$. 
 
When we measure $\hat{q}$ on the $1^{st}$ mode with the outcome $q$, we obtain on the second mode: 
\begin{equation}
\begin{split}
&|\Psi^{q}_{\textrm{out}}\rangle\\
=&
\langle q|\Psi_{12}(\xi,\beta,\theta,\gamma)\rangle \\
=& \frac{1}{\pi}
\int\, \,dq_2\,\,d\alpha
 \Psi_{\textrm{in}}(q\cos{\theta}+q_2\sin{\theta})
  \\ &
 \Psi_{{\xi,\beta}}(-q\sin{\theta}+q_2\cos{\theta})\,\, \langle\alpha|q_2\rangle |\alpha e^{-i\gamma}\rangle.
 \label{app:eq:output2}
\end{split}
\end{equation}

We now introduce the finitely resolved homodyne operator~\cite{paris2003, douce2017}
\begin{equation}
 \hat{Q}_n=\int_{q_n-\delta}^{q_n+\delta}dq |q\rangle\langle q|.
\end{equation}

The density matrix operator $\hat{\rho}_{n,{\rm cond}}$ on mode 2 conditioned on the measurement outcome $q_n$ on mode 1 can be expressed in terms of \eqref{app:eq:output2} and is given by
\begin{equation}
\begin{split}
&\hat{\rho}_{n,{\rm cond}}
\\
=&\frac{\textrm{Tr}_1[\hat{Q}_n\otimes I_2|\Psi_{12}(\xi,\beta,\theta,\gamma)\rangle\langle\{\Psi_{12}(\xi,\beta,\theta,\gamma)|\hat{Q}_n\otimes I_2]}{\textrm{Prob}[q_n]}\\
=&\frac{\int^{q_n+\delta}_{q_n-\delta} dq \quad  _1\langle q| \Psi_{12}(\xi,\beta,\theta,\gamma)\rangle\langle\Psi_{12}(\xi,\beta,\theta,\gamma)|q\rangle _1}{\textrm{Prob}[q_n]}\\
=&\frac{1}{\textrm{Prob}[q_n]}\int^{q_n+\delta}_{q_n-\delta} dq  |\Psi^{q}_{\textrm{out}}\rangle\langle\Psi^{q}_{\textrm{out}}| ,
 \label{eq:prob}
 \end{split}
\end{equation}
where $\textrm{Tr}_1$ is the partial trace over mode 1, and the probability of obtaining an outcome $q_n$ is expressed as:
\begin{equation}
\begin{split}
&\textrm{Prob}[q_n]\\
=&\bra{\Psi_{12}(\xi,\beta,\theta,\gamma)}\hat{Q_n}\otimes I_2|\Psi_{12}(\xi,\beta,\theta,\gamma)\rangle\\
=&\int^{q_n+\delta}_{q_n-\delta}dq\langle\Psi_{12}(\xi,\beta,\theta,\gamma)|q\rangle\langle q |\Psi_{12}(\xi,\beta,\theta,\gamma)\rangle\\
=&\frac{1}{\pi^2}  \int^{q_n+\delta}_{q_n-\delta} d q\int\,  dq_2 dq_2' d\alpha d\alpha' \, \Psi_{\textrm{in}}^{\ast}(q\cos{\theta}+q_2'\sin{\theta})\\
& \Psi_{\textrm{in}}(q\cos{\theta}+q_2\sin{\theta})
\Psi_{\xi,\beta}(q_2\cos{\theta}-q\sin{\theta})
\\&
\Psi_{\xi,\beta}^{\ast}(q_2'\cos{\theta}-q\sin{\theta})
\langle q_2'|\alpha'\rangle
\langle \alpha|q_2\rangle
\langle\alpha'
|\alpha\rangle \\
=&\int^{q_n+\delta}_{q_n-\delta} d q\int\,  dq_2 \, |\Psi_{\textrm{in}}(q\cos{\theta}+q_2\sin{\theta})|^2 \\
&|\Psi_{\xi,\beta} (q_2\cos{\theta}-q\sin{\theta})
|^2,
 \label{prob}\
 \end{split}
\end{equation}
where we have used that
\begin{equation}\begin{split}
& \int d\alpha d\alpha'\langle q_2'|\alpha'\rangle
\langle \alpha|q_2\rangle
\langle\alpha'
|\alpha\rangle  
\\
=&
\int d\alpha d\alpha'  \left(\frac{2}{\pi}\right)^{1/2}e^{iab}e^{-2ibq_2}e^{-(q_2-a)^2}
\\&
e^{-ia'b'}e^{2ib'q_2'}e^{-(q_2'-a')^2}\\
&\exp{(-\frac{|\alpha'|^2}{2}-\frac{|\alpha|^2}{2}+\alpha'^{\ast}\alpha)}\\
=&
\pi^2e^{-\frac{1}{2}(q_2-q_2')^2}\delta(q_2-q_2').
\end{split}
\end{equation}

Our fidelity can be written as
\begin{equation}
\begin{split}
F_{q_n}=&\langle\Psi_{\textrm{target}}| \hat{\rho}_{n,{\rm cond}}  |\Psi_{\textrm{target}}\rangle
\\
=&\frac{1}{\textrm{Prob}[q_n]}
\int^{q_n+\delta}_{q_n-\delta}\, \,dq \,|\langle\Psi_{\textrm{target}}|\Psi^q_{\textrm{out}}\rangle|^2,
 \label{fidelity}
 \end{split}
\end{equation}
where the output state is given in Eq.(\ref{app:eq:output2}). We can write the overlap as:
\begin{equation}
\begin{split}
&
\langle\Psi_{\textrm{target}}|\Psi^q_{\textrm{out}}\rangle \\ 
= &
\frac{1}{\pi}\int dq_0 \, dq_2 \,d\alpha \, \Psi_{\textrm{target}}^*(q_0)\Psi_{\textrm{in}}(q\cos{\theta}+q_2\sin{\theta})
\\&
\Psi_{\xi,\beta}(-q\sin{\theta}+q_2\cos{\theta})
\langle q_2|\alpha\rangle^{\ast}\langle q_0|\alpha e^{-i\gamma}\rangle ,\\
\label{eq:overlap}
\end{split}
\end{equation}
where 
\begin{equation}
\begin{split}
&\int d\alpha\langle q_2|\alpha\rangle^{\ast}\langle q_0|\alpha e^{-i\gamma}\rangle\\
=&
\sqrt{\frac{2}{\pi}}\pi\frac{1}{\sqrt{1-e^{-2i\gamma}}}
\\&
\exp(i\csc{\gamma}(-2q_0q_2+(q_0^2+q_2^2)\cos{\gamma})) \, (\textrm{for } \gamma\ne0),
\label{eq:F}
\end{split}
\end{equation}
and where the wave function of the target cubic phase state Eq.(\ref{eq:target_state}) is easily computed as
\begin{equation}
\begin{split}
\label{eq:target}
\Psi_{\textrm{target}}(q) = \langle q \ket{\Psi_{\textrm{target}}} = \left( \frac{2}{\pi}\right)^\frac{1}{4}e^{\xi_{\textrm{target}}/2} e^{-e^{2\xi_{\textrm{target}}}q^2}e^{irq^3}.
\end{split}    
\end{equation}
The wave function of the displaced target cubic phase state is then expressed as
\begin{equation}
\begin{split}
\Psi_{\textrm{target}}(q,d) = \left( \frac{2}{\pi}\right)^\frac{1}{4}e^{\xi_{\textrm{target}}/2} e^{-e^{2\xi_{\textrm{target}}}q^2}e^{irq^3}e^{-iqd}.
\end{split}    
\end{equation}
Finally, we obtain the expression for the fidelity by combining Eqs.~(\ref{eq:prob}), (\ref{eq:overlap}) and (\ref{eq:F}):
\begin{equation}
\begin{split}
&F_{q_n}(q_n,\delta,\gamma,\theta,\xi,\beta)=\frac{\int^{q_n+\delta}_{q_n-\delta}\, \,dq \,|\langle\Psi_{\textrm{target}}|\Psi^q_{\textrm{out}} \rangle|^2}{\textrm{Prob}[q_n]},
 \label{eq:fidelity_result}
\end{split}    
\end{equation}
where we have explicated the dependence on the squeezed state parameters, and where
\begin{equation}
\begin{split}
&\int^{q_n+\delta}_{q_n-\delta}\, \,dq \,|\langle\Psi_{\textrm{target}}|\Psi^q_{\textrm{out}}\rangle|^2\\
=&
\int^{q_n+\delta}_{q_n-\delta} dq|\int dq_0dq_2 \Psi_{\textrm{target}}^*(q_0) \Psi_{\textrm{in}}(q\cos{\theta}+q_2\sin{\theta})
\\&
\Psi_{\xi,\beta}(-q\sin{\theta}+q_2\cos{\theta})\\
&
\sqrt{\frac{2}{\pi}}e^{i\csc{\gamma}(-2q_0q_2+(q_0^2+q_2^2)\cos{\gamma})}/\sqrt{1-e^{-2i\gamma}}|^2.
 \label{Eq:up}
\end{split}    
\end{equation}

\section{Which parameters control which property of the output state?}
\label{parameters-control}

\begin{figure}[h]
\begin{subfigure}{.235\textwidth}
  \centering
  \includegraphics[width=1\linewidth]{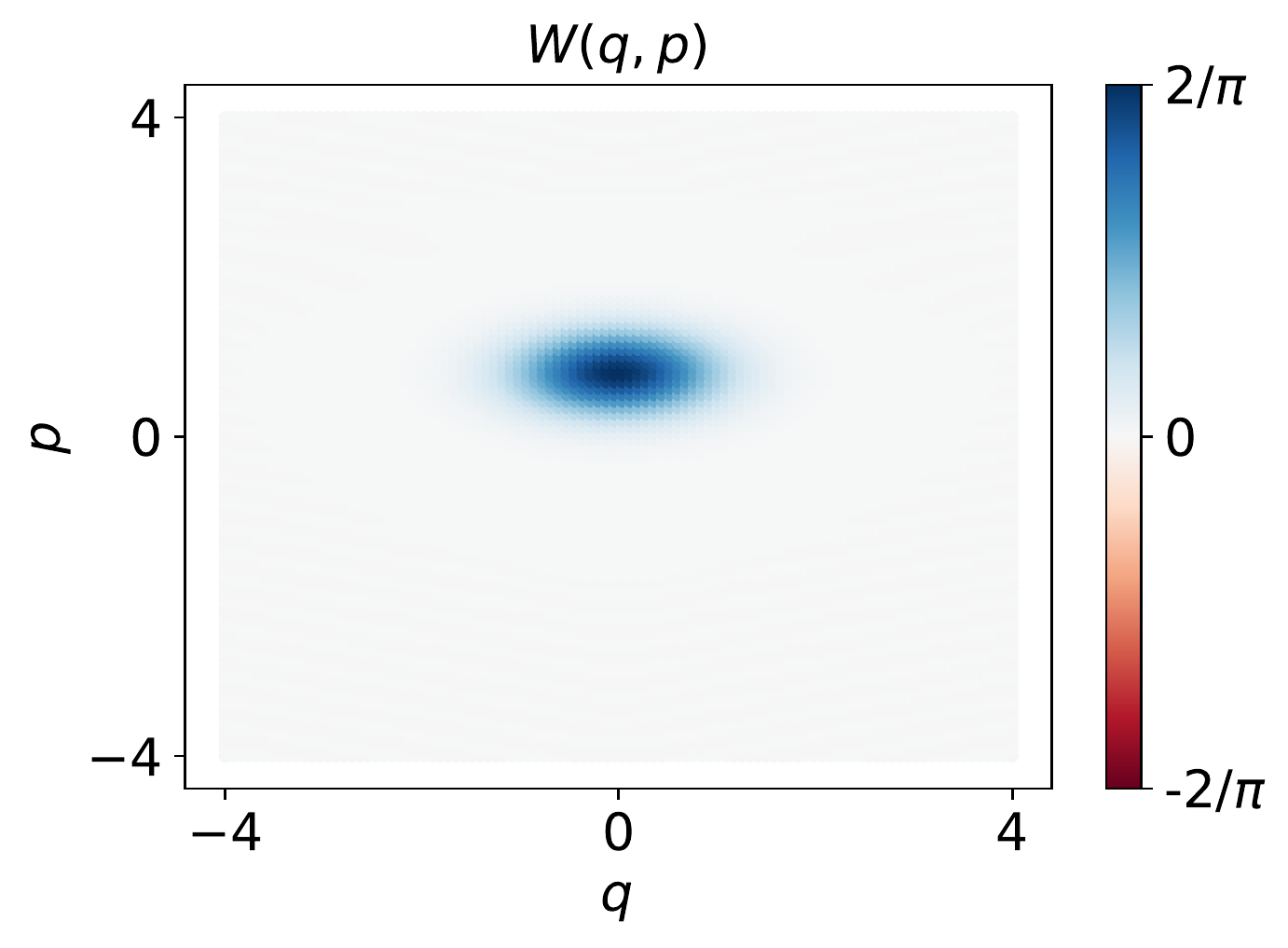}  
  \caption{ \centering{$\theta=0$}}
\end{subfigure}
\begin{subfigure}{.235\textwidth}
  \centering
  \includegraphics[width=1\linewidth]{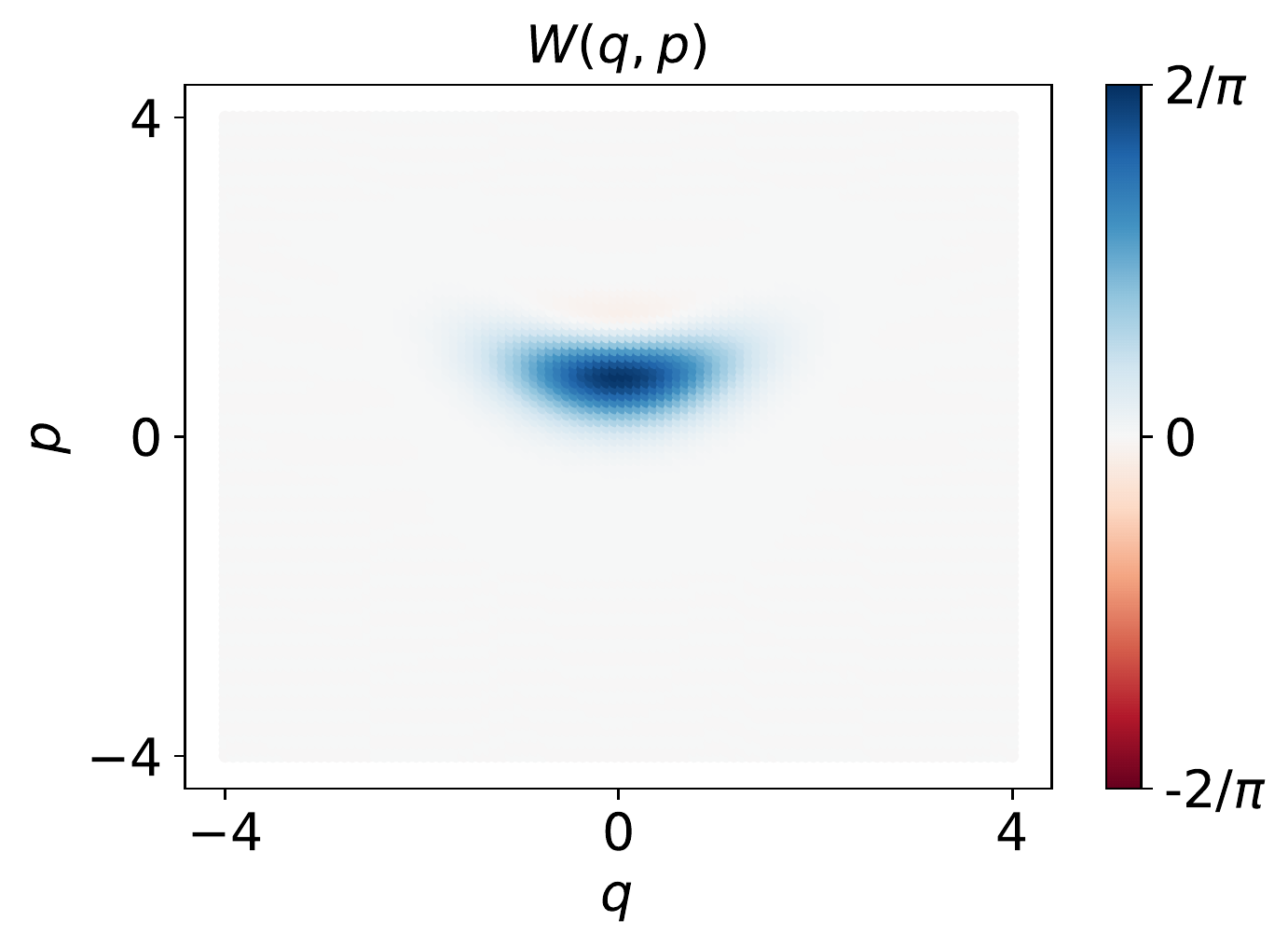} 
    \caption{ \centering{$\theta=\pi/4$}}
\end{subfigure}
\begin{subfigure}{.235\textwidth}
  \centering
  \includegraphics[width=1\linewidth]{output_pro.pdf}
    \caption{ \centering{$\theta=0.8987$ }}
    \label{fig:theta_op}
\end{subfigure}
\begin{subfigure}{.235\textwidth}
  \centering
  \includegraphics[width=1\linewidth]{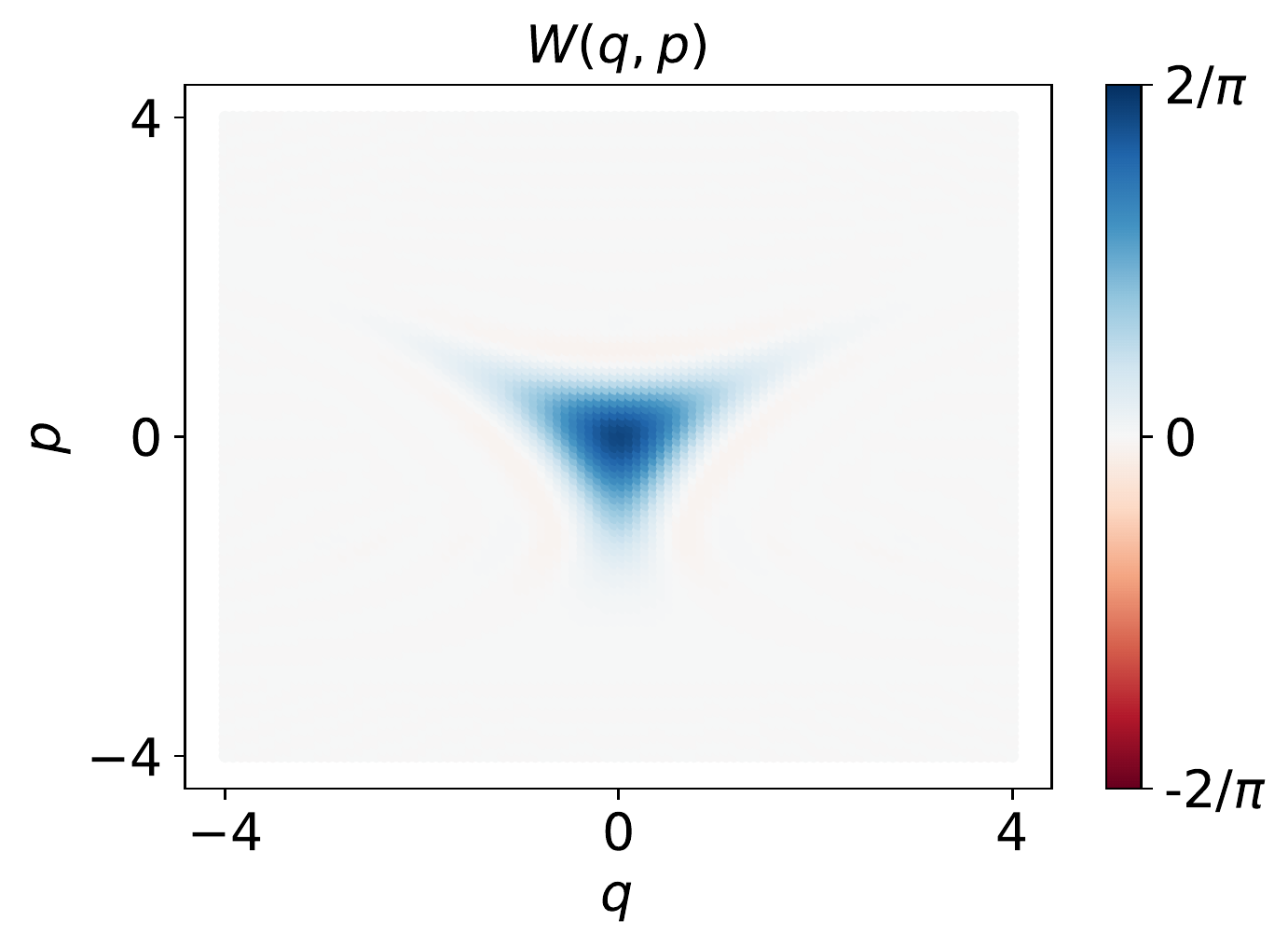} 
    \caption{ \centering{$\theta=\pi/2$}}
\end{subfigure}
\caption{
Wigner function of the output state while changing $\theta$ from 0 to 90 degrees. The other parameters are shown in the first row of Table.~\ref{table:probabilistic}. Notice that Fig.~\ref{fig:theta_op} corresponds to the optimal result.}
\label{fig:theta}
\end{figure}

\begin{figure}[h]
\begin{subfigure}{.235\textwidth}
  \centering
  \includegraphics[width=1\linewidth]{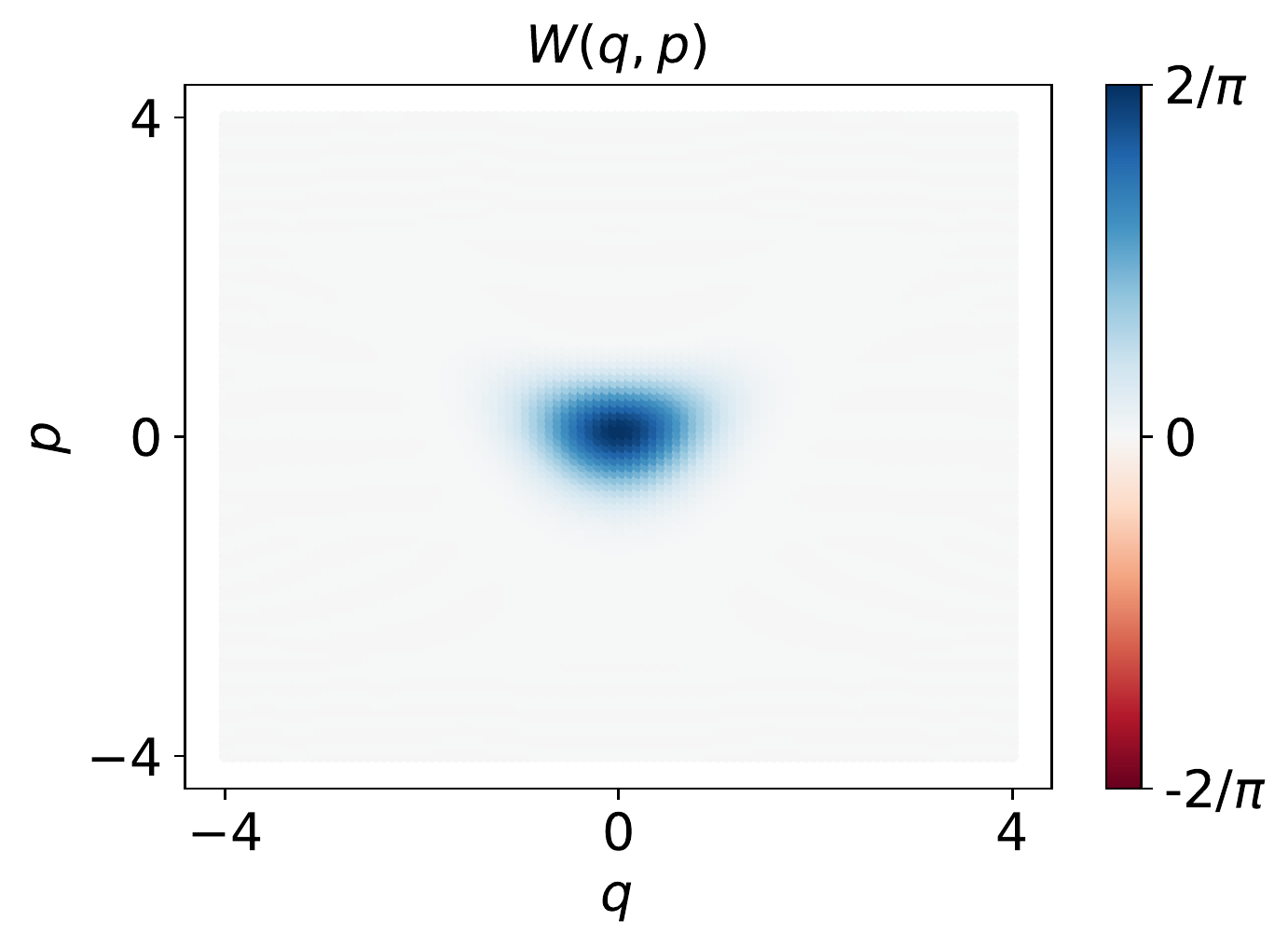} 
    \caption{ \centering{$q_{\beta}=0$}}
\end{subfigure}
\begin{subfigure}{.235\textwidth}
  \centering
  \includegraphics[width=1\linewidth]{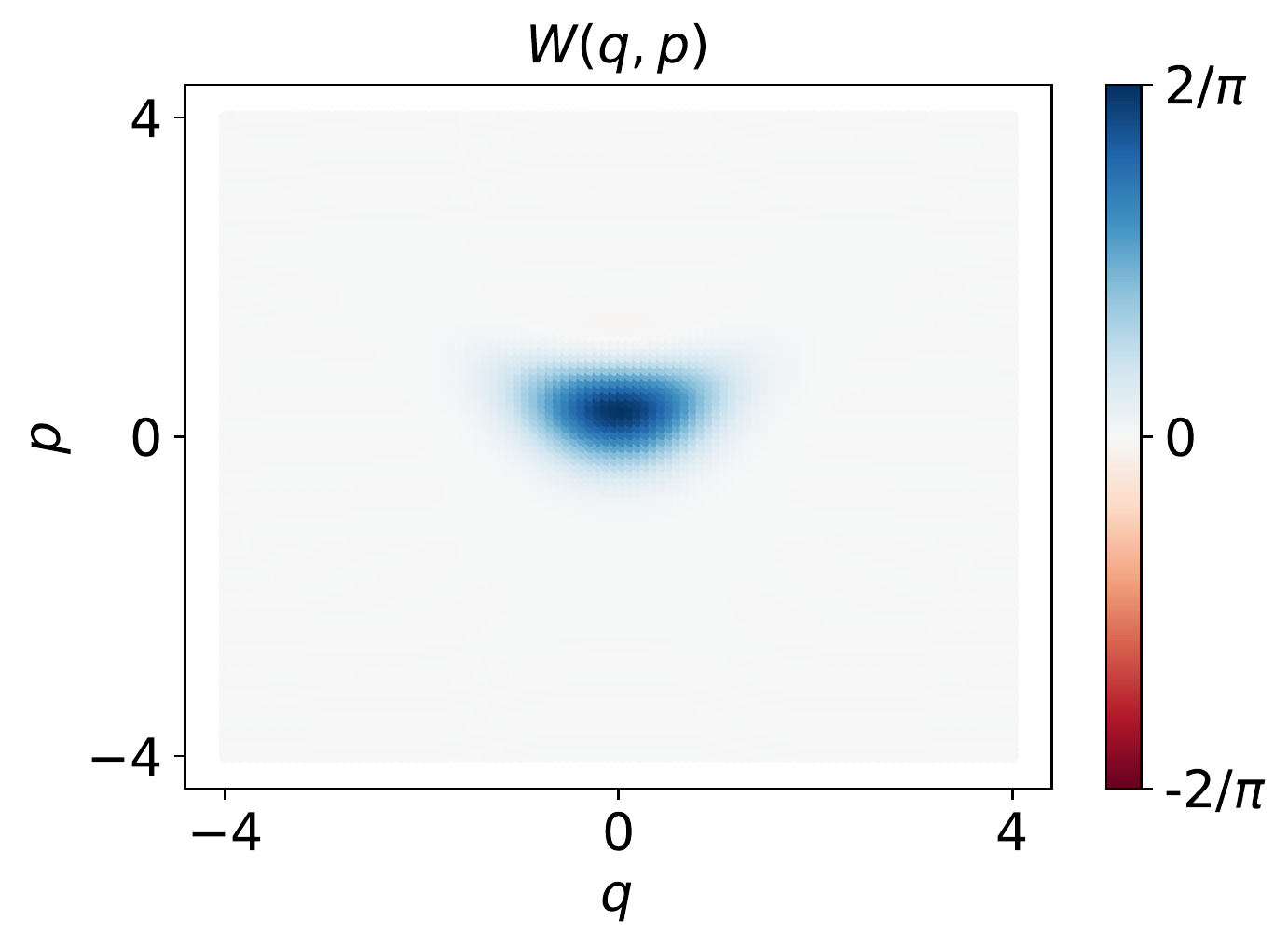}
      \caption{ \centering{$q_{\beta}=0.3$}}
\end{subfigure}
\begin{subfigure}{.235\textwidth}
  \centering
  \includegraphics[width=1\linewidth]{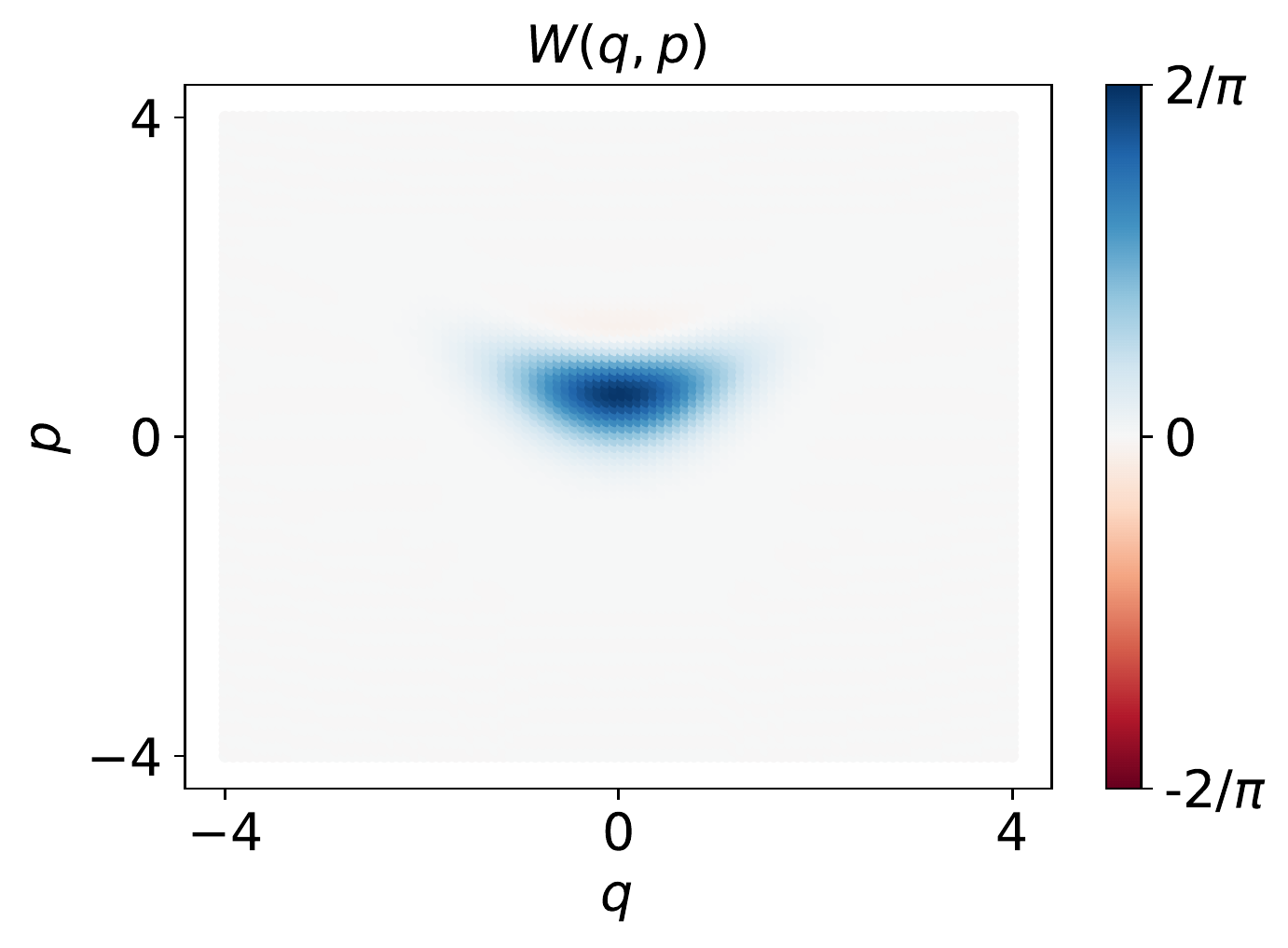}  
        \caption{ \centering{$q_{\beta}=0.6$}}
\end{subfigure}
\begin{subfigure}{.235\textwidth}
  \centering
  \includegraphics[width=1\linewidth]{output_pro.pdf} 
        \caption{ \centering{$q_{\beta}=0.95$}}
        \label{fig:displacement_op}
\end{subfigure}
\caption{
The Wigner function of the output states 
for different values of the parameter
$q_\beta$, corresponding to the displacement in position of the input ancillary displaced squeezed state. The other parameters correspond to those in the first row of Table.~\ref{table:probabilistic}. Also notice that Fig.~\ref{fig:displacement_op} corresponds to the optimal result. }
\label{fig:q3}
\end{figure}

\begin{figure}[h]
\begin{subfigure}{.235\textwidth}
  \centering
  \includegraphics[width=1\linewidth]{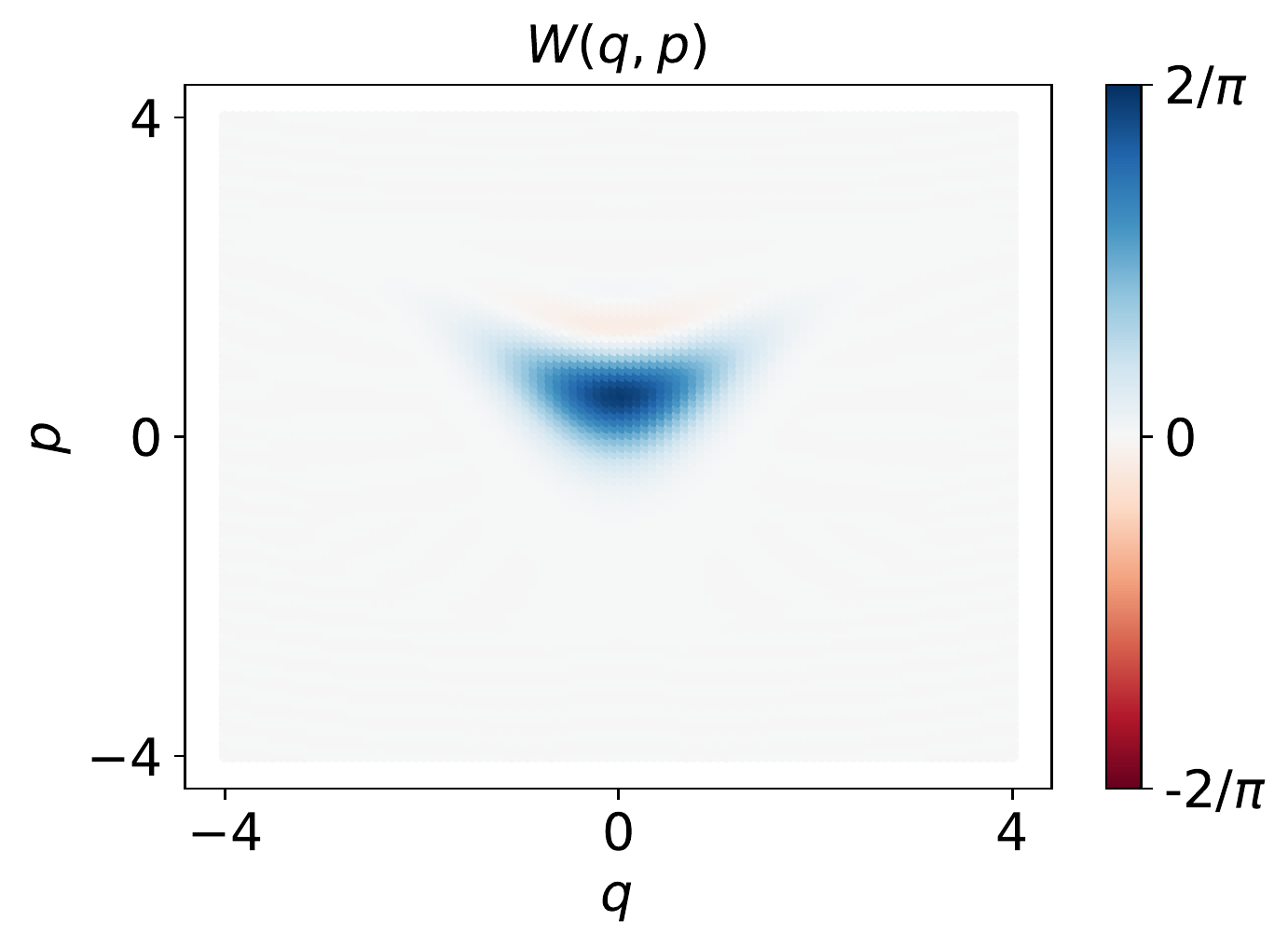} 
    \caption{ \centering{$\xi=0.1\, \textrm{dB}$}}
\end{subfigure}
\begin{subfigure}{.235\textwidth}
  \centering
  \includegraphics[width=1\linewidth]{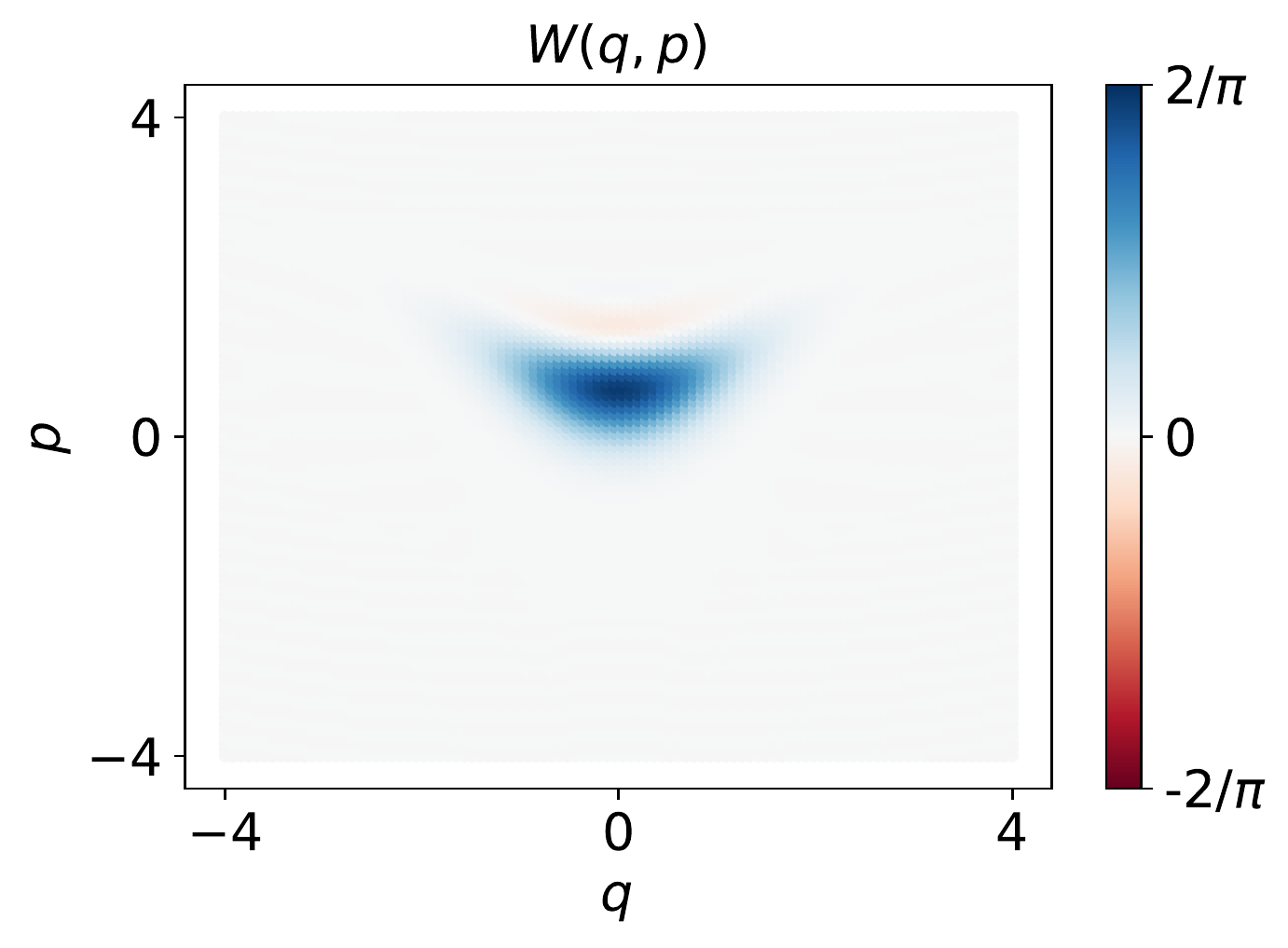}
      \caption{ \centering{$\xi=0.14\, \textrm{dB}$}}
\end{subfigure}
\begin{subfigure}{.235\textwidth}
  \centering
  \includegraphics[width=1\linewidth]{output_pro.pdf}  
        \caption{ \centering{$\xi=2.83\, \textrm{dB}$}}
        \label{fig:xi_op}
\end{subfigure}
\begin{subfigure}{.235\textwidth}
  \centering
  \includegraphics[width=1\linewidth]{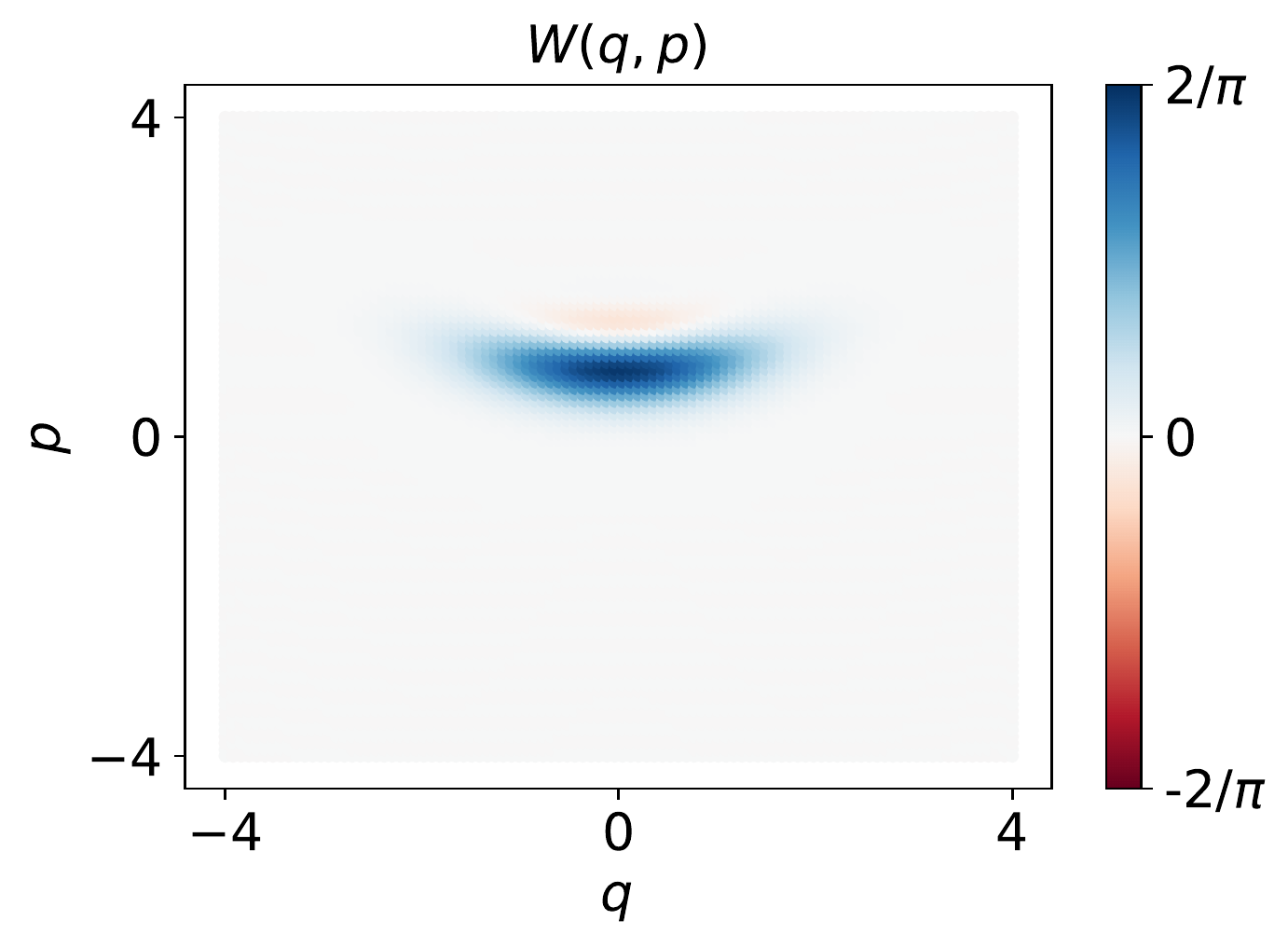} 
        \caption{ \centering{$\xi=5.66\, \textrm{dB}$}}
        
\end{subfigure}
\caption{
Wigner function of the output states 
for different values of the parameter
$\xi$, corresponding to the squeezing strength of the input ancillary displaced squeezed state. The other parameters correspond to those in the first row of Table.~\ref{table:probabilistic}. Also notice that Fig.~\ref{fig:xi_op} corresponds to the optimal result.} 
\label{fig:xi}
\end{figure}

The symmetry of the Wigner function plays an important role  in this optimization problem, and has a lot of applications, for instance in the design of rotationally-symmetric bosonic codes~\cite{grimsmo2019}.
We now discuss, relative to the second protocol that we have introduced, the relationship between the tunable parameters of our protocol and the Wigner function of the output state.

Figure~\ref{fig:theta} shows that the amount of negativity in the output state increases with increasing $\theta$. This happens because 
 $\cos{\theta}$ and $\sin{\theta}$ are the transmission and reflection coefficients of the beam-splitter, respectively. If the trisqueezed state is mostly reflected, then it is expected that the negativity in the output will be negligible. 
 The cubicity in the output state on the lower rail, therefore, will be proportional to $\sin{\theta}$.  
 
 Another important property is revealed by Fig.~\ref{fig:q3}. The Wigner negativity 
 varies  with 
 the displacement parameter of the ancillary displaced squeezed state: it becomes larger when the squeezed state is displaced further from the original point in the position direction.  Similarly, the direction of the squeezing in the ancillary displaced squeezed state, Arg$(\xi)$, affects the curvature of the main negative fringes. The width of the Wigner function is instead influenced by the strength of the parameter $\xi$.

Finally, Fig.~\ref{fig:xi} shows how the squeezing parameter in the ancillary displaced squeezed state impacts the output state. By increasing the squeezing parameter $\xi$, the output state becomes more squeezed.


\section{Interpretation of the probabilistic protocol}
\label{ap:sec:interpre}
\label{app:mbqc-interpretation}
In this Appendix, we aim at showing that the effect of the probabilistic protocol is to implement a deformed (filtered) squeezing on the input state, extending the findings of Sec.\ref{protocol1}.
For simplicity, we ignore the final phase rotation  and the output state in Eq.(\ref{app:eq:output2}) can thus be written as 
\begin{equation}
\begin{split}
&|\Psi^{q}_{\textrm{out}}\rangle = \langle q|\Psi_{12}(\xi,\beta,\theta,\gamma)\rangle \\
=&  
\int\, \,dq_2\,\,
 \Psi_{{\xi,\beta}}(-q\sin{\theta}+q_2\cos{\theta}) \Psi_{\textrm{in}}(q\cos{\theta}+q_2\sin{\theta})\,\,|q_2\rangle\\
 =&  
\int\, \,dq_2\,\,
 \Psi_{{\xi,\beta}}(-q\sin{\theta}+ q_2\cos{\theta}) \Psi_{\textrm{in}}(q\cos{\theta}+q_2\sin{\theta} )
\,\,|q_2\rangle \\ 
 =&   \Psi_{{\xi,\beta}}(\cos{\theta} \hat{q}_2-q\sin{\theta})\,\,
  \\
 &
\int\, \,dq_2\,\,
\Psi_{\textrm{in}}(\sin{\theta} (q_2+\frac{q\cos{\theta}}{\sin{\theta}}))\ket{q_2}  \\ 
 =& \Psi_{{\xi,\beta}}(\cos{\theta} \hat{q}_2-q\sin{\theta})\,\,
 \\
 &
 \hat{D} (d_c)\hat{S}(s_c)
\int\, \,dq_2\,\,
\Psi_{\textrm{in}}(q_2)\ket{q_2}
\\
 =& \Psi_{{\xi,\beta}}(\cos{\theta} \hat{q}_2-q\sin{\theta})\,\,
 \hat{D} 
 (d_c) \hat{S}(s_c)
 \label{ap:eq:interpretation of pro}
\end{split}
\end{equation}
where $d_c \propto \frac{\cos{\theta}}{\sin{\theta}}$, $|s_c| \propto |(\sin{\theta}-1)|$, and $\Psi_{{\xi,\beta}}(\cos{\theta} \hat{q}_2-q\sin{\theta})$ is a Gaussian filter~\cite{Sabapathy2018,sabapathy2019production,gu2009quantum}.
Note that for $q = 0$ we obtain a zero displacement. This situation is reminiscent of gate teleportation, with however an additional filtering factor.

\section{Derivation of the gate error from the infidelity of the ancillary cubic phase state}
\label{app:gate-error}

We want to calculate the wavefunction of a GKP state $\ket{+_L}$ after the teleportation gadget shown in Fig. \ref{fig:gadget-simp}. Consider first the case where the ancillary state is the cubic phase state given in Eq.(\ref{eq:target_state}), i.e. the target of our Gaussian conversion protocol. The wave function after the $\hat{C}_Z$-gate is the given as
\begin{align}
\begin{split}
    \ket{\Psi} &= \hat{C}_Z \ket{+_L} \ket{\Psi_{\textrm{target}}}\\
    &=\int dq_1 dq_2 \bra{q_1}\ket{+_L} \bra{q_2}\ket{\Psi_{\textrm{target}}} e^{i q_1 q_2}\ket{q_1} \ket{q_2}\\
    &= \int dq_1 dq_2 dp  \bra{q_1}\ket{+_L} \bra{q_2}\ket{\Psi_{\textrm{target}}} e^{i q_1 q_2} \frac{e^{ip q_1}}{\sqrt{2\pi}} \ket{p} \ket{q_2}. \nonumber
    \end{split}
\end{align}
Afterwards we have to measure $\hat{p}$ on the first rail. For simplicity we post-select on $p = 0$, so that there is no need for Gaussian corrections $\hat{C}$. Thus we obtain:
\begin{align}
    \ket{\Tilde{\Psi}} & = \bra{p = 0}\ket{\Psi}\\  &=
     \frac{1}{\sqrt{2\pi}} \int dq_1 dq_2   \bra{q_1}\ket{+_L} \bra{q_2}\ket{\Psi_{\textrm{target}}} e^{i q_1 q_2}   \ket{q_2}
\end{align}
with the wave function in the position representation given by:
\begin{align}
\label{eq:appendix-gate}
    \Tilde{\Psi}(q) = \bra{q}\ket{\Tilde{\Psi}} = \frac{1}{\sqrt{2\pi}} \bra{q}\ket{\Psi_{\textrm{target}}} \int dq' \bra{q'}\ket{+_L} e^{i q' q}. \nonumber
\end{align}
By replacing the explicit expression of the GKP state given in Eq.(\ref{eq:gkp-plus}), we arrive at
 \begin{align}
 \begin{split}
 & \Tilde{\Psi}(q) \\
     &= \frac{1}{N} \bra{q}\ket{\Psi_{\textrm{target}}} \int dq' \bra{q'}\ket{+_L} e^{i q' q} \\
     &=\frac{1}{\Tilde{N}} \bra{q}\ket{\Psi_{\textrm{target}}} \int dq'      \int dp \\
     &
     \sum_{s=-\infty}^\infty e^{-\frac{\Delta^2}{2} (2s)^2 \pi}  e^{-\frac{1}{2\Delta^2} (p-2s \sqrt{\pi})^2 } e^{i q' (q-p)}     \\
     &=\frac{1}{\Tilde{N}} \bra{q}\ket{\Psi_{\textrm{target}}}      \int dp
     \\
     &
     \sum_{s=-\infty}^\infty e^{-\frac{\Delta^2}{2} (2s)^2 \pi}  e^{-\frac{1}{2\Delta^2} (p-2s \sqrt{\pi})^2 } \delta(q-p)     \\
     &=\frac{1}{\Tilde{N}} \bra{q}\ket{\Psi_{\textrm{target}}}       \sum_{s=-\infty}^\infty e^{-\frac{\Delta^2}{2} (2s)^2 \pi}  e^{-\frac{1}{2\Delta^2} (q-2s \sqrt{\pi})^2 },
     \end{split}
 \end{align}
where $\Tilde{N}$ is another normalisation constant.

When instead we use as ancillary state the output state of our probabilistic conversion protocol $\hat{\rho}_{\textrm{cond}}$ (corresponding to Eq.(\ref{eq:cond-out}) with $n = 0$), we have to generalise to density matrix operators. The state after the  $\hat{C}_Z$-gate is then obtained as:
 \begin{align}
 \begin{split}
     \hat{\rho} &=  \hat{C}_Z  \qty(  \ket{+_L}\bra{+_L} \otimes \hat{\rho}_{\textrm{cond}} )  \hat{C}_Z^\dagger\\
     &= \int dq_1 dq_2 dq_3 dq_4 \hat{C}_Z  \big(  \ket{q_1}\bra{q_1}\ket{+_L}\bra{+_L}\ket{q_2}\bra{q_2}\\
     &\otimes \bra{q_3}\hat{\rho}_{\textrm{cond}}\ket{q_4} \ket{q_3}\bra{q_4} \big)  \hat{C}_Z^\dagger\\
     &= \int dq_1 dq_2 dq_3 dq_4 e^{i(q_1q_3-q_2q_4)}  \big(  \ket{q_1}\bra{q_1}\ket{+_L}\bra{+_L}\ket{q_2}\bra{q_2} \\
     &\otimes \bra{q_3}\hat{\rho}_{\textrm{cond}}\ket{q_4} \ket{q_3}\bra{q_4} \big)  \\ \nonumber
     \end{split}
 \end{align}
 and after the measurement including postselection on $ p=0$:
 \begin{align}
  \begin{split}
    \hat{\Tilde{\rho}} &= \bra{p=0} \hat{\rho} \ket{p=0} =  \int dq_1 dq_2 dq_3 dq_4 e^{i(q_1q_3-q_2q_4)} \\
     &\qty(  \bra{q_1}\ket{+_L}\bra{+_L}\ket{q_2}  \bra{q3}\hat{\rho}_{\textrm{cond}}\ket{q_4} \ket{q_3}\bra{q_4} ). \nonumber
      \end{split}
 \end{align}
So we obtain for the density matrix in position space:
 \begin{align}
  \begin{split}
  &\bra{q} \hat{\Tilde{\rho}} \ket{q'} \\ 
     &= \bra{q} \hat{\rho}_{\textrm{cond}} \ket{q'} \int dq_1 dq_2 e^{i( q_1 q- q_2 q')} \bra{q_1}\ket{+_L} \bra{+_L}\ket{q_2}\\ 
     &= \bra{q} \hat{\rho}_{\textrm{cond}} \ket{q'} \qty(\int dq_1  e^{i q_1 q} \bra{q_1}\ket{+_L}) \qty(\int dq_2  e^{i q_2 q'} \bra{q_2}\ket{+_L})^*\\
     &= \bra{q} \hat{\rho}_{\textrm{cond}} \ket{q'} \qty(\sum_{s=-\infty}^\infty e^{-\frac{\Delta^2}{2} (2s)^2 \pi}  e^{-\frac{1}{2\Delta^2} (q-2s \sqrt{\pi})^2 } ) \\
     &\times \qty(\sum_{m=-\infty}^\infty e^{-\frac{\Delta^2}{2} (2m)^2 \pi}  e^{-\frac{1}{2\Delta^2} (q'-2m \sqrt{\pi})^2 }).\\
     \end{split} \nonumber
 \end{align}
 
The fidelity between the two GKP states after the teleportation gadget when using the perfect cubic-phase state as input for the first and our state after the probabilistic protocol as input for the second is then given as
 \begin{align}
  \begin{split}
     F &= \bra{\Tilde{\Psi}} \Tilde{\rho}  \ket{\Tilde{\Psi}}\\
     &=\int dq dq'\bra{\Tilde{\Psi}}\ket{q} \bra{q} \Tilde{\rho}\ket{q'}  \bra{q'}\ket{\Tilde{\Psi}}
     \end{split}
     \label{final-fidelity}
 \end{align}
 where using Eq.(\ref{eq:cond-out}) we have
 \begin{align}
      \bra{q} \hat{\rho}_{\textrm{cond}} \ket{q'} = \frac{1}{\text{Prob}[0]}\int_{-\delta}^\delta dq''  \bra{q}\ket{\Psi_{\textrm{out}}^{q''}} \bra{\Psi_{\textrm{out}}^{q''}} \ket{q'},
 \end{align}
with $\ket{\Psi_{\textrm{out}}^q}$ given by Eq.(\ref{eq:output2}). The gate error can then be finally defined as $\epsilon = 1 - F$, with $F$ given in Eq.(\ref{final-fidelity}).

\section{Inefficient homodyne detection and performance of the probabilistic protocol}
\label{app:ineff}

We now study the effects of inefficient homodyne detection in the probabilistic protocol. 

\subsection{Partial measurements and inefficient homodyne detection }

We first model the effect of the inefficient homodyne measurement on one of the two modes of a generic two-mode state.
Given an initial state $\rho$ and the POVM $\{ \hat \Pi_j = \hat E^\dagger_j \hat E_j \}$, we will denote
\begin{equation}
\widetilde \rho = \frac{ \hat E_j \rho \hat E^\dagger_j }{ {\rm Tr}\left( \hat \Pi_j \rho \right) },
\end{equation}
the normalized state after the measurement outcome $j$ is obtained. 
Note that the
denominator corresponds to the probability of obtaining the measurement outcome $j$
\begin{equation}
{\rm Tr} \left( \hat E_j \rho \hat E_j^\dagger \right) = 
 {\rm Tr} \left(  \hat E_j^\dagger \hat E_j \rho \right) = {\rm Tr}\left( \hat \Pi_j \rho \right) .
\end{equation}

In our case, we are dealing with a \emph{partial measurement}, i.e., we consider a bipartite system $\rho_{12}$ and the measurement is only applied to mode 1. In this case the state after the measurement (with outcome $j$) is given by
\begin{equation}
\widetilde \rho_2 = \frac{  {\rm Tr}_1 \left[  (\hat E_j \otimes I) \rho_{12} (\hat E_j^\dagger \otimes I)  \right]   }{  {\rm Tr} \left[ ( \hat \Pi_j \otimes I  )  \rho_{12}   \right] } .
\end{equation}
Now, the numerator can be rewritten as
\begin{equation}
\begin{split}
 {\rm Tr}_1 \left[  (\hat E_j \otimes I) \rho_{12} (\hat E_j^\dagger \otimes I)  \right] & = 
 {\rm Tr}_1 \left[ (\hat E_j^\dagger \otimes I) (\hat E_j \otimes I) \rho_{12}  \right]
\\&=
 {\rm Tr}_1 \left[ (\hat \Pi_j \otimes I) \rho_{12}  \right] .
 \end{split}
\end{equation}
It is easy to verify that the cyclic property holds for the partial trace in this case (see Appendix~\ref{app:partial}).

Following \cite{dall2010purification}, the POVM corresponding to homodyne detection with efficiency $\eta$ is given by
\begin{equation}
\hat \Pi_\eta (q) = \frac{1}{ \sqrt{2\pi \Delta_\eta^2} } \int_{-\infty}^{+ \infty} {\rm d}q' \, {\rm e}^{- \frac{ (q' - q)^2 }{ 2 \Delta_\eta^2 }} \vert q' \rangle \langle q' \vert ,
\end{equation}
with
\begin{equation}\label{eq:app_delta}
\Delta_\eta^2 = \frac{1 - \eta}{ 4 \eta} .
\end{equation}
Note that for $\eta \to 1$, $\Delta_\eta^2  \to 0$ and the Gaussian function in the integrand approaches a Dirac delta. In this case, we recover the projector $\hat \Pi (q) = \vert q \rangle \langle q \vert$.

As the probability of obtaining a single continuous-variable outcome $q_n$ is negligible, we are going to consider all possible results in a bin of half-width $\delta$ around $q_n$. The corresponding POVM is
\begin{align}
\hat Q_n(\delta, \eta) &= \int_{q_n - \delta}^{q_n + \delta} {\rm d}q' \, \hat \Pi_\eta (q') \\
&= \frac{1}{ \sqrt{2\pi \Delta_\eta^2} } \int_{q_n - \delta}^{ q_n + \delta} {\rm d}q' \,
 \int_{-\infty}^{+ \infty} {\rm d}q'' \, \\
&
 {\rm e}^{- \frac{ (q'' - q')^2 }{ 2 \Delta_\eta^2 }} \vert q'' \rangle \langle q'' \vert .
\end{align}

\subsection{Inefficient homodyne detection for probabilistic cubic phase state protocol}
Following the above sections, the conditioned state $\rho_{{\rm cond}, 2}$ upon an inefficient homodyne measurement with measurement outcome in $[q_n - \delta, q_n + \delta]$
is given by
\begin{align}
\rho_{{\rm cond}, 2} &= \frac{ {\rm Tr}_1 \left[ \left( \hat Q(\eta, \delta) \otimes I \right) \rho_{12}  \right]  }{ {\rm Tr} \left[ \left( \hat Q(\eta, \delta) \otimes I \right) \rho_{12}  \right]  } \\&
=   \frac{ {\rm Tr}_1 \left[ \left( \hat Q(\eta, \delta) \otimes I \right) \rho_{12}  \right]  }{  {\rm Prob}[q_n]   },
\end{align}
with $\rho_{12} = \vert \Psi_{12}(\xi, \beta, \theta, \gamma) \rangle \langle \Psi_{12} (\xi, \beta, \theta, \gamma)  \vert $.

The numerator corresponds to
\begin{align}
&{\rm Tr}_1 \left[ \left( \hat Q(\eta, \delta) \otimes I \right) \rho_{12}  \right]  \\
&= \frac{1}{\pi^2 \sqrt{2\pi \delta_\eta^2} } \int_{q_n - \delta}^{q_n + \delta} {\rm d}q' \, \int_{- \infty}^{+\infty}
{\rm d} q \, {\rm d}q_2 \, {\rm d}q_4 \, {\rm d} \alpha \, {\rm d} \epsilon \\
&\, \Psi_{\rm in} (q \cos \theta + q_2 \sin \theta ) \nonumber \Psi_{\xi, \beta} (- q \sin \theta +q_2 \cos \theta ) 
\\
&\times 
\Psi^*_{\rm in} (q \cos \theta + q_4 \sin \theta )
\Psi^*_{\xi, \beta} (- q \sin \theta +q_4 \cos \theta ) \nonumber\\
&\times \langle \alpha \vert q_2 \rangle \langle \epsilon \vert q_4 \rangle^* {\rm e}^{-\frac{(q - q')^2}{2 \Delta_\eta^2} } \vert \alpha {\rm e}^{-i \gamma}  \rangle  \langle \epsilon {\rm e}^{-i \gamma}  \vert .
\end{align}
For the denominator we have instead
\begin{align}
{\rm Prob}[q_n] &=  \frac{1}{ \sqrt{2 \pi \Delta^2_\eta}} \int_{q_n - \delta}^{q_n + \delta} {\rm d}q \, \int {\rm d}q_1 \, {\rm d}q_2 \,
\\
&\times 
\vert \Psi_{\rm in} (q_1\cos \theta + q_2 \sin \theta)   \vert^2
\nonumber\\ 
&\times \vert \Psi_{\xi, \beta} (- q_1 \sin \theta + q_2 \cos \theta)  \vert^2 
 {\rm e}^{- \frac{ (q_1 - q)^2 }{ 2\Delta_\eta^2 }}  .
\end{align}

Finally, from these results we can calculate the fidelity
\begin{equation}
F = \langle \Psi_{\rm target}  \vert \rho_{{\rm cond}, 2} \vert \Psi_{\rm target}   \rangle .
\end{equation}
This can be reduced to
\begin{align}
F & = \frac{1}{{\rm Prob}[q_n] } \frac{1}{\pi^2 \sqrt{2\pi \Delta^2_\eta} } \int_{q_n - \delta}^{q_n + \delta} {\rm d}q' \,
\int_{- \infty}^{+ \infty} {\rm d}q \,
\\
& 
{\rm e}^{-\frac{(q - q')^2}{2 \Delta_\eta^2} }  I_1 \times I_1^*,
\end{align}
where
\begin{align}
I_1 &= \int_{-\infty}^{+ \infty} {\rm d}q'' \, {\rm d} q_2 \, {\rm d} \alpha \, \Psi_{\rm in} (q \cos \theta + q_2 \sin \theta) 
\\& 
\Psi_{\xi, \beta} (-q \sin \theta + q_2 \cos \theta ) \langle \alpha \vert q_2 \rangle \nonumber\\
&\times \langle q'' \vert \alpha {\rm e}^{-i \gamma} \rangle \Psi^*_{\rm target} (q'') .
\end{align}
Notice that this corresponds to the overlap integral Eq. (D17).

\begin{figure}[h]
\centering
\includegraphics[width=0.99\linewidth]{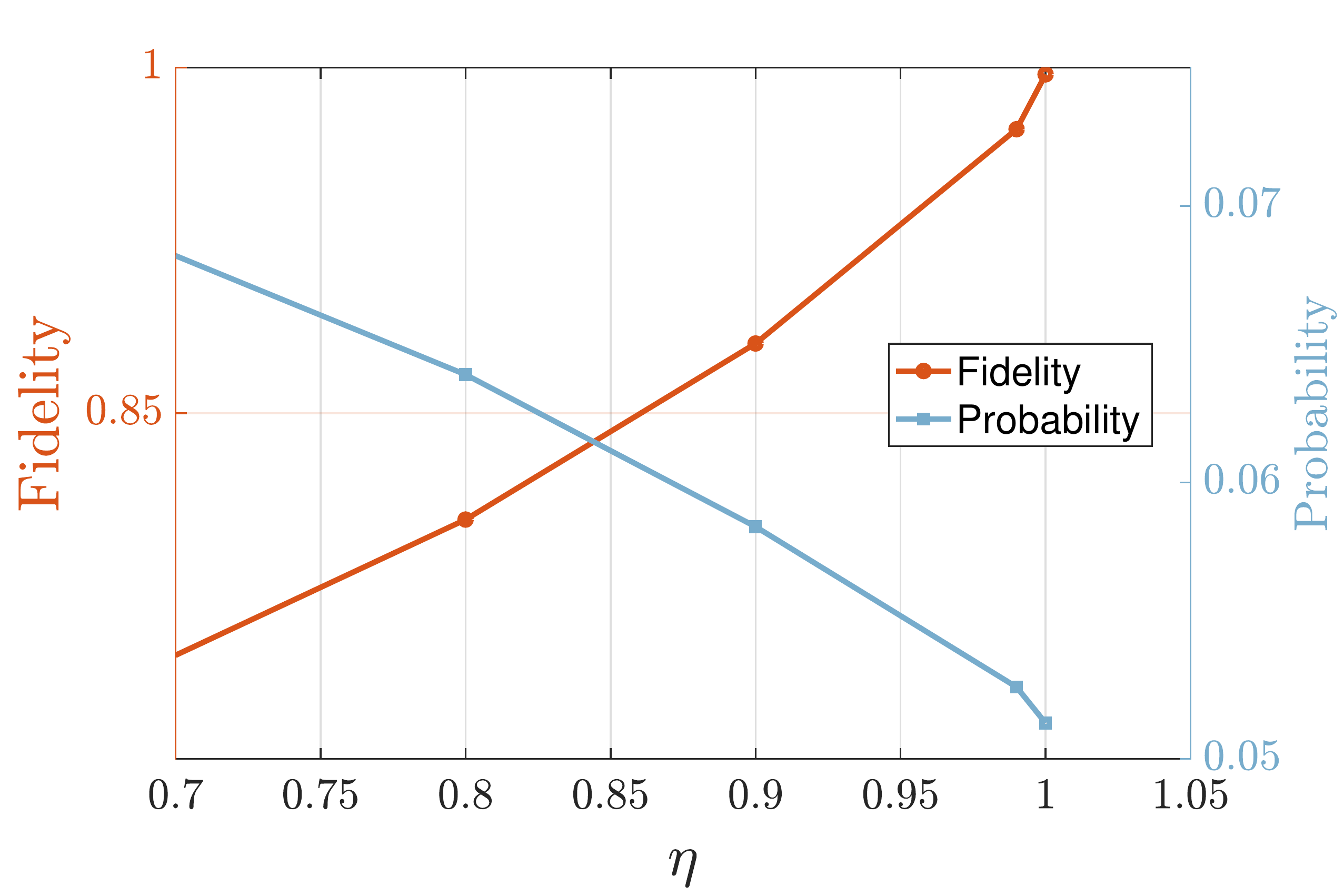}  
\caption{Fidelity and success probability as a function of the efficiency of the homodyne detection $\eta$. The parameters in the circuit, including the value of the mana, correspond to the first row of Table.~\ref{table:probabilistic}. Here, we set the half-width of the measurement bin $\delta=0.1$.}
\label{fig:efficiency}
\end{figure}
In Fig.~\ref{fig:efficiency}, we show the effect of the efficiency $\eta$ of the homodyne detection on the fidelity as well as on the success probability of our protocol for a fixed value of $\delta$. For $\eta \to 1$, we recover our previous results (see Fig.~\ref{fig:eta}). As expected, the fidelity decreases with a decreasing efficiency while, on the other hand, the success probability increases with it. The latter is expected as,  from Eq. \eqref{eq:app_delta}, by decreasing $\eta$ 
we project on a larger domain of quadrature eigenstates with equal weights. This effectively translates into an increased acceptance region. The added noise of the inefficient homodyne detector could be removed by  the phase sensitive amplification of the signal~\cite{dall2010purification}.  
Homodyne detection efficiencies as high as 0.98 have been reported for which our probabilistic protocol still achieves a very high fidelity.

\section{Partial trace and cyclic property}\label{app:partial}

We begin by writing $\rho_{12}$ in the Fock basis
\begin{equation}
\rho_{12} = \sum_{mnpq} \rho_{mnpq} \left( \vert m \rangle  \langle n  \vert  \right)_1 \otimes \left(  \vert  p \rangle \langle q \vert \right)_2 .
\end{equation}
Now, consider the following:
\begin{align}
\left( \hat A \otimes I  \right) \rho_{12} &= \sum_{mnpq} \rho_{mnpq} \left(  \hat A \vert m \rangle  \langle n  \vert  \right)_1 \otimes \left(  \vert  p \rangle \langle q \vert \right)_2 \\
 \rho_{12} \left( \hat A \otimes I  \right) &= \sum_{mnpq} \rho_{mnpq} \left( \vert m \rangle  \langle n  \vert  \hat A   \right)_1 \otimes \left(  \vert  p \rangle \langle q \vert \right)_2
\end{align}
where $A$ is an arbitrary operator acting on mode 1. From here, it is straightforward to see that the partial trace on mode 1 yields  the matrix element $\langle n  \vert \hat A \vert m   \rangle$ in both cases. Therefore, 
\begin{equation}
{\rm Tr}_1  \left[ \left( \hat A \otimes I  \right) \rho_{12}   \right] =  
{\rm Tr}_1  \left[  \rho_{12} \left( \hat A \otimes I  \right) \right] .
\end{equation}

\vspace{10cm}

\bibliography{newlink}

\providecommand{\noopsort}[1]{}\providecommand{\singleletter}[1]{#1}%
\begin{thebibliography}{116}%
\makeatletter
\providecommand \@ifxundefined [1]{%
 \@ifx{#1\undefined}
}%
\providecommand \@ifnum [1]{%
 \ifnum #1\expandafter \@firstoftwo
 \else \expandafter \@secondoftwo
 \fi
}%
\providecommand \@ifx [1]{%
 \ifx #1\expandafter \@firstoftwo
 \else \expandafter \@secondoftwo
 \fi
}%
\providecommand \natexlab [1]{#1}%
\providecommand \enquote  [1]{``#1''}%
\providecommand \bibnamefont  [1]{#1}%
\providecommand \bibfnamefont [1]{#1}%
\providecommand \citenamefont [1]{#1}%
\providecommand \href@noop [0]{\@secondoftwo}%
\providecommand \href [0]{\begingroup \@sanitize@url \@href}%
\providecommand \@href[1]{\@@startlink{#1}\@@href}%
\providecommand \@@href[1]{\endgroup#1\@@endlink}%
\providecommand \@sanitize@url [0]{\catcode `\\12\catcode `\$12\catcode
  `\&12\catcode `\#12\catcode `\^12\catcode `\_12\catcode `\%12\relax}%
\providecommand \@@startlink[1]{}%
\providecommand \@@endlink[0]{}%
\providecommand \url  [0]{\begingroup\@sanitize@url \@url }%
\providecommand \@url [1]{\endgroup\@href {#1}{\urlprefix }}%
\providecommand \urlprefix  [0]{URL }%
\providecommand \Eprint [0]{\href }%
\providecommand \doibase [0]{http://dx.doi.org/}%
\providecommand \selectlanguage [0]{\@gobble}%
\providecommand \bibinfo  [0]{\@secondoftwo}%
\providecommand \bibfield  [0]{\@secondoftwo}%
\providecommand \translation [1]{[#1]}%
\providecommand \BibitemOpen [0]{}%
\providecommand \bibitemStop [0]{}%
\providecommand \bibitemNoStop [0]{.\EOS\space}%
\providecommand \EOS [0]{\spacefactor3000\relax}%
\providecommand \BibitemShut  [1]{\csname bibitem#1\endcsname}%
\let\auto@bib@innerbib\@empty
\bibitem [{\citenamefont {Serafini}(2017)}]{serafini2017quantum}%
  \BibitemOpen
  \bibfield  {author} {\bibinfo {author} {\bibfnamefont {A.}~\bibnamefont
  {Serafini}},\ }\href@noop {} {\emph {\bibinfo {title} {Quantum Continuous
  Variables: A Primer of Theoretical Methods}}}\ (\bibinfo  {publisher} {CRC
  press},\ \bibinfo {year} {2017})\BibitemShut {NoStop}%
\bibitem [{\citenamefont {Pfister}(2020)}]{pfister2019continuous}%
  \BibitemOpen
  \bibfield  {author} {\bibinfo {author} {\bibfnamefont {O.}~\bibnamefont
  {Pfister}},\ }\href {\doibase 10.1088/1361-6455/ab526f} {\bibfield  {journal}
  {\bibinfo  {journal} {Journal of Physics B: Atomic, Molecular and Optical
  Physics}\ }\textbf {\bibinfo {volume} {53}},\ \bibinfo {pages} {012001}
  (\bibinfo {year} {2020})}\BibitemShut {NoStop}%
\bibitem [{\citenamefont {Blais}\ \emph {et~al.}(2019)\citenamefont {Blais},
  \citenamefont {Grimsmo}, \citenamefont {Girvin},\ and\ \citenamefont
  {Wallraff}}]{blais2020circuit}%
  \BibitemOpen
  \bibfield  {author} {\bibinfo {author} {\bibfnamefont {A.}~\bibnamefont
  {Blais}}, \bibinfo {author} {\bibfnamefont {A.~L.}\ \bibnamefont {Grimsmo}},
  \bibinfo {author} {\bibfnamefont {S.~M.}\ \bibnamefont {Girvin}}, \ and\
  \bibinfo {author} {\bibfnamefont {A.}~\bibnamefont {Wallraff}},\ }in\ \href
  {\doibase 10.1142/9789811201400_0006} {\emph {\bibinfo {booktitle}
  {Mesoscopic Physics meets Quantum Engineering}}}\ (\bibinfo  {publisher}
  {WORLD SCIENTIFIC},\ \bibinfo {year} {2019})\ pp.\ \bibinfo {pages}
  {135--153}\BibitemShut {NoStop}%
\bibitem [{\citenamefont {Grimsmo}\ and\ \citenamefont
  {Blais}(2017)}]{grimsmo2017squeezing}%
  \BibitemOpen
  \bibfield  {author} {\bibinfo {author} {\bibfnamefont {A.~L.}\ \bibnamefont
  {Grimsmo}}\ and\ \bibinfo {author} {\bibfnamefont {A.}~\bibnamefont
  {Blais}},\ }\href {\doibase 10.1038/s41534-017-0020-8} {\bibfield  {journal}
  {\bibinfo  {journal} {npj Quantum Information}\ }\textbf {\bibinfo {volume}
  {3}},\ \bibinfo {pages} {20} (\bibinfo {year} {2017})}\BibitemShut {NoStop}%
\bibitem [{\citenamefont {Hillmann}\ \emph {et~al.}(2020)\citenamefont
  {Hillmann}, \citenamefont {Quijandr{\'{i}}a}, \citenamefont {Johansson},
  \citenamefont {Ferraro}, \citenamefont {Gasparinetti},\ and\ \citenamefont
  {Ferrini}}]{hillmannUniversalGateSet2020}%
  \BibitemOpen
  \bibfield  {author} {\bibinfo {author} {\bibfnamefont {T.}~\bibnamefont
  {Hillmann}}, \bibinfo {author} {\bibfnamefont {F.}~\bibnamefont
  {Quijandr{\'{i}}a}}, \bibinfo {author} {\bibfnamefont {G.}~\bibnamefont
  {Johansson}}, \bibinfo {author} {\bibfnamefont {A.}~\bibnamefont {Ferraro}},
  \bibinfo {author} {\bibfnamefont {S.}~\bibnamefont {Gasparinetti}}, \ and\
  \bibinfo {author} {\bibfnamefont {G.}~\bibnamefont {Ferrini}},\ }\href
  {\doibase 10.1103/PhysRevLett.125.160501} {\bibfield  {journal} {\bibinfo
  {journal} {Physical review letters}\ }\textbf {\bibinfo {volume} {125}},\
  \bibinfo {pages} {160501} (\bibinfo {year} {2020})}\BibitemShut {NoStop}%
\bibitem [{\citenamefont {Serafini}\ \emph {et~al.}(2009)\citenamefont
  {Serafini}, \citenamefont {Retzker},\ and\ \citenamefont
  {Plenio}}]{serafini2009manipulating}%
  \BibitemOpen
  \bibfield  {author} {\bibinfo {author} {\bibfnamefont {A.}~\bibnamefont
  {Serafini}}, \bibinfo {author} {\bibfnamefont {A.}~\bibnamefont {Retzker}}, \
  and\ \bibinfo {author} {\bibfnamefont {M.~B.}\ \bibnamefont {Plenio}},\
  }\href {\doibase 10.1088/1367-2630/11/2/023007} {\bibfield  {journal}
  {\bibinfo  {journal} {New Journal of Physics}\ }\textbf {\bibinfo {volume}
  {11}},\ \bibinfo {pages} {023007} (\bibinfo {year} {2009})}\BibitemShut
  {NoStop}%
\bibitem [{\citenamefont {Fl{\"{u}}hmann}\ \emph {et~al.}(2019)\citenamefont
  {Fl{\"{u}}hmann}, \citenamefont {Nguyen}, \citenamefont {Marinelli},
  \citenamefont {Negnevitsky}, \citenamefont {Mehta},\ and\ \citenamefont
  {Home}}]{fluhmann2019encoding}%
  \BibitemOpen
  \bibfield  {author} {\bibinfo {author} {\bibfnamefont {C.}~\bibnamefont
  {Fl{\"{u}}hmann}}, \bibinfo {author} {\bibfnamefont {T.~L.}\ \bibnamefont
  {Nguyen}}, \bibinfo {author} {\bibfnamefont {M.}~\bibnamefont {Marinelli}},
  \bibinfo {author} {\bibfnamefont {V.}~\bibnamefont {Negnevitsky}}, \bibinfo
  {author} {\bibfnamefont {K.}~\bibnamefont {Mehta}}, \ and\ \bibinfo {author}
  {\bibfnamefont {J.~P.}\ \bibnamefont {Home}},\ }\href {\doibase
  10.1038/s41586-019-0960-6} {\bibfield  {journal} {\bibinfo  {journal}
  {Nature}\ }\textbf {\bibinfo {volume} {566}},\ \bibinfo {pages} {513}
  (\bibinfo {year} {2019})}\BibitemShut {NoStop}%
\bibitem [{\citenamefont {Schmidt}\ \emph {et~al.}(2012)\citenamefont
  {Schmidt}, \citenamefont {Ludwig},\ and\ \citenamefont
  {Marquardt}}]{schmidt2012optomechanical}%
  \BibitemOpen
  \bibfield  {author} {\bibinfo {author} {\bibfnamefont {M.}~\bibnamefont
  {Schmidt}}, \bibinfo {author} {\bibfnamefont {M.}~\bibnamefont {Ludwig}}, \
  and\ \bibinfo {author} {\bibfnamefont {F.}~\bibnamefont {Marquardt}},\ }\href
  {\doibase 10.1088/1367-2630/14/12/125005} {\bibfield  {journal} {\bibinfo
  {journal} {New Journal of Physics}\ }\textbf {\bibinfo {volume} {14}},\
  \bibinfo {pages} {125005} (\bibinfo {year} {2012})}\BibitemShut {NoStop}%
\bibitem [{\citenamefont {Houhou}\ \emph {et~al.}(2015)\citenamefont {Houhou},
  \citenamefont {Aissaoui},\ and\ \citenamefont
  {Ferraro}}]{houhou2015generation}%
  \BibitemOpen
  \bibfield  {author} {\bibinfo {author} {\bibfnamefont {O.}~\bibnamefont
  {Houhou}}, \bibinfo {author} {\bibfnamefont {H.}~\bibnamefont {Aissaoui}}, \
  and\ \bibinfo {author} {\bibfnamefont {A.}~\bibnamefont {Ferraro}},\ }\href
  {\doibase 10.1103/PhysRevA.92.063843} {\bibfield  {journal} {\bibinfo
  {journal} {Physical Review A}\ }\textbf {\bibinfo {volume} {92}},\ \bibinfo
  {pages} {063843} (\bibinfo {year} {2015})}\BibitemShut {NoStop}%
\bibitem [{\citenamefont {Nielsen}\ \emph {et~al.}(2017)\citenamefont
  {Nielsen}, \citenamefont {Tsaturyan}, \citenamefont {M{\o}ller},
  \citenamefont {Polzik},\ and\ \citenamefont
  {Schliesser}}]{nielsen2017multimode}%
  \BibitemOpen
  \bibfield  {author} {\bibinfo {author} {\bibfnamefont {W.~H.~P.}\
  \bibnamefont {Nielsen}}, \bibinfo {author} {\bibfnamefont {Y.}~\bibnamefont
  {Tsaturyan}}, \bibinfo {author} {\bibfnamefont {C.~B.}\ \bibnamefont
  {M{\o}ller}}, \bibinfo {author} {\bibfnamefont {E.~S.}\ \bibnamefont
  {Polzik}}, \ and\ \bibinfo {author} {\bibfnamefont {A.}~\bibnamefont
  {Schliesser}},\ }\href {\doibase 10.1073/pnas.1608412114} {\bibfield
  {journal} {\bibinfo  {journal} {Proceedings of the National Academy of
  Sciences}\ }\textbf {\bibinfo {volume} {114}},\ \bibinfo {pages} {62}
  (\bibinfo {year} {2017})}\BibitemShut {NoStop}%
\bibitem [{\citenamefont {Stasi{\'{n}}ska}\ \emph {et~al.}(2009)\citenamefont
  {Stasi{\'{n}}ska}, \citenamefont {Rod{\'{o}}}, \citenamefont {Paganelli},
  \citenamefont {Birkl},\ and\ \citenamefont
  {Sanpera}}]{stasinska2009manipulating}%
  \BibitemOpen
  \bibfield  {author} {\bibinfo {author} {\bibfnamefont {J.}~\bibnamefont
  {Stasi{\'{n}}ska}}, \bibinfo {author} {\bibfnamefont {C.}~\bibnamefont
  {Rod{\'{o}}}}, \bibinfo {author} {\bibfnamefont {S.}~\bibnamefont
  {Paganelli}}, \bibinfo {author} {\bibfnamefont {G.}~\bibnamefont {Birkl}}, \
  and\ \bibinfo {author} {\bibfnamefont {A.}~\bibnamefont {Sanpera}},\ }\href
  {\doibase 10.1103/PhysRevA.80.062304} {\bibfield  {journal} {\bibinfo
  {journal} {Physical Review A}\ }\textbf {\bibinfo {volume} {80}},\ \bibinfo
  {pages} {062304} (\bibinfo {year} {2009})}\BibitemShut {NoStop}%
\bibitem [{\citenamefont {Milne}\ and\ \citenamefont
  {Korolkova}(2012)}]{milne2012composite}%
  \BibitemOpen
  \bibfield  {author} {\bibinfo {author} {\bibfnamefont {D.~F.}\ \bibnamefont
  {Milne}}\ and\ \bibinfo {author} {\bibfnamefont {N.~V.}\ \bibnamefont
  {Korolkova}},\ }\href {\doibase 10.1103/PhysRevA.85.032310} {\bibfield
  {journal} {\bibinfo  {journal} {Physical Review A}\ }\textbf {\bibinfo
  {volume} {85}},\ \bibinfo {pages} {032310} (\bibinfo {year}
  {2012})}\BibitemShut {NoStop}%
\bibitem [{\citenamefont {Ikeda}\ and\ \citenamefont
  {Yamamoto}(2013)}]{ikeda2013deterministic}%
  \BibitemOpen
  \bibfield  {author} {\bibinfo {author} {\bibfnamefont {Y.}~\bibnamefont
  {Ikeda}}\ and\ \bibinfo {author} {\bibfnamefont {N.}~\bibnamefont
  {Yamamoto}},\ }\href {\doibase 10.1103/PhysRevA.87.033802} {\bibfield
  {journal} {\bibinfo  {journal} {Physical Review A}\ }\textbf {\bibinfo
  {volume} {87}},\ \bibinfo {pages} {033802} (\bibinfo {year}
  {2013})}\BibitemShut {NoStop}%
\bibitem [{\citenamefont {Motes}\ \emph {et~al.}(2017)\citenamefont {Motes},
  \citenamefont {Baragiola}, \citenamefont {Gilchrist},\ and\ \citenamefont
  {Menicucci}}]{motes2017encoding}%
  \BibitemOpen
  \bibfield  {author} {\bibinfo {author} {\bibfnamefont {K.~R.}\ \bibnamefont
  {Motes}}, \bibinfo {author} {\bibfnamefont {B.~Q.}\ \bibnamefont
  {Baragiola}}, \bibinfo {author} {\bibfnamefont {A.}~\bibnamefont
  {Gilchrist}}, \ and\ \bibinfo {author} {\bibfnamefont {N.~C.}\ \bibnamefont
  {Menicucci}},\ }\href {\doibase 10.1103/PhysRevA.95.053819} {\bibfield
  {journal} {\bibinfo  {journal} {Physical Review A}\ }\textbf {\bibinfo
  {volume} {95}},\ \bibinfo {pages} {053819} (\bibinfo {year}
  {2017})}\BibitemShut {NoStop}%
\bibitem [{\citenamefont {Aolita}\ \emph {et~al.}(2011)\citenamefont {Aolita},
  \citenamefont {Roncaglia}, \citenamefont {Ferraro},\ and\ \citenamefont
  {Ac{\'{i}}n}}]{aolita2011gapped}%
  \BibitemOpen
  \bibfield  {author} {\bibinfo {author} {\bibfnamefont {L.}~\bibnamefont
  {Aolita}}, \bibinfo {author} {\bibfnamefont {A.~J.}\ \bibnamefont
  {Roncaglia}}, \bibinfo {author} {\bibfnamefont {A.}~\bibnamefont {Ferraro}},
  \ and\ \bibinfo {author} {\bibfnamefont {A.}~\bibnamefont {Ac{\'{i}}n}},\
  }\href {\doibase 10.1103/PhysRevLett.106.090501} {\bibfield  {journal}
  {\bibinfo  {journal} {Physical Review Letters}\ }\textbf {\bibinfo {volume}
  {106}},\ \bibinfo {pages} {090501} (\bibinfo {year} {2011})}\BibitemShut
  {NoStop}%
\bibitem [{\citenamefont {Yoshikawa}\ \emph {et~al.}(2016)\citenamefont
  {Yoshikawa}, \citenamefont {Yokoyama}, \citenamefont {Kaji}, \citenamefont
  {Sornphiphatphong}, \citenamefont {Shiozawa}, \citenamefont {Makino},\ and\
  \citenamefont {Furusawa}}]{yoshikawa2016}%
  \BibitemOpen
  \bibfield  {author} {\bibinfo {author} {\bibfnamefont {J.}~\bibnamefont
  {Yoshikawa}}, \bibinfo {author} {\bibfnamefont {S.}~\bibnamefont {Yokoyama}},
  \bibinfo {author} {\bibfnamefont {T.}~\bibnamefont {Kaji}}, \bibinfo {author}
  {\bibfnamefont {C.}~\bibnamefont {Sornphiphatphong}}, \bibinfo {author}
  {\bibfnamefont {Y.}~\bibnamefont {Shiozawa}}, \bibinfo {author}
  {\bibfnamefont {K.}~\bibnamefont {Makino}}, \ and\ \bibinfo {author}
  {\bibfnamefont {A.}~\bibnamefont {Furusawa}},\ }\href {\doibase
  10.1063/1.4962732} {\bibfield  {journal} {\bibinfo  {journal} {APL
  Photonics}\ }\textbf {\bibinfo {volume} {1}},\ \bibinfo {pages} {060801}
  (\bibinfo {year} {2016})}\BibitemShut {NoStop}%
\bibitem [{\citenamefont {Larsen}\ \emph {et~al.}(2019)\citenamefont {Larsen},
  \citenamefont {Guo}, \citenamefont {Breum}, \citenamefont
  {Neergaard-Nielsen},\ and\ \citenamefont
  {Andersen}}]{larsen2019deterministic}%
  \BibitemOpen
  \bibfield  {author} {\bibinfo {author} {\bibfnamefont {M.~V.}\ \bibnamefont
  {Larsen}}, \bibinfo {author} {\bibfnamefont {X.}~\bibnamefont {Guo}},
  \bibinfo {author} {\bibfnamefont {C.~R.}\ \bibnamefont {Breum}}, \bibinfo
  {author} {\bibfnamefont {J.~S.}\ \bibnamefont {Neergaard-Nielsen}}, \ and\
  \bibinfo {author} {\bibfnamefont {U.~L.}\ \bibnamefont {Andersen}},\ }\href
  {\doibase 10.1126/science.aay4354} {\bibfield  {journal} {\bibinfo  {journal}
  {Science}\ }\textbf {\bibinfo {volume} {366}},\ \bibinfo {pages} {369}
  (\bibinfo {year} {2019})}\BibitemShut {NoStop}%
\bibitem [{\citenamefont {Asavanant}\ \emph {et~al.}(2019)\citenamefont
  {Asavanant}, \citenamefont {Shiozawa}, \citenamefont {Yokoyama},
  \citenamefont {Charoensombutamon}, \citenamefont {Emura}, \citenamefont
  {Alexander}, \citenamefont {Takeda}, \citenamefont {Yoshikawa}, \citenamefont
  {Menicucci}, \citenamefont {Yonezawa},\ and\ \citenamefont
  {Furusawa}}]{asavanant2019generation}%
  \BibitemOpen
  \bibfield  {author} {\bibinfo {author} {\bibfnamefont {W.}~\bibnamefont
  {Asavanant}}, \bibinfo {author} {\bibfnamefont {Y.}~\bibnamefont {Shiozawa}},
  \bibinfo {author} {\bibfnamefont {S.}~\bibnamefont {Yokoyama}}, \bibinfo
  {author} {\bibfnamefont {B.}~\bibnamefont {Charoensombutamon}}, \bibinfo
  {author} {\bibfnamefont {H.}~\bibnamefont {Emura}}, \bibinfo {author}
  {\bibfnamefont {R.~N.}\ \bibnamefont {Alexander}}, \bibinfo {author}
  {\bibfnamefont {S.}~\bibnamefont {Takeda}}, \bibinfo {author} {\bibfnamefont
  {J.-i.}\ \bibnamefont {Yoshikawa}}, \bibinfo {author} {\bibfnamefont {N.~C.}\
  \bibnamefont {Menicucci}}, \bibinfo {author} {\bibfnamefont {H.}~\bibnamefont
  {Yonezawa}}, \ and\ \bibinfo {author} {\bibfnamefont {A.}~\bibnamefont
  {Furusawa}},\ }\href {\doibase 10.1126/science.aay2645} {\bibfield  {journal}
  {\bibinfo  {journal} {Science}\ }\textbf {\bibinfo {volume} {366}},\ \bibinfo
  {pages} {373} (\bibinfo {year} {2019})}\BibitemShut {NoStop}%
\bibitem [{\citenamefont {Lenzini}\ \emph {et~al.}(2018)\citenamefont
  {Lenzini}, \citenamefont {Janousek}, \citenamefont {Thearle}, \citenamefont
  {Villa}, \citenamefont {Haylock}, \citenamefont {Kasture}, \citenamefont
  {Cui}, \citenamefont {Phan}, \citenamefont {Dao}, \citenamefont {Yonezawa},
  \citenamefont {Lam}, \citenamefont {Huntington},\ and\ \citenamefont
  {Lobino}}]{lenzini2018integrated}%
  \BibitemOpen
  \bibfield  {author} {\bibinfo {author} {\bibfnamefont {F.}~\bibnamefont
  {Lenzini}}, \bibinfo {author} {\bibfnamefont {J.}~\bibnamefont {Janousek}},
  \bibinfo {author} {\bibfnamefont {O.}~\bibnamefont {Thearle}}, \bibinfo
  {author} {\bibfnamefont {M.}~\bibnamefont {Villa}}, \bibinfo {author}
  {\bibfnamefont {B.}~\bibnamefont {Haylock}}, \bibinfo {author} {\bibfnamefont
  {S.}~\bibnamefont {Kasture}}, \bibinfo {author} {\bibfnamefont
  {L.}~\bibnamefont {Cui}}, \bibinfo {author} {\bibfnamefont {H.-P.}\
  \bibnamefont {Phan}}, \bibinfo {author} {\bibfnamefont {D.~V.}\ \bibnamefont
  {Dao}}, \bibinfo {author} {\bibfnamefont {H.}~\bibnamefont {Yonezawa}},
  \bibinfo {author} {\bibfnamefont {P.~K.}\ \bibnamefont {Lam}}, \bibinfo
  {author} {\bibfnamefont {E.~H.}\ \bibnamefont {Huntington}}, \ and\ \bibinfo
  {author} {\bibfnamefont {M.}~\bibnamefont {Lobino}},\ }\href {\doibase
  10.1126/sciadv.aat9331} {\bibfield  {journal} {\bibinfo  {journal} {Science
  Advances}\ }\textbf {\bibinfo {volume} {4}},\ \bibinfo {pages} {eaat9331}
  (\bibinfo {year} {2018})}\BibitemShut {NoStop}%
\bibitem [{\citenamefont {Ofek}\ \emph {et~al.}(2016)\citenamefont {Ofek},
  \citenamefont {Petrenko}, \citenamefont {Heeres}, \citenamefont {Reinhold},
  \citenamefont {Leghtas}, \citenamefont {Vlastakis}, \citenamefont {Liu},
  \citenamefont {Frunzio}, \citenamefont {Girvin}, \citenamefont {Jiang},
  \citenamefont {Mirrahimi}, \citenamefont {Devoret},\ and\ \citenamefont
  {Schoelkopf}}]{ofek2016}%
  \BibitemOpen
  \bibfield  {author} {\bibinfo {author} {\bibfnamefont {N.}~\bibnamefont
  {Ofek}}, \bibinfo {author} {\bibfnamefont {A.}~\bibnamefont {Petrenko}},
  \bibinfo {author} {\bibfnamefont {R.}~\bibnamefont {Heeres}}, \bibinfo
  {author} {\bibfnamefont {P.}~\bibnamefont {Reinhold}}, \bibinfo {author}
  {\bibfnamefont {Z.}~\bibnamefont {Leghtas}}, \bibinfo {author} {\bibfnamefont
  {B.}~\bibnamefont {Vlastakis}}, \bibinfo {author} {\bibfnamefont
  {Y.}~\bibnamefont {Liu}}, \bibinfo {author} {\bibfnamefont {L.}~\bibnamefont
  {Frunzio}}, \bibinfo {author} {\bibfnamefont {S.~M.}\ \bibnamefont {Girvin}},
  \bibinfo {author} {\bibfnamefont {L.}~\bibnamefont {Jiang}}, \bibinfo
  {author} {\bibfnamefont {M.}~\bibnamefont {Mirrahimi}}, \bibinfo {author}
  {\bibfnamefont {M.~H.}\ \bibnamefont {Devoret}}, \ and\ \bibinfo {author}
  {\bibfnamefont {R.~J.}\ \bibnamefont {Schoelkopf}},\ }\href {\doibase
  10.1038/nature18949} {\bibfield  {journal} {\bibinfo  {journal} {Nature}\
  }\textbf {\bibinfo {volume} {536}},\ \bibinfo {pages} {441} (\bibinfo {year}
  {2016})}\BibitemShut {NoStop}%
\bibitem [{\citenamefont {Romanenko}\ \emph {et~al.}(2020)\citenamefont
  {Romanenko}, \citenamefont {Pilipenko}, \citenamefont {Zorzetti},
  \citenamefont {Frolov}, \citenamefont {Awida}, \citenamefont {Belomestnykh},
  \citenamefont {Posen},\ and\ \citenamefont
  {Grassellino}}]{romanenko2020three}%
  \BibitemOpen
  \bibfield  {author} {\bibinfo {author} {\bibfnamefont {A.}~\bibnamefont
  {Romanenko}}, \bibinfo {author} {\bibfnamefont {R.}~\bibnamefont
  {Pilipenko}}, \bibinfo {author} {\bibfnamefont {S.}~\bibnamefont {Zorzetti}},
  \bibinfo {author} {\bibfnamefont {D.}~\bibnamefont {Frolov}}, \bibinfo
  {author} {\bibfnamefont {M.}~\bibnamefont {Awida}}, \bibinfo {author}
  {\bibfnamefont {S.}~\bibnamefont {Belomestnykh}}, \bibinfo {author}
  {\bibfnamefont {S.}~\bibnamefont {Posen}}, \ and\ \bibinfo {author}
  {\bibfnamefont {A.}~\bibnamefont {Grassellino}},\ }\href {\doibase
  10.1103/PhysRevApplied.13.034032} {\bibfield  {journal} {\bibinfo  {journal}
  {Physical Review Applied}\ }\textbf {\bibinfo {volume} {13}},\ \bibinfo
  {pages} {1} (\bibinfo {year} {2020})}\BibitemShut {NoStop}%
\bibitem [{\citenamefont {Gottesman}\ \emph {et~al.}(2001)\citenamefont
  {Gottesman}, \citenamefont {Kitaev},\ and\ \citenamefont
  {Preskill}}]{gottesman2001}%
  \BibitemOpen
  \bibfield  {author} {\bibinfo {author} {\bibfnamefont {D.}~\bibnamefont
  {Gottesman}}, \bibinfo {author} {\bibfnamefont {A.}~\bibnamefont {Kitaev}}, \
  and\ \bibinfo {author} {\bibfnamefont {J.}~\bibnamefont {Preskill}},\ }\href
  {\doibase 10.1103/PhysRevA.64.012310} {\bibfield  {journal} {\bibinfo
  {journal} {Physical Review A}\ }\textbf {\bibinfo {volume} {64}},\ \bibinfo
  {pages} {012310} (\bibinfo {year} {2001})}\BibitemShut {NoStop}%
\bibitem [{\citenamefont {Menicucci}(2014)}]{menicucci2014fault}%
  \BibitemOpen
  \bibfield  {author} {\bibinfo {author} {\bibfnamefont {N.~C.}\ \bibnamefont
  {Menicucci}},\ }\href {\doibase 10.1103/PhysRevLett.112.120504} {\bibfield
  {journal} {\bibinfo  {journal} {Physical Review Letters}\ }\textbf {\bibinfo
  {volume} {112}},\ \bibinfo {pages} {120504} (\bibinfo {year}
  {2014})}\BibitemShut {NoStop}%
\bibitem [{\citenamefont {Campagne-Ibarcq}\ \emph {et~al.}(2020)\citenamefont
  {Campagne-Ibarcq}, \citenamefont {Eickbusch}, \citenamefont {Touzard},
  \citenamefont {Zalys-Geller}, \citenamefont {Frattini}, \citenamefont
  {Sivak}, \citenamefont {Reinhold}, \citenamefont {Puri}, \citenamefont
  {Shankar}, \citenamefont {Schoelkopf}, \citenamefont {Frunzio}, \citenamefont
  {Mirrahimi},\ and\ \citenamefont {Devoret}}]{campagne-ibarcq2019}%
  \BibitemOpen
  \bibfield  {author} {\bibinfo {author} {\bibfnamefont {P.}~\bibnamefont
  {Campagne-Ibarcq}}, \bibinfo {author} {\bibfnamefont {A.}~\bibnamefont
  {Eickbusch}}, \bibinfo {author} {\bibfnamefont {S.}~\bibnamefont {Touzard}},
  \bibinfo {author} {\bibfnamefont {E.}~\bibnamefont {Zalys-Geller}}, \bibinfo
  {author} {\bibfnamefont {N.~E.}\ \bibnamefont {Frattini}}, \bibinfo {author}
  {\bibfnamefont {V.~V.}\ \bibnamefont {Sivak}}, \bibinfo {author}
  {\bibfnamefont {P.}~\bibnamefont {Reinhold}}, \bibinfo {author}
  {\bibfnamefont {S.}~\bibnamefont {Puri}}, \bibinfo {author} {\bibfnamefont
  {S.}~\bibnamefont {Shankar}}, \bibinfo {author} {\bibfnamefont {R.~J.}\
  \bibnamefont {Schoelkopf}}, \bibinfo {author} {\bibfnamefont
  {L.}~\bibnamefont {Frunzio}}, \bibinfo {author} {\bibfnamefont
  {M.}~\bibnamefont {Mirrahimi}}, \ and\ \bibinfo {author} {\bibfnamefont
  {M.~H.}\ \bibnamefont {Devoret}},\ }\href {\doibase
  10.1038/s41586-020-2603-3} {\bibfield  {journal} {\bibinfo  {journal}
  {Nature}\ }\textbf {\bibinfo {volume} {584}},\ \bibinfo {pages} {368}
  (\bibinfo {year} {2020})}\BibitemShut {NoStop}%
\bibitem [{\citenamefont {Ferraro}\ \emph {et~al.}(2005)\citenamefont
  {Ferraro}, \citenamefont {Olivares},\ and\ \citenamefont
  {Paris}}]{ferraro2005}%
  \BibitemOpen
  \bibfield  {author} {\bibinfo {author} {\bibfnamefont {A.}~\bibnamefont
  {Ferraro}}, \bibinfo {author} {\bibfnamefont {S.}~\bibnamefont {Olivares}}, \
  and\ \bibinfo {author} {\bibfnamefont {M.~G.~A.}\ \bibnamefont {Paris}},\
  }\href {http://arxiv.org/abs/quant-ph/0503237} {\emph {\bibinfo {title}
  {Gaussian States in Quantum Information}}}\ (\bibinfo  {publisher}
  {Bibliopolis},\ \bibinfo {address} {Napoli},\ \bibinfo {year}
  {2005})\BibitemShut {NoStop}%
\bibitem [{\citenamefont {Weedbrook}\ \emph {et~al.}(2012)\citenamefont
  {Weedbrook}, \citenamefont {Pirandola}, \citenamefont {Garc\'{\i}a-Patr\'on},
  \citenamefont {Cerf}, \citenamefont {Ralph}, \citenamefont {Shapiro},\ and\
  \citenamefont {Lloyd}}]{weedbrook2012}%
  \BibitemOpen
  \bibfield  {author} {\bibinfo {author} {\bibfnamefont {C.}~\bibnamefont
  {Weedbrook}}, \bibinfo {author} {\bibfnamefont {S.}~\bibnamefont
  {Pirandola}}, \bibinfo {author} {\bibfnamefont {R.}~\bibnamefont
  {Garc\'{\i}a-Patr\'on}}, \bibinfo {author} {\bibfnamefont {N.~J.}\
  \bibnamefont {Cerf}}, \bibinfo {author} {\bibfnamefont {T.~C.}\ \bibnamefont
  {Ralph}}, \bibinfo {author} {\bibfnamefont {J.~H.}\ \bibnamefont {Shapiro}},
  \ and\ \bibinfo {author} {\bibfnamefont {S.}~\bibnamefont {Lloyd}},\ }\href
  {\doibase 10.1103/RevModPhys.84.621} {\bibfield  {journal} {\bibinfo
  {journal} {Rev. Mod. Phys.}\ }\textbf {\bibinfo {volume} {84}},\ \bibinfo
  {pages} {621} (\bibinfo {year} {2012})}\BibitemShut {NoStop}%
\bibitem [{\citenamefont {Adesso}\ \emph {et~al.}(2014)\citenamefont {Adesso},
  \citenamefont {Ragy},\ and\ \citenamefont {Lee}}]{adesso2014continuous}%
  \BibitemOpen
  \bibfield  {author} {\bibinfo {author} {\bibfnamefont {G.}~\bibnamefont
  {Adesso}}, \bibinfo {author} {\bibfnamefont {S.}~\bibnamefont {Ragy}}, \ and\
  \bibinfo {author} {\bibfnamefont {A.~R.}\ \bibnamefont {Lee}},\ }\href
  {\doibase 10.1142/S1230161214400010} {\bibfield  {journal} {\bibinfo
  {journal} {Open Systems {\&} Information Dynamics}\ }\textbf {\bibinfo
  {volume} {21}},\ \bibinfo {pages} {1440001} (\bibinfo {year}
  {2014})}\BibitemShut {NoStop}%
\bibitem [{\citenamefont {Lloyd}\ and\ \citenamefont
  {Braunstein}(1999)}]{lloyd1999quantum}%
  \BibitemOpen
  \bibfield  {author} {\bibinfo {author} {\bibfnamefont {S.}~\bibnamefont
  {Lloyd}}\ and\ \bibinfo {author} {\bibfnamefont {S.~L.}\ \bibnamefont
  {Braunstein}},\ }in\ \href@noop {} {\emph {\bibinfo {booktitle} {Quantum
  Information with Continuous Variables}}}\ (\bibinfo  {publisher} {Springer},\
  \bibinfo {year} {1999})\ pp.\ \bibinfo {pages} {9--17}\BibitemShut {NoStop}%
\bibitem [{\citenamefont {Albarelli}\ \emph {et~al.}(2018)\citenamefont
  {Albarelli}, \citenamefont {Genoni}, \citenamefont {Paris},\ and\
  \citenamefont {Ferraro}}]{albarelli2018resource}%
  \BibitemOpen
  \bibfield  {author} {\bibinfo {author} {\bibfnamefont {F.}~\bibnamefont
  {Albarelli}}, \bibinfo {author} {\bibfnamefont {M.~G.}\ \bibnamefont
  {Genoni}}, \bibinfo {author} {\bibfnamefont {M.~G.~A.}\ \bibnamefont
  {Paris}}, \ and\ \bibinfo {author} {\bibfnamefont {A.}~\bibnamefont
  {Ferraro}},\ }\href {\doibase 10.1103/PhysRevA.98.052350} {\bibfield
  {journal} {\bibinfo  {journal} {Physical Review A}\ }\textbf {\bibinfo
  {volume} {98}},\ \bibinfo {pages} {052350} (\bibinfo {year}
  {2018})}\BibitemShut {NoStop}%
\bibitem [{\citenamefont {Takagi}\ and\ \citenamefont
  {Zhuang}(2018)}]{takagi2018}%
  \BibitemOpen
  \bibfield  {author} {\bibinfo {author} {\bibfnamefont {R.}~\bibnamefont
  {Takagi}}\ and\ \bibinfo {author} {\bibfnamefont {Q.}~\bibnamefont
  {Zhuang}},\ }\href {\doibase 10.1103/PhysRevA.97.062337} {\bibfield
  {journal} {\bibinfo  {journal} {Physical Review A}\ }\textbf {\bibinfo
  {volume} {97}},\ \bibinfo {pages} {062337} (\bibinfo {year}
  {2018})}\BibitemShut {NoStop}%
\bibitem [{\citenamefont {Baragiola}\ \emph {et~al.}(2019)\citenamefont
  {Baragiola}, \citenamefont {Pantaleoni}, \citenamefont {Alexander},
  \citenamefont {Karanjai},\ and\ \citenamefont {Menicucci}}]{baragiola2019}%
  \BibitemOpen
  \bibfield  {author} {\bibinfo {author} {\bibfnamefont {B.~Q.}\ \bibnamefont
  {Baragiola}}, \bibinfo {author} {\bibfnamefont {G.}~\bibnamefont
  {Pantaleoni}}, \bibinfo {author} {\bibfnamefont {R.~N.}\ \bibnamefont
  {Alexander}}, \bibinfo {author} {\bibfnamefont {A.}~\bibnamefont {Karanjai}},
  \ and\ \bibinfo {author} {\bibfnamefont {N.~C.}\ \bibnamefont {Menicucci}},\
  }\href {\doibase 10.1103/PhysRevLett.123.200502} {\bibfield  {journal}
  {\bibinfo  {journal} {Phys. Rev. Lett.}\ }\textbf {\bibinfo {volume} {123}},\
  \bibinfo {pages} {200502} (\bibinfo {year} {2019})}\BibitemShut {NoStop}%
\bibitem [{\citenamefont {Yamasaki}\ \emph {et~al.}(2020)\citenamefont
  {Yamasaki}, \citenamefont {Matsuura},\ and\ \citenamefont
  {Koashi}}]{yamasaki2020cost}%
  \BibitemOpen
  \bibfield  {author} {\bibinfo {author} {\bibfnamefont {H.}~\bibnamefont
  {Yamasaki}}, \bibinfo {author} {\bibfnamefont {T.}~\bibnamefont {Matsuura}},
  \ and\ \bibinfo {author} {\bibfnamefont {M.}~\bibnamefont {Koashi}},\ }\href
  {\doibase 10.1103/PhysRevResearch.2.023270} {\bibfield  {journal} {\bibinfo
  {journal} {Physical Review Research}\ }\textbf {\bibinfo {volume} {2}},\
  \bibinfo {pages} {023270} (\bibinfo {year} {2020})}\BibitemShut {NoStop}%
\bibitem [{\citenamefont {Gu}\ \emph {et~al.}(2009)\citenamefont {Gu},
  \citenamefont {Weedbrook}, \citenamefont {Menicucci}, \citenamefont {Ralph},\
  and\ \citenamefont {van Loock}}]{gu2009quantum}%
  \BibitemOpen
  \bibfield  {author} {\bibinfo {author} {\bibfnamefont {M.}~\bibnamefont
  {Gu}}, \bibinfo {author} {\bibfnamefont {C.}~\bibnamefont {Weedbrook}},
  \bibinfo {author} {\bibfnamefont {N.~C.}\ \bibnamefont {Menicucci}}, \bibinfo
  {author} {\bibfnamefont {T.~C.}\ \bibnamefont {Ralph}}, \ and\ \bibinfo
  {author} {\bibfnamefont {P.}~\bibnamefont {van Loock}},\ }\href {\doibase
  10.1103/PhysRevA.79.062318} {\bibfield  {journal} {\bibinfo  {journal} {Phys.
  Rev. A}\ }\textbf {\bibinfo {volume} {79}},\ \bibinfo {pages} {062318}
  (\bibinfo {year} {2009})}\BibitemShut {NoStop}%
\bibitem [{\citenamefont {Douce}\ \emph {et~al.}(2019)\citenamefont {Douce},
  \citenamefont {Markham}, \citenamefont {Kashefi}, \citenamefont {van Loock},\
  and\ \citenamefont {Ferrini}}]{douce2019}%
  \BibitemOpen
  \bibfield  {author} {\bibinfo {author} {\bibfnamefont {T.}~\bibnamefont
  {Douce}}, \bibinfo {author} {\bibfnamefont {D.}~\bibnamefont {Markham}},
  \bibinfo {author} {\bibfnamefont {E.}~\bibnamefont {Kashefi}}, \bibinfo
  {author} {\bibfnamefont {P.}~\bibnamefont {van Loock}}, \ and\ \bibinfo
  {author} {\bibfnamefont {G.}~\bibnamefont {Ferrini}},\ }\href {\doibase
  10.1103/PhysRevA.99.012344} {\bibfield  {journal} {\bibinfo  {journal}
  {Physical Review A}\ }\textbf {\bibinfo {volume} {99}},\ \bibinfo {pages}
  {012344} (\bibinfo {year} {2019})}\BibitemShut {NoStop}%
\bibitem [{\citenamefont {Fl\"uhmann}\ \emph {et~al.}(2018)\citenamefont
  {Fl\"uhmann}, \citenamefont {Negnevitsky}, \citenamefont {Marinelli},\ and\
  \citenamefont {Home}}]{fluhmann2018}%
  \BibitemOpen
  \bibfield  {author} {\bibinfo {author} {\bibfnamefont {C.}~\bibnamefont
  {Fl\"uhmann}}, \bibinfo {author} {\bibfnamefont {V.}~\bibnamefont
  {Negnevitsky}}, \bibinfo {author} {\bibfnamefont {M.}~\bibnamefont
  {Marinelli}}, \ and\ \bibinfo {author} {\bibfnamefont {J.~P.}\ \bibnamefont
  {Home}},\ }\href {\doibase 10.1103/PhysRevX.8.021001} {\bibfield  {journal}
  {\bibinfo  {journal} {Phys. Rev. X}\ }\textbf {\bibinfo {volume} {8}},\
  \bibinfo {pages} {021001} (\bibinfo {year} {2018})}\BibitemShut {NoStop}%
\bibitem [{\citenamefont {Ghose}\ and\ \citenamefont
  {Sanders}(2007)}]{ghose2007}%
  \BibitemOpen
  \bibfield  {author} {\bibinfo {author} {\bibfnamefont {S.}~\bibnamefont
  {Ghose}}\ and\ \bibinfo {author} {\bibfnamefont {B.~C.}\ \bibnamefont
  {Sanders}},\ }\href {\doibase 10.1080/09500340601101575} {\bibfield
  {journal} {\bibinfo  {journal} {Journal of Modern Optics}\ }\textbf {\bibinfo
  {volume} {54}},\ \bibinfo {pages} {855} (\bibinfo {year} {2007})}\BibitemShut
  {NoStop}%
\bibitem [{\citenamefont {Miyata}\ \emph {et~al.}(2016)\citenamefont {Miyata},
  \citenamefont {Ogawa}, \citenamefont {Marek}, \citenamefont {Filip},
  \citenamefont {Yonezawa}, \citenamefont {Yoshikawa},\ and\ \citenamefont
  {Furusawa}}]{miyata2016}%
  \BibitemOpen
  \bibfield  {author} {\bibinfo {author} {\bibfnamefont {K.}~\bibnamefont
  {Miyata}}, \bibinfo {author} {\bibfnamefont {H.}~\bibnamefont {Ogawa}},
  \bibinfo {author} {\bibfnamefont {P.}~\bibnamefont {Marek}}, \bibinfo
  {author} {\bibfnamefont {R.}~\bibnamefont {Filip}}, \bibinfo {author}
  {\bibfnamefont {H.}~\bibnamefont {Yonezawa}}, \bibinfo {author}
  {\bibfnamefont {J.-i.}\ \bibnamefont {Yoshikawa}}, \ and\ \bibinfo {author}
  {\bibfnamefont {A.}~\bibnamefont {Furusawa}},\ }\href {\doibase
  10.1103/PhysRevA.93.022301} {\bibfield  {journal} {\bibinfo  {journal}
  {Physical Review A}\ }\textbf {\bibinfo {volume} {93}},\ \bibinfo {pages}
  {022301} (\bibinfo {year} {2016})}\BibitemShut {NoStop}%
\bibitem [{\citenamefont {Arzani}\ \emph {et~al.}(2017)\citenamefont {Arzani},
  \citenamefont {Treps},\ and\ \citenamefont {Ferrini}}]{arzani2017}%
  \BibitemOpen
  \bibfield  {author} {\bibinfo {author} {\bibfnamefont {F.}~\bibnamefont
  {Arzani}}, \bibinfo {author} {\bibfnamefont {N.}~\bibnamefont {Treps}}, \
  and\ \bibinfo {author} {\bibfnamefont {G.}~\bibnamefont {Ferrini}},\ }\href
  {\doibase 10.1103/PhysRevA.95.052352} {\bibfield  {journal} {\bibinfo
  {journal} {Physical Review A}\ }\textbf {\bibinfo {volume} {95}},\ \bibinfo
  {pages} {052352} (\bibinfo {year} {2017})}\BibitemShut {NoStop}%
\bibitem [{\citenamefont {Sabapathy}\ \emph {et~al.}(2019)\citenamefont
  {Sabapathy}, \citenamefont {Qi}, \citenamefont {Izaac},\ and\ \citenamefont
  {Weedbrook}}]{sabapathy2019production}%
  \BibitemOpen
  \bibfield  {author} {\bibinfo {author} {\bibfnamefont {K.~K.}\ \bibnamefont
  {Sabapathy}}, \bibinfo {author} {\bibfnamefont {H.}~\bibnamefont {Qi}},
  \bibinfo {author} {\bibfnamefont {J.}~\bibnamefont {Izaac}}, \ and\ \bibinfo
  {author} {\bibfnamefont {C.}~\bibnamefont {Weedbrook}},\ }\href {\doibase
  10.1103/PhysRevA.100.012326} {\bibfield  {journal} {\bibinfo  {journal}
  {Physical Review A}\ }\textbf {\bibinfo {volume} {100}},\ \bibinfo {pages}
  {012326} (\bibinfo {year} {2019})}\BibitemShut {NoStop}%
\bibitem [{\citenamefont {Yanagimoto}\ \emph {et~al.}(2019)\citenamefont
  {Yanagimoto}, \citenamefont {Onodera}, \citenamefont {Ng}, \citenamefont
  {Wright}, \citenamefont {McMahon},\ and\ \citenamefont
  {Mabuchi}}]{yanagimoto2019engineering}%
  \BibitemOpen
  \bibfield  {author} {\bibinfo {author} {\bibfnamefont {R.}~\bibnamefont
  {Yanagimoto}}, \bibinfo {author} {\bibfnamefont {T.}~\bibnamefont {Onodera}},
  \bibinfo {author} {\bibfnamefont {E.}~\bibnamefont {Ng}}, \bibinfo {author}
  {\bibfnamefont {L.~G.}\ \bibnamefont {Wright}}, \bibinfo {author}
  {\bibfnamefont {P.~L.}\ \bibnamefont {McMahon}}, \ and\ \bibinfo {author}
  {\bibfnamefont {H.}~\bibnamefont {Mabuchi}},\ }\href {\doibase
  10.1103/PhysRevLett.124.240503} {\bibfield  {journal} {\bibinfo  {journal}
  {Physical Review Letters}\ }\textbf {\bibinfo {volume} {124}},\ \bibinfo
  {pages} {240503} (\bibinfo {year} {2019})}\BibitemShut {NoStop}%
\bibitem [{\citenamefont {Yukawa}\ \emph {et~al.}(2013)\citenamefont {Yukawa},
  \citenamefont {Miyata}, \citenamefont {Yonezawa}, \citenamefont {Marek},
  \citenamefont {Filip},\ and\ \citenamefont {Furusawa}}]{yukawa2013}%
  \BibitemOpen
  \bibfield  {author} {\bibinfo {author} {\bibfnamefont {M.}~\bibnamefont
  {Yukawa}}, \bibinfo {author} {\bibfnamefont {K.}~\bibnamefont {Miyata}},
  \bibinfo {author} {\bibfnamefont {H.}~\bibnamefont {Yonezawa}}, \bibinfo
  {author} {\bibfnamefont {P.}~\bibnamefont {Marek}}, \bibinfo {author}
  {\bibfnamefont {R.}~\bibnamefont {Filip}}, \ and\ \bibinfo {author}
  {\bibfnamefont {A.}~\bibnamefont {Furusawa}},\ }\href {\doibase
  10.1103/PhysRevA.88.053816} {\bibfield  {journal} {\bibinfo  {journal}
  {Physical Review A}\ }\textbf {\bibinfo {volume} {88}},\ \bibinfo {pages}
  {053816} (\bibinfo {year} {2013})}\BibitemShut {NoStop}%
\bibitem [{\citenamefont {Dakna}\ \emph {et~al.}(1997)\citenamefont {Dakna},
  \citenamefont {Anhut}, \citenamefont {Opatrn{\'{y}}}, \citenamefont
  {Kn{\"{o}}ll},\ and\ \citenamefont {Welsch}}]{dakna1997generating}%
  \BibitemOpen
  \bibfield  {author} {\bibinfo {author} {\bibfnamefont {M.}~\bibnamefont
  {Dakna}}, \bibinfo {author} {\bibfnamefont {T.}~\bibnamefont {Anhut}},
  \bibinfo {author} {\bibfnamefont {T.}~\bibnamefont {Opatrn{\'{y}}}}, \bibinfo
  {author} {\bibfnamefont {L.}~\bibnamefont {Kn{\"{o}}ll}}, \ and\ \bibinfo
  {author} {\bibfnamefont {D.-G.}\ \bibnamefont {Welsch}},\ }\href {\doibase
  10.1103/PhysRevA.55.3184} {\bibfield  {journal} {\bibinfo  {journal}
  {Physical Review A}\ }\textbf {\bibinfo {volume} {55}},\ \bibinfo {pages}
  {3184} (\bibinfo {year} {1997})}\BibitemShut {NoStop}%
\bibitem [{\citenamefont {Wenger}\ \emph {et~al.}(2004)\citenamefont {Wenger},
  \citenamefont {Tualle-Brouri},\ and\ \citenamefont
  {Grangier}}]{wenger2004non}%
  \BibitemOpen
  \bibfield  {author} {\bibinfo {author} {\bibfnamefont {J.}~\bibnamefont
  {Wenger}}, \bibinfo {author} {\bibfnamefont {R.}~\bibnamefont
  {Tualle-Brouri}}, \ and\ \bibinfo {author} {\bibfnamefont {P.}~\bibnamefont
  {Grangier}},\ }\href {\doibase 10.1103/PhysRevLett.92.153601} {\bibfield
  {journal} {\bibinfo  {journal} {Physical Review Letters}\ }\textbf {\bibinfo
  {volume} {92}},\ \bibinfo {pages} {153601} (\bibinfo {year}
  {2004})}\BibitemShut {NoStop}%
\bibitem [{\citenamefont {Parigi}\ \emph {et~al.}(2007)\citenamefont {Parigi},
  \citenamefont {Zavatta}, \citenamefont {Kim},\ and\ \citenamefont
  {Bellini}}]{parigi2007probing}%
  \BibitemOpen
  \bibfield  {author} {\bibinfo {author} {\bibfnamefont {V.}~\bibnamefont
  {Parigi}}, \bibinfo {author} {\bibfnamefont {A.}~\bibnamefont {Zavatta}},
  \bibinfo {author} {\bibfnamefont {M.}~\bibnamefont {Kim}}, \ and\ \bibinfo
  {author} {\bibfnamefont {M.}~\bibnamefont {Bellini}},\ }\href {\doibase
  10.1126/science.1146204} {\bibfield  {journal} {\bibinfo  {journal}
  {Science}\ }\textbf {\bibinfo {volume} {317}},\ \bibinfo {pages} {1890}
  (\bibinfo {year} {2007})}\BibitemShut {NoStop}%
\bibitem [{\citenamefont {Hofheinz}\ \emph {et~al.}(2008)\citenamefont
  {Hofheinz}, \citenamefont {Weig}, \citenamefont {Ansmann}, \citenamefont
  {Bialczak}, \citenamefont {Lucero}, \citenamefont {Neeley}, \citenamefont
  {O'Connell}, \citenamefont {Wang}, \citenamefont {Martinis},\ and\
  \citenamefont {Cleland}}]{hofheinz2008generation}%
  \BibitemOpen
  \bibfield  {author} {\bibinfo {author} {\bibfnamefont {M.}~\bibnamefont
  {Hofheinz}}, \bibinfo {author} {\bibfnamefont {E.~M.}\ \bibnamefont {Weig}},
  \bibinfo {author} {\bibfnamefont {M.}~\bibnamefont {Ansmann}}, \bibinfo
  {author} {\bibfnamefont {R.~C.}\ \bibnamefont {Bialczak}}, \bibinfo {author}
  {\bibfnamefont {E.}~\bibnamefont {Lucero}}, \bibinfo {author} {\bibfnamefont
  {M.}~\bibnamefont {Neeley}}, \bibinfo {author} {\bibfnamefont {A.~D.}\
  \bibnamefont {O'Connell}}, \bibinfo {author} {\bibfnamefont {H.}~\bibnamefont
  {Wang}}, \bibinfo {author} {\bibfnamefont {J.~M.}\ \bibnamefont {Martinis}},
  \ and\ \bibinfo {author} {\bibfnamefont {A.~N.}\ \bibnamefont {Cleland}},\
  }\href {\doibase 10.1038/nature07136} {\bibfield  {journal} {\bibinfo
  {journal} {Nature}\ }\textbf {\bibinfo {volume} {454}},\ \bibinfo {pages}
  {310} (\bibinfo {year} {2008})}\BibitemShut {NoStop}%
\bibitem [{\citenamefont {Heeres}\ \emph {et~al.}(2015)\citenamefont {Heeres},
  \citenamefont {Vlastakis}, \citenamefont {Holland}, \citenamefont
  {Krastanov}, \citenamefont {Albert}, \citenamefont {Frunzio}, \citenamefont
  {Jiang},\ and\ \citenamefont {Schoelkopf}}]{Heeres2015}%
  \BibitemOpen
  \bibfield  {author} {\bibinfo {author} {\bibfnamefont {R.~W.}\ \bibnamefont
  {Heeres}}, \bibinfo {author} {\bibfnamefont {B.}~\bibnamefont {Vlastakis}},
  \bibinfo {author} {\bibfnamefont {E.}~\bibnamefont {Holland}}, \bibinfo
  {author} {\bibfnamefont {S.}~\bibnamefont {Krastanov}}, \bibinfo {author}
  {\bibfnamefont {V.~V.}\ \bibnamefont {Albert}}, \bibinfo {author}
  {\bibfnamefont {L.}~\bibnamefont {Frunzio}}, \bibinfo {author} {\bibfnamefont
  {L.}~\bibnamefont {Jiang}}, \ and\ \bibinfo {author} {\bibfnamefont {R.~J.}\
  \bibnamefont {Schoelkopf}},\ }\href {\doibase 10.1103/PhysRevLett.115.137002}
  {\bibfield  {journal} {\bibinfo  {journal} {Phys. Rev. Lett.}\ }\textbf
  {\bibinfo {volume} {115}},\ \bibinfo {pages} {137002} (\bibinfo {year}
  {2015})}\BibitemShut {NoStop}%
\bibitem [{\citenamefont {Kirchmair}\ \emph {et~al.}(2013)\citenamefont
  {Kirchmair}, \citenamefont {Vlastakis}, \citenamefont {Leghtas},
  \citenamefont {Nigg}, \citenamefont {Paik}, \citenamefont {Ginossar},
  \citenamefont {Mirrahimi}, \citenamefont {Frunzio}, \citenamefont {Girvin},\
  and\ \citenamefont {Schoelkopf}}]{Kirchmair2013}%
  \BibitemOpen
  \bibfield  {author} {\bibinfo {author} {\bibfnamefont {G.}~\bibnamefont
  {Kirchmair}}, \bibinfo {author} {\bibfnamefont {B.}~\bibnamefont
  {Vlastakis}}, \bibinfo {author} {\bibfnamefont {Z.}~\bibnamefont {Leghtas}},
  \bibinfo {author} {\bibfnamefont {S.~E.}\ \bibnamefont {Nigg}}, \bibinfo
  {author} {\bibfnamefont {H.}~\bibnamefont {Paik}}, \bibinfo {author}
  {\bibfnamefont {E.}~\bibnamefont {Ginossar}}, \bibinfo {author}
  {\bibfnamefont {M.}~\bibnamefont {Mirrahimi}}, \bibinfo {author}
  {\bibfnamefont {L.}~\bibnamefont {Frunzio}}, \bibinfo {author} {\bibfnamefont
  {S.~M.}\ \bibnamefont {Girvin}}, \ and\ \bibinfo {author} {\bibfnamefont
  {R.~J.}\ \bibnamefont {Schoelkopf}},\ }\href {\doibase 10.1038/nature11902}
  {\bibfield  {journal} {\bibinfo  {journal} {Nature}\ }\textbf {\bibinfo
  {volume} {495}},\ \bibinfo {pages} {205} (\bibinfo {year}
  {2013})}\BibitemShut {NoStop}%
\bibitem [{\citenamefont {Svensson}\ \emph {et~al.}(2018)\citenamefont
  {Svensson}, \citenamefont {Bengtsson}, \citenamefont {Bylander},
  \citenamefont {Shumeiko},\ and\ \citenamefont
  {Delsing}}]{svensson_multiplication_2018}%
  \BibitemOpen
  \bibfield  {author} {\bibinfo {author} {\bibfnamefont {I.-M.}\ \bibnamefont
  {Svensson}}, \bibinfo {author} {\bibfnamefont {A.}~\bibnamefont {Bengtsson}},
  \bibinfo {author} {\bibfnamefont {J.}~\bibnamefont {Bylander}}, \bibinfo
  {author} {\bibfnamefont {V.}~\bibnamefont {Shumeiko}}, \ and\ \bibinfo
  {author} {\bibfnamefont {P.}~\bibnamefont {Delsing}},\ }\href {\doibase
  10.1063/1.5026974} {\bibfield  {journal} {\bibinfo  {journal} {Applied
  Physics Letters}\ }\textbf {\bibinfo {volume} {113}},\ \bibinfo {pages}
  {022602} (\bibinfo {year} {2018})}\BibitemShut {NoStop}%
\bibitem [{\citenamefont {Touzard}\ \emph {et~al.}(2018)\citenamefont
  {Touzard}, \citenamefont {Grimm}, \citenamefont {Leghtas}, \citenamefont
  {Mundhada}, \citenamefont {Reinhold}, \citenamefont {Axline}, \citenamefont
  {Reagor}, \citenamefont {Chou}, \citenamefont {Blumoff}, \citenamefont
  {Sliwa}, \citenamefont {Shankar}, \citenamefont {Frunzio}, \citenamefont
  {Schoelkopf}, \citenamefont {Mirrahimi},\ and\ \citenamefont
  {Devoret}}]{Touzard2018}%
  \BibitemOpen
  \bibfield  {author} {\bibinfo {author} {\bibfnamefont {S.}~\bibnamefont
  {Touzard}}, \bibinfo {author} {\bibfnamefont {A.}~\bibnamefont {Grimm}},
  \bibinfo {author} {\bibfnamefont {Z.}~\bibnamefont {Leghtas}}, \bibinfo
  {author} {\bibfnamefont {S.~O.}\ \bibnamefont {Mundhada}}, \bibinfo {author}
  {\bibfnamefont {P.}~\bibnamefont {Reinhold}}, \bibinfo {author}
  {\bibfnamefont {C.}~\bibnamefont {Axline}}, \bibinfo {author} {\bibfnamefont
  {M.}~\bibnamefont {Reagor}}, \bibinfo {author} {\bibfnamefont
  {K.}~\bibnamefont {Chou}}, \bibinfo {author} {\bibfnamefont {J.}~\bibnamefont
  {Blumoff}}, \bibinfo {author} {\bibfnamefont {K.~M.}\ \bibnamefont {Sliwa}},
  \bibinfo {author} {\bibfnamefont {S.}~\bibnamefont {Shankar}}, \bibinfo
  {author} {\bibfnamefont {L.}~\bibnamefont {Frunzio}}, \bibinfo {author}
  {\bibfnamefont {R.~J.}\ \bibnamefont {Schoelkopf}}, \bibinfo {author}
  {\bibfnamefont {M.}~\bibnamefont {Mirrahimi}}, \ and\ \bibinfo {author}
  {\bibfnamefont {M.~H.}\ \bibnamefont {Devoret}},\ }\href {\doibase
  10.1103/PhysRevX.8.021005} {\bibfield  {journal} {\bibinfo  {journal} {Phys.
  Rev. X}\ }\textbf {\bibinfo {volume} {8}},\ \bibinfo {pages} {021005}
  (\bibinfo {year} {2018})}\BibitemShut {NoStop}%
\bibitem [{\citenamefont {Grimm}\ \emph {et~al.}(2020)\citenamefont {Grimm},
  \citenamefont {Frattini}, \citenamefont {Puri}, \citenamefont {Mundhada},
  \citenamefont {Touzard}, \citenamefont {Mirrahimi}, \citenamefont {Girvin},
  \citenamefont {Shankar},\ and\ \citenamefont {Devoret}}]{Grimm2019}%
  \BibitemOpen
  \bibfield  {author} {\bibinfo {author} {\bibfnamefont {A.}~\bibnamefont
  {Grimm}}, \bibinfo {author} {\bibfnamefont {N.~E.}\ \bibnamefont {Frattini}},
  \bibinfo {author} {\bibfnamefont {S.}~\bibnamefont {Puri}}, \bibinfo {author}
  {\bibfnamefont {S.~O.}\ \bibnamefont {Mundhada}}, \bibinfo {author}
  {\bibfnamefont {S.}~\bibnamefont {Touzard}}, \bibinfo {author} {\bibfnamefont
  {M.}~\bibnamefont {Mirrahimi}}, \bibinfo {author} {\bibfnamefont {S.~M.}\
  \bibnamefont {Girvin}}, \bibinfo {author} {\bibfnamefont {S.}~\bibnamefont
  {Shankar}}, \ and\ \bibinfo {author} {\bibfnamefont {M.~H.}\ \bibnamefont
  {Devoret}},\ }\href {\doibase 10.1038/s41586-020-2587-z} {\bibfield
  {journal} {\bibinfo  {journal} {Nature}\ }\textbf {\bibinfo {volume} {584}},\
  \bibinfo {pages} {205} (\bibinfo {year} {2020})}\BibitemShut {NoStop}%
\bibitem [{\citenamefont {Chang}\ \emph {et~al.}(2020)\citenamefont {Chang},
  \citenamefont {Sab\'{\i}n}, \citenamefont {Forn-D\'{\i}az}, \citenamefont
  {Quijandr\'{\i}a}, \citenamefont {Vadiraj}, \citenamefont {Nsanzineza},
  \citenamefont {Johansson},\ and\ \citenamefont
  {Wilson}}]{chang_observation_2019}%
  \BibitemOpen
  \bibfield  {author} {\bibinfo {author} {\bibfnamefont {C.~W.~S.}\
  \bibnamefont {Chang}}, \bibinfo {author} {\bibfnamefont {C.}~\bibnamefont
  {Sab\'{\i}n}}, \bibinfo {author} {\bibfnamefont {P.}~\bibnamefont
  {Forn-D\'{\i}az}}, \bibinfo {author} {\bibfnamefont {F.}~\bibnamefont
  {Quijandr\'{\i}a}}, \bibinfo {author} {\bibfnamefont {A.~M.}\ \bibnamefont
  {Vadiraj}}, \bibinfo {author} {\bibfnamefont {I.}~\bibnamefont {Nsanzineza}},
  \bibinfo {author} {\bibfnamefont {G.}~\bibnamefont {Johansson}}, \ and\
  \bibinfo {author} {\bibfnamefont {C.~M.}\ \bibnamefont {Wilson}},\ }\href
  {\doibase 10.1103/PhysRevX.10.011011} {\bibfield  {journal} {\bibinfo
  {journal} {Phys. Rev. X}\ }\textbf {\bibinfo {volume} {10}},\ \bibinfo
  {pages} {011011} (\bibinfo {year} {2020})}\BibitemShut {NoStop}%
\bibitem [{\citenamefont {{Lescanne}}\ \emph {et~al.}(2020)\citenamefont
  {{Lescanne}}, \citenamefont {{Villiers}}, \citenamefont {{Peronnin}},
  \citenamefont {{Sarlette}}, \citenamefont {{Delbecq}}, \citenamefont
  {{Huard}}, \citenamefont {{Kontos}}, \citenamefont {{Mirrahimi}},\ and\
  \citenamefont {{Leghtas}}}]{Lescanne2020}%
  \BibitemOpen
  \bibfield  {author} {\bibinfo {author} {\bibfnamefont {R.}~\bibnamefont
  {{Lescanne}}}, \bibinfo {author} {\bibfnamefont {M.}~\bibnamefont
  {{Villiers}}}, \bibinfo {author} {\bibfnamefont {T.}~\bibnamefont
  {{Peronnin}}}, \bibinfo {author} {\bibfnamefont {A.}~\bibnamefont
  {{Sarlette}}}, \bibinfo {author} {\bibfnamefont {M.}~\bibnamefont
  {{Delbecq}}}, \bibinfo {author} {\bibfnamefont {B.}~\bibnamefont {{Huard}}},
  \bibinfo {author} {\bibfnamefont {T.}~\bibnamefont {{Kontos}}}, \bibinfo
  {author} {\bibfnamefont {M.}~\bibnamefont {{Mirrahimi}}}, \ and\ \bibinfo
  {author} {\bibfnamefont {Z.}~\bibnamefont {{Leghtas}}},\ }\href {\doibase
  10.1038/s41567-020-0824-x} {\bibfield  {journal} {\bibinfo  {journal} {Nature
  Physics}\ }\textbf {\bibinfo {volume} {16}},\ \bibinfo {pages} {509}
  (\bibinfo {year} {2020})}\BibitemShut {NoStop}%
\bibitem [{\citenamefont {Braunstein}\ and\ \citenamefont
  {McLachlan}(1987)}]{braunstein1986generalized}%
  \BibitemOpen
  \bibfield  {author} {\bibinfo {author} {\bibfnamefont {S.~L.}\ \bibnamefont
  {Braunstein}}\ and\ \bibinfo {author} {\bibfnamefont {R.~I.}\ \bibnamefont
  {McLachlan}},\ }\href {\doibase 10.1103/PhysRevA.35.1659} {\bibfield
  {journal} {\bibinfo  {journal} {Physical Review A}\ }\textbf {\bibinfo
  {volume} {35}},\ \bibinfo {pages} {1659} (\bibinfo {year}
  {1987})}\BibitemShut {NoStop}%
\bibitem [{\citenamefont {Banaszek}\ and\ \citenamefont
  {Knight}(1997)}]{banaszek1997quantum}%
  \BibitemOpen
  \bibfield  {author} {\bibinfo {author} {\bibfnamefont {K.}~\bibnamefont
  {Banaszek}}\ and\ \bibinfo {author} {\bibfnamefont {P.~L.}\ \bibnamefont
  {Knight}},\ }\href {\doibase 10.1103/PhysRevA.55.2368} {\bibfield  {journal}
  {\bibinfo  {journal} {Physical Review A}\ }\textbf {\bibinfo {volume} {55}},\
  \bibinfo {pages} {2368} (\bibinfo {year} {1997})}\BibitemShut {NoStop}%
\bibitem [{\citenamefont {Albert}\ and\ \citenamefont
  {Jiang}(2014)}]{albert2014symmetries}%
  \BibitemOpen
  \bibfield  {author} {\bibinfo {author} {\bibfnamefont {V.~V.}\ \bibnamefont
  {Albert}}\ and\ \bibinfo {author} {\bibfnamefont {L.}~\bibnamefont {Jiang}},\
  }\href {\doibase 10.1103/PhysRevA.89.022118} {\bibfield  {journal} {\bibinfo
  {journal} {Physical Review A}\ }\textbf {\bibinfo {volume} {89}},\ \bibinfo
  {pages} {022118} (\bibinfo {year} {2014})}\BibitemShut {NoStop}%
\bibitem [{Note1()}]{Note1}%
  \BibitemOpen
  \bibinfo {note} {Note that the value of the cubicity depends on the
  convention used. In our case, the values refer to $\hbar = 1/2$.}\BibitemShut
  {Stop}%
\bibitem [{Note2()}]{Note2}%
  \BibitemOpen
  \bibinfo {note} {The relation between the value of a quantity with respect to
  a reference value and its counterpart in decibel (dB) is $\Delta _{\protect
  \text {dB}}= -10 \protect \textrm {log}_{10}(\protect \frac {\Delta
  _0^2}{\Delta ^2})$, where here for instance $\Delta _0^2$ and $\Delta ^2$ are
  the variances of the vacuum and a squeezed state respectively. Here, we have
  $\hbar =1/2$, $\Delta _0^2=1/4$ and after applying the squeezing operator
  Eq.(\ref {eq:squeezing}) to the vacuum the variance of the $\protect \hat q$
  quadrature is given by $\Delta ^2 =\exp [-2 \abs {\xi } ]/4$.}\BibitemShut
  {Stop}%
\bibitem [{\citenamefont {Bencheikh}\ \emph {et~al.}(2007)\citenamefont
  {Bencheikh}, \citenamefont {Gravier}, \citenamefont {Douady}, \citenamefont
  {Levenson},\ and\ \citenamefont {Boulanger}}]{bencheikh2007triple}%
  \BibitemOpen
  \bibfield  {author} {\bibinfo {author} {\bibfnamefont {K.}~\bibnamefont
  {Bencheikh}}, \bibinfo {author} {\bibfnamefont {F.}~\bibnamefont {Gravier}},
  \bibinfo {author} {\bibfnamefont {J.}~\bibnamefont {Douady}}, \bibinfo
  {author} {\bibfnamefont {A.}~\bibnamefont {Levenson}}, \ and\ \bibinfo
  {author} {\bibfnamefont {B.}~\bibnamefont {Boulanger}},\ }\href {\doibase
  10.1016/j.crhy.2006.07.014} {\bibfield  {journal} {\bibinfo  {journal}
  {Comptes Rendus Physique}\ }\textbf {\bibinfo {volume} {8}},\ \bibinfo
  {pages} {206} (\bibinfo {year} {2007})}\BibitemShut {NoStop}%
\bibitem [{\citenamefont {Brunelli}\ and\ \citenamefont
  {Houhou}(2019)}]{brunelli2019linear}%
  \BibitemOpen
  \bibfield  {author} {\bibinfo {author} {\bibfnamefont {M.}~\bibnamefont
  {Brunelli}}\ and\ \bibinfo {author} {\bibfnamefont {O.}~\bibnamefont
  {Houhou}},\ }\href {\doibase 10.1103/PhysRevA.100.013831} {\bibfield
  {journal} {\bibinfo  {journal} {Physical Review A}\ }\textbf {\bibinfo
  {volume} {100}},\ \bibinfo {pages} {013831} (\bibinfo {year}
  {2019})}\BibitemShut {NoStop}%
\bibitem [{\citenamefont {Mari}\ and\ \citenamefont
  {Eisert}(2012)}]{mari2012positive}%
  \BibitemOpen
  \bibfield  {author} {\bibinfo {author} {\bibfnamefont {A.}~\bibnamefont
  {Mari}}\ and\ \bibinfo {author} {\bibfnamefont {J.}~\bibnamefont {Eisert}},\
  }\href {\doibase 10.1103/PhysRevLett.109.230503} {\bibfield  {journal}
  {\bibinfo  {journal} {Physical Review Letters}\ }\textbf {\bibinfo {volume}
  {109}},\ \bibinfo {pages} {230503} (\bibinfo {year} {2012})}\BibitemShut
  {NoStop}%
\bibitem [{\citenamefont {Veitch}\ \emph {et~al.}(2014)\citenamefont {Veitch},
  \citenamefont {Mousavian}, \citenamefont {Gottesman},\ and\ \citenamefont
  {Emerson}}]{veitch2014}%
  \BibitemOpen
  \bibfield  {author} {\bibinfo {author} {\bibfnamefont {V.}~\bibnamefont
  {Veitch}}, \bibinfo {author} {\bibfnamefont {S.~A.~H.}\ \bibnamefont
  {Mousavian}}, \bibinfo {author} {\bibfnamefont {D.}~\bibnamefont
  {Gottesman}}, \ and\ \bibinfo {author} {\bibfnamefont {J.}~\bibnamefont
  {Emerson}},\ }\href {\doibase 10.1088/1367-2630/16/1/013009} {\bibfield
  {journal} {\bibinfo  {journal} {New Journal of Physics}\ }\textbf {\bibinfo
  {volume} {16}},\ \bibinfo {pages} {013009} (\bibinfo {year}
  {2014})}\BibitemShut {NoStop}%
\bibitem [{\citenamefont {Jozsa}(1994)}]{jozsa1994fidelity}%
  \BibitemOpen
  \bibfield  {author} {\bibinfo {author} {\bibfnamefont {R.}~\bibnamefont
  {Jozsa}},\ }\href {\doibase 10.1080/09500349414552171} {\bibfield  {journal}
  {\bibinfo  {journal} {Journal of Modern Optics}\ }\textbf {\bibinfo {volume}
  {41}},\ \bibinfo {pages} {2315} (\bibinfo {year} {1994})}\BibitemShut
  {NoStop}%
\bibitem [{\citenamefont {De~Palma}\ \emph {et~al.}(2015)\citenamefont
  {De~Palma}, \citenamefont {Mari}, \citenamefont {Giovannetti},\ and\
  \citenamefont {Holevo}}]{depalmaNormalFormDecomposition2015}%
  \BibitemOpen
  \bibfield  {author} {\bibinfo {author} {\bibfnamefont {G.}~\bibnamefont
  {De~Palma}}, \bibinfo {author} {\bibfnamefont {A.}~\bibnamefont {Mari}},
  \bibinfo {author} {\bibfnamefont {V.}~\bibnamefont {Giovannetti}}, \ and\
  \bibinfo {author} {\bibfnamefont {A.~S.}\ \bibnamefont {Holevo}},\ }\href
  {\doibase 10.1063/1.4921265} {\bibfield  {journal} {\bibinfo  {journal}
  {Journal of Mathematical Physics}\ }\textbf {\bibinfo {volume} {56}},\
  \bibinfo {pages} {052202} (\bibinfo {year} {2015})}\BibitemShut {NoStop}%
\bibitem [{\citenamefont {Dopico}\ and\ \citenamefont
  {Johnson}()}]{dopicoParametrizationMatrixSymplectic2009}%
  \BibitemOpen
  \bibfield  {author} {\bibinfo {author} {\bibfnamefont {F.~M.}\ \bibnamefont
  {Dopico}}\ and\ \bibinfo {author} {\bibfnamefont {C.~R.}\ \bibnamefont
  {Johnson}},\ }\href {\doibase 10.1137/060678221} {\bibfield  {journal}
  {\bibinfo  {journal} {SIAM J. Matrix Anal. Appl.}\ }\textbf {\bibinfo
  {volume} {31}},\ \bibinfo {pages} {650}}\BibitemShut {NoStop}%
\bibitem [{\citenamefont {Cahill}\ and\ \citenamefont
  {Glauber}(1969{\natexlab{a}})}]{cahill1969density}%
  \BibitemOpen
  \bibfield  {author} {\bibinfo {author} {\bibfnamefont {K.~E.}\ \bibnamefont
  {Cahill}}\ and\ \bibinfo {author} {\bibfnamefont {R.~J.}\ \bibnamefont
  {Glauber}},\ }\href {\doibase 10.1103/PhysRev.177.1882} {\bibfield  {journal}
  {\bibinfo  {journal} {Physical Review}\ }\textbf {\bibinfo {volume} {177}},\
  \bibinfo {pages} {1882} (\bibinfo {year} {1969}{\natexlab{a}})}\BibitemShut
  {NoStop}%
\bibitem [{\citenamefont {Filip}\ \emph {et~al.}(2005)\citenamefont {Filip},
  \citenamefont {Marek},\ and\ \citenamefont
  {Andersen}}]{filip2005measurement}%
  \BibitemOpen
  \bibfield  {author} {\bibinfo {author} {\bibfnamefont {R.}~\bibnamefont
  {Filip}}, \bibinfo {author} {\bibfnamefont {P.}~\bibnamefont {Marek}}, \ and\
  \bibinfo {author} {\bibfnamefont {U.~L.}\ \bibnamefont {Andersen}},\ }\href
  {\doibase 10.1103/PhysRevA.71.042308} {\bibfield  {journal} {\bibinfo
  {journal} {Physical Review A}\ }\textbf {\bibinfo {volume} {71}},\ \bibinfo
  {pages} {042308} (\bibinfo {year} {2005})}\BibitemShut {NoStop}%
\bibitem [{\citenamefont {Miwa}\ \emph {et~al.}(2014)\citenamefont {Miwa},
  \citenamefont {Yoshikawa}, \citenamefont {Iwata}, \citenamefont {Endo},
  \citenamefont {Marek}, \citenamefont {Filip}, \citenamefont {van Loock},\
  and\ \citenamefont {Furusawa}}]{miwa2014exploring}%
  \BibitemOpen
  \bibfield  {author} {\bibinfo {author} {\bibfnamefont {Y.}~\bibnamefont
  {Miwa}}, \bibinfo {author} {\bibfnamefont {J.-i.}\ \bibnamefont {Yoshikawa}},
  \bibinfo {author} {\bibfnamefont {N.}~\bibnamefont {Iwata}}, \bibinfo
  {author} {\bibfnamefont {M.}~\bibnamefont {Endo}}, \bibinfo {author}
  {\bibfnamefont {P.}~\bibnamefont {Marek}}, \bibinfo {author} {\bibfnamefont
  {R.}~\bibnamefont {Filip}}, \bibinfo {author} {\bibfnamefont
  {P.}~\bibnamefont {van Loock}}, \ and\ \bibinfo {author} {\bibfnamefont
  {A.}~\bibnamefont {Furusawa}},\ }\href {\doibase
  10.1103/PhysRevLett.113.013601} {\bibfield  {journal} {\bibinfo  {journal}
  {Physical Review Letters}\ }\textbf {\bibinfo {volume} {113}},\ \bibinfo
  {pages} {013601} (\bibinfo {year} {2014})}\BibitemShut {NoStop}%
\bibitem [{\citenamefont {Nielsen}\ and\ \citenamefont
  {Chuang}(1997)}]{nielsen1997programmable}%
  \BibitemOpen
  \bibfield  {author} {\bibinfo {author} {\bibfnamefont {M.~A.}\ \bibnamefont
  {Nielsen}}\ and\ \bibinfo {author} {\bibfnamefont {I.~L.}\ \bibnamefont
  {Chuang}},\ }\href {\doibase 10.1103/PhysRevLett.79.321} {\bibfield
  {journal} {\bibinfo  {journal} {Physical Review Letters}\ }\textbf {\bibinfo
  {volume} {79}},\ \bibinfo {pages} {321} (\bibinfo {year} {1997})}\BibitemShut
  {NoStop}%
\bibitem [{\citenamefont {Gottesman}\ and\ \citenamefont
  {Chuang}(1999)}]{gottesman1999b}%
  \BibitemOpen
  \bibfield  {author} {\bibinfo {author} {\bibfnamefont {D.}~\bibnamefont
  {Gottesman}}\ and\ \bibinfo {author} {\bibfnamefont {I.~L.}\ \bibnamefont
  {Chuang}},\ }\href {\doibase 10.1038/46503} {\bibfield  {journal} {\bibinfo
  {journal} {Nature}\ }\textbf {\bibinfo {volume} {402}},\ \bibinfo {pages}
  {390} (\bibinfo {year} {1999})}\BibitemShut {NoStop}%
\bibitem [{\citenamefont {Bartlett}\ and\ \citenamefont
  {Munro}(2003)}]{bartlett2003quantum}%
  \BibitemOpen
  \bibfield  {author} {\bibinfo {author} {\bibfnamefont {S.~D.}\ \bibnamefont
  {Bartlett}}\ and\ \bibinfo {author} {\bibfnamefont {W.~J.}\ \bibnamefont
  {Munro}},\ }\href {\doibase 10.1103/PhysRevLett.90.117901} {\bibfield
  {journal} {\bibinfo  {journal} {Physical Review Letters}\ }\textbf {\bibinfo
  {volume} {90}},\ \bibinfo {pages} {117901} (\bibinfo {year}
  {2003})}\BibitemShut {NoStop}%
\bibitem [{\citenamefont {Yoshikawa}\ \emph {et~al.}(2007)\citenamefont
  {Yoshikawa}, \citenamefont {Hayashi}, \citenamefont {Akiyama}, \citenamefont
  {Takei}, \citenamefont {Huck}, \citenamefont {Andersen},\ and\ \citenamefont
  {Furusawa}}]{yoshikawa2007demonstration}%
  \BibitemOpen
  \bibfield  {author} {\bibinfo {author} {\bibfnamefont {J.-i.}\ \bibnamefont
  {Yoshikawa}}, \bibinfo {author} {\bibfnamefont {T.}~\bibnamefont {Hayashi}},
  \bibinfo {author} {\bibfnamefont {T.}~\bibnamefont {Akiyama}}, \bibinfo
  {author} {\bibfnamefont {N.}~\bibnamefont {Takei}}, \bibinfo {author}
  {\bibfnamefont {A.}~\bibnamefont {Huck}}, \bibinfo {author} {\bibfnamefont
  {U.~L.}\ \bibnamefont {Andersen}}, \ and\ \bibinfo {author} {\bibfnamefont
  {A.}~\bibnamefont {Furusawa}},\ }\href {\doibase 10.1103/PhysRevA.76.060301}
  {\bibfield  {journal} {\bibinfo  {journal} {Physical Review A}\ }\textbf
  {\bibinfo {volume} {76}},\ \bibinfo {pages} {060301} (\bibinfo {year}
  {2007})}\BibitemShut {NoStop}%
\bibitem [{\citenamefont {Fuwa}\ \emph {et~al.}(2014)\citenamefont {Fuwa},
  \citenamefont {Toba}, \citenamefont {Takeda}, \citenamefont {Marek},
  \citenamefont {Mi{\v{s}}ta}, \citenamefont {Filip}, \citenamefont {van
  Loock}, \citenamefont {Yoshikawa},\ and\ \citenamefont
  {Furusawa}}]{fuwa2014noiseless}%
  \BibitemOpen
  \bibfield  {author} {\bibinfo {author} {\bibfnamefont {M.}~\bibnamefont
  {Fuwa}}, \bibinfo {author} {\bibfnamefont {S.}~\bibnamefont {Toba}}, \bibinfo
  {author} {\bibfnamefont {S.}~\bibnamefont {Takeda}}, \bibinfo {author}
  {\bibfnamefont {P.}~\bibnamefont {Marek}}, \bibinfo {author} {\bibfnamefont
  {L.}~\bibnamefont {Mi{\v{s}}ta}}, \bibinfo {author} {\bibfnamefont
  {R.}~\bibnamefont {Filip}}, \bibinfo {author} {\bibfnamefont
  {P.}~\bibnamefont {van Loock}}, \bibinfo {author} {\bibfnamefont {J.-i.}\
  \bibnamefont {Yoshikawa}}, \ and\ \bibinfo {author} {\bibfnamefont
  {A.}~\bibnamefont {Furusawa}},\ }\href {\doibase
  10.1103/PhysRevLett.113.223602} {\bibfield  {journal} {\bibinfo  {journal}
  {Physical Review Letters}\ }\textbf {\bibinfo {volume} {113}},\ \bibinfo
  {pages} {223602} (\bibinfo {year} {2014})}\BibitemShut {NoStop}%
\bibitem [{\citenamefont {{NVIDIA Corporation}}(2019)}]{CUDA2019}%
  \BibitemOpen
  \bibfield  {author} {\bibinfo {author} {\bibnamefont {{NVIDIA
  Corporation}}},\ }\href
  {https://docs.nvidia.com/cuda/cuda-c-programming-guide/index.html} {\enquote
  {\bibinfo {title} {{NVIDIA CUDA C} programming guide},}\ } (\bibinfo {year}
  {2019}),\ \bibinfo {note} {version 10.2}\BibitemShut {NoStop}%
\bibitem [{\citenamefont {{Matthews}}(2018)}]{Matthews2018}%
  \BibitemOpen
  \bibfield  {author} {\bibinfo {author} {\bibfnamefont {D.}~\bibnamefont
  {{Matthews}}},\ }\href {\doibase 10.1038/d41586-018-06870-8} {\bibfield
  {journal} {\bibinfo  {journal} {Nature}\ }\textbf {\bibinfo {volume} {562}},\
  \bibinfo {pages} {151} (\bibinfo {year} {2018})}\BibitemShut {NoStop}%
\bibitem [{\citenamefont {Devoret}\ and\ \citenamefont
  {Martinis}(2004)}]{Devoret2004}%
  \BibitemOpen
  \bibfield  {author} {\bibinfo {author} {\bibfnamefont {M.~H.}\ \bibnamefont
  {Devoret}}\ and\ \bibinfo {author} {\bibfnamefont {J.~M.}\ \bibnamefont
  {Martinis}},\ }\href {\doibase 10.1007/s11128-004-3101-5} {\bibfield
  {journal} {\bibinfo  {journal} {Quantum Inf. Process.}\ }\textbf {\bibinfo
  {volume} {3}},\ \bibinfo {pages} {163–203} (\bibinfo {year}
  {2004})}\BibitemShut {NoStop}%
\bibitem [{\citenamefont {Gu}\ \emph {et~al.}(2017)\citenamefont {Gu},
  \citenamefont {Kockum}, \citenamefont {Miranowicz}, \citenamefont {Liu},\
  and\ \citenamefont {Nori}}]{gu2017}%
  \BibitemOpen
  \bibfield  {author} {\bibinfo {author} {\bibfnamefont {X.}~\bibnamefont
  {Gu}}, \bibinfo {author} {\bibfnamefont {A.~F.}\ \bibnamefont {Kockum}},
  \bibinfo {author} {\bibfnamefont {A.}~\bibnamefont {Miranowicz}}, \bibinfo
  {author} {\bibfnamefont {Y.-x.}\ \bibnamefont {Liu}}, \ and\ \bibinfo
  {author} {\bibfnamefont {F.}~\bibnamefont {Nori}},\ }\href {\doibase
  10.1016/j.physrep.2017.10.002} {\bibfield  {journal} {\bibinfo  {journal}
  {Physics Reports}\ }\textbf {\bibinfo {volume} {718-719}},\ \bibinfo {pages}
  {1} (\bibinfo {year} {2017})}\BibitemShut {NoStop}%
\bibitem [{\citenamefont {Wustmann}\ and\ \citenamefont
  {Shumeiko}(2013)}]{vitaly-parametric-2013}%
  \BibitemOpen
  \bibfield  {author} {\bibinfo {author} {\bibfnamefont {W.}~\bibnamefont
  {Wustmann}}\ and\ \bibinfo {author} {\bibfnamefont {V.}~\bibnamefont
  {Shumeiko}},\ }\href {\doibase 10.1103/PhysRevB.87.184501} {\bibfield
  {journal} {\bibinfo  {journal} {Physical Review B}\ }\textbf {\bibinfo
  {volume} {87}},\ \bibinfo {pages} {184501} (\bibinfo {year}
  {2013})}\BibitemShut {NoStop}%
\bibitem [{\citenamefont {Wustmann}\ and\ \citenamefont
  {Shumeiko}(2017)}]{vitaly-parametric-2017}%
  \BibitemOpen
  \bibfield  {author} {\bibinfo {author} {\bibfnamefont {W.}~\bibnamefont
  {Wustmann}}\ and\ \bibinfo {author} {\bibfnamefont {V.}~\bibnamefont
  {Shumeiko}},\ }\href {\doibase 10.1103/PhysRevApplied.8.024018} {\bibfield
  {journal} {\bibinfo  {journal} {Phys. Rev. Applied}\ }\textbf {\bibinfo
  {volume} {8}},\ \bibinfo {pages} {024018} (\bibinfo {year}
  {2017})}\BibitemShut {NoStop}%
\bibitem [{\citenamefont {Gao}\ \emph {et~al.}(2018)\citenamefont {Gao},
  \citenamefont {Lester}, \citenamefont {Zhang}, \citenamefont {Wang},
  \citenamefont {Rosenblum}, \citenamefont {Frunzio}, \citenamefont {Jiang},
  \citenamefont {Girvin},\ and\ \citenamefont
  {Schoelkopf}}]{gao2018programmable}%
  \BibitemOpen
  \bibfield  {author} {\bibinfo {author} {\bibfnamefont {Y.~Y.}\ \bibnamefont
  {Gao}}, \bibinfo {author} {\bibfnamefont {B.~J.}\ \bibnamefont {Lester}},
  \bibinfo {author} {\bibfnamefont {Y.}~\bibnamefont {Zhang}}, \bibinfo
  {author} {\bibfnamefont {C.}~\bibnamefont {Wang}}, \bibinfo {author}
  {\bibfnamefont {S.}~\bibnamefont {Rosenblum}}, \bibinfo {author}
  {\bibfnamefont {L.}~\bibnamefont {Frunzio}}, \bibinfo {author} {\bibfnamefont
  {L.}~\bibnamefont {Jiang}}, \bibinfo {author} {\bibfnamefont {S.~M.}\
  \bibnamefont {Girvin}}, \ and\ \bibinfo {author} {\bibfnamefont {R.~J.}\
  \bibnamefont {Schoelkopf}},\ }\href {\doibase 10.1103/PhysRevX.8.021073}
  {\bibfield  {journal} {\bibinfo  {journal} {Physical Review X}\ }\textbf
  {\bibinfo {volume} {8}},\ \bibinfo {pages} {021073} (\bibinfo {year}
  {2018})}\BibitemShut {NoStop}%
\bibitem [{\citenamefont {Eichler}\ \emph {et~al.}(2012)\citenamefont
  {Eichler}, \citenamefont {Bozyigit},\ and\ \citenamefont
  {Wallraff}}]{eichler-linear-2012}%
  \BibitemOpen
  \bibfield  {author} {\bibinfo {author} {\bibfnamefont {C.}~\bibnamefont
  {Eichler}}, \bibinfo {author} {\bibfnamefont {D.}~\bibnamefont {Bozyigit}}, \
  and\ \bibinfo {author} {\bibfnamefont {A.}~\bibnamefont {Wallraff}},\ }\href
  {\doibase 10.1103/PhysRevA.86.032106} {\bibfield  {journal} {\bibinfo
  {journal} {Phys. Rev. A}\ }\textbf {\bibinfo {volume} {86}},\ \bibinfo
  {pages} {032106} (\bibinfo {year} {2012})}\BibitemShut {NoStop}%
\bibitem [{\citenamefont {Walter}\ \emph {et~al.}(2017)\citenamefont {Walter},
  \citenamefont {Kurpiers}, \citenamefont {Gasparinetti}, \citenamefont
  {Magnard}, \citenamefont {Poto{\v{c}}nik}, \citenamefont {Salath{\'{e}}},
  \citenamefont {Pechal}, \citenamefont {Mondal}, \citenamefont {Oppliger},
  \citenamefont {Eichler},\ and\ \citenamefont {Wallraff}}]{walter2017rapid}%
  \BibitemOpen
  \bibfield  {author} {\bibinfo {author} {\bibfnamefont {T.}~\bibnamefont
  {Walter}}, \bibinfo {author} {\bibfnamefont {P.}~\bibnamefont {Kurpiers}},
  \bibinfo {author} {\bibfnamefont {S.}~\bibnamefont {Gasparinetti}}, \bibinfo
  {author} {\bibfnamefont {P.}~\bibnamefont {Magnard}}, \bibinfo {author}
  {\bibfnamefont {A.}~\bibnamefont {Poto{\v{c}}nik}}, \bibinfo {author}
  {\bibfnamefont {Y.}~\bibnamefont {Salath{\'{e}}}}, \bibinfo {author}
  {\bibfnamefont {M.}~\bibnamefont {Pechal}}, \bibinfo {author} {\bibfnamefont
  {M.}~\bibnamefont {Mondal}}, \bibinfo {author} {\bibfnamefont
  {M.}~\bibnamefont {Oppliger}}, \bibinfo {author} {\bibfnamefont
  {C.}~\bibnamefont {Eichler}}, \ and\ \bibinfo {author} {\bibfnamefont
  {A.}~\bibnamefont {Wallraff}},\ }\href {\doibase
  10.1103/PhysRevApplied.7.054020} {\bibfield  {journal} {\bibinfo  {journal}
  {Physical Review Applied}\ }\textbf {\bibinfo {volume} {7}},\ \bibinfo
  {pages} {054020} (\bibinfo {year} {2017})}\BibitemShut {NoStop}%
\bibitem [{\citenamefont {Pfaff}\ \emph {et~al.}(2017)\citenamefont {Pfaff},
  \citenamefont {Axline}, \citenamefont {Burkhart}, \citenamefont {Vool},
  \citenamefont {Reinhold}, \citenamefont {Frunzio}, \citenamefont {Jiang},
  \citenamefont {Devoret},\ and\ \citenamefont
  {Schoelkopf}}]{pfaff2017controlled}%
  \BibitemOpen
  \bibfield  {author} {\bibinfo {author} {\bibfnamefont {W.}~\bibnamefont
  {Pfaff}}, \bibinfo {author} {\bibfnamefont {C.~J.}\ \bibnamefont {Axline}},
  \bibinfo {author} {\bibfnamefont {L.~D.}\ \bibnamefont {Burkhart}}, \bibinfo
  {author} {\bibfnamefont {U.}~\bibnamefont {Vool}}, \bibinfo {author}
  {\bibfnamefont {P.}~\bibnamefont {Reinhold}}, \bibinfo {author}
  {\bibfnamefont {L.}~\bibnamefont {Frunzio}}, \bibinfo {author} {\bibfnamefont
  {L.}~\bibnamefont {Jiang}}, \bibinfo {author} {\bibfnamefont {M.~H.}\
  \bibnamefont {Devoret}}, \ and\ \bibinfo {author} {\bibfnamefont {R.~J.}\
  \bibnamefont {Schoelkopf}},\ }\href {\doibase 10.1038/nphys4143} {\bibfield
  {journal} {\bibinfo  {journal} {Nature Physics}\ }\textbf {\bibinfo {volume}
  {13}},\ \bibinfo {pages} {882} (\bibinfo {year} {2017})}\BibitemShut
  {NoStop}%
\bibitem [{\citenamefont {Hastrup}\ \emph {et~al.}(2020)\citenamefont
  {Hastrup}, \citenamefont {Larsen}, \citenamefont {Neergaard-Nielsen},
  \citenamefont {Menicucci},\ and\ \citenamefont {Andersen}}]{hastrup2020}%
  \BibitemOpen
  \bibfield  {author} {\bibinfo {author} {\bibfnamefont {J.}~\bibnamefont
  {Hastrup}}, \bibinfo {author} {\bibfnamefont {M.~V.}\ \bibnamefont {Larsen}},
  \bibinfo {author} {\bibfnamefont {J.~S.}\ \bibnamefont {Neergaard-Nielsen}},
  \bibinfo {author} {\bibfnamefont {N.~C.}\ \bibnamefont {Menicucci}}, \ and\
  \bibinfo {author} {\bibfnamefont {U.~L.}\ \bibnamefont {Andersen}},\ }\href
  {http://arxiv.org/abs/2009.05309} {\bibfield  {journal} {\bibinfo  {journal}
  {arXiv:2009.05309}\ } (\bibinfo {year} {2020})}\BibitemShut {NoStop}%
\bibitem [{\citenamefont {Vahlbruch}\ \emph {et~al.}(2016)\citenamefont
  {Vahlbruch}, \citenamefont {Mehmet}, \citenamefont {Danzmann},\ and\
  \citenamefont {Schnabel}}]{vahlbruch2016detection}%
  \BibitemOpen
  \bibfield  {author} {\bibinfo {author} {\bibfnamefont {H.}~\bibnamefont
  {Vahlbruch}}, \bibinfo {author} {\bibfnamefont {M.}~\bibnamefont {Mehmet}},
  \bibinfo {author} {\bibfnamefont {K.}~\bibnamefont {Danzmann}}, \ and\
  \bibinfo {author} {\bibfnamefont {R.}~\bibnamefont {Schnabel}},\ }\href
  {\doibase 10.1103/PhysRevLett.117.110801} {\bibfield  {journal} {\bibinfo
  {journal} {Physical Review Letters}\ }\textbf {\bibinfo {volume} {117}},\
  \bibinfo {pages} {110801} (\bibinfo {year} {2016})}\BibitemShut {NoStop}%
\bibitem [{\citenamefont {Cai}\ \emph {et~al.}(2017)\citenamefont {Cai},
  \citenamefont {Roslund}, \citenamefont {Ferrini}, \citenamefont {Arzani},
  \citenamefont {Xu}, \citenamefont {Fabre},\ and\ \citenamefont
  {Treps}}]{cai2017}%
  \BibitemOpen
  \bibfield  {author} {\bibinfo {author} {\bibfnamefont {Y.}~\bibnamefont
  {Cai}}, \bibinfo {author} {\bibfnamefont {J.}~\bibnamefont {Roslund}},
  \bibinfo {author} {\bibfnamefont {G.}~\bibnamefont {Ferrini}}, \bibinfo
  {author} {\bibfnamefont {F.}~\bibnamefont {Arzani}}, \bibinfo {author}
  {\bibfnamefont {X.}~\bibnamefont {Xu}}, \bibinfo {author} {\bibfnamefont
  {C.}~\bibnamefont {Fabre}}, \ and\ \bibinfo {author} {\bibfnamefont
  {N.}~\bibnamefont {Treps}},\ }\href {\doibase 10.1038/ncomms15645} {\bibfield
   {journal} {\bibinfo  {journal} {Nature Communications}\ }\textbf {\bibinfo
  {volume} {8}},\ \bibinfo {pages} {15645} (\bibinfo {year}
  {2017})}\BibitemShut {NoStop}%
\bibitem [{\citenamefont {Gonz{\'{a}}lez}\ \emph {et~al.}(2018)\citenamefont
  {Gonz{\'{a}}lez}, \citenamefont {Borne}, \citenamefont {Boulanger},
  \citenamefont {Levenson},\ and\ \citenamefont
  {Bencheikh}}]{gonzalez2018continuous}%
  \BibitemOpen
  \bibfield  {author} {\bibinfo {author} {\bibfnamefont {E.~A.~R.}\
  \bibnamefont {Gonz{\'{a}}lez}}, \bibinfo {author} {\bibfnamefont
  {A.}~\bibnamefont {Borne}}, \bibinfo {author} {\bibfnamefont
  {B.}~\bibnamefont {Boulanger}}, \bibinfo {author} {\bibfnamefont {J.~A.}\
  \bibnamefont {Levenson}}, \ and\ \bibinfo {author} {\bibfnamefont
  {K.}~\bibnamefont {Bencheikh}},\ }\href {\doibase
  10.1103/PhysRevLett.120.043601} {\bibfield  {journal} {\bibinfo  {journal}
  {Physical Review Letters}\ }\textbf {\bibinfo {volume} {120}},\ \bibinfo
  {pages} {043601} (\bibinfo {year} {2018})}\BibitemShut {NoStop}%
\bibitem [{\citenamefont {Okoth}\ \emph {et~al.}(2019)\citenamefont {Okoth},
  \citenamefont {Cavanna}, \citenamefont {Joly},\ and\ \citenamefont
  {Chekhova}}]{okoth2019seeded}%
  \BibitemOpen
  \bibfield  {author} {\bibinfo {author} {\bibfnamefont {C.}~\bibnamefont
  {Okoth}}, \bibinfo {author} {\bibfnamefont {A.}~\bibnamefont {Cavanna}},
  \bibinfo {author} {\bibfnamefont {N.~Y.}\ \bibnamefont {Joly}}, \ and\
  \bibinfo {author} {\bibfnamefont {M.~V.}\ \bibnamefont {Chekhova}},\ }\href
  {\doibase 10.1103/PhysRevA.99.043809} {\bibfield  {journal} {\bibinfo
  {journal} {Physical Review A}\ }\textbf {\bibinfo {volume} {99}},\ \bibinfo
  {pages} {043809} (\bibinfo {year} {2019})}\BibitemShut {NoStop}%
\bibitem [{\citenamefont {Zhang}\ \emph {et~al.}(2020)\citenamefont {Zhang},
  \citenamefont {Cai}, \citenamefont {Zheng}, \citenamefont {Barral},
  \citenamefont {Zhang}, \citenamefont {Xiao},\ and\ \citenamefont
  {Bencheikh}}]{zhang2020non}%
  \BibitemOpen
  \bibfield  {author} {\bibinfo {author} {\bibfnamefont {D.}~\bibnamefont
  {Zhang}}, \bibinfo {author} {\bibfnamefont {Y.}~\bibnamefont {Cai}}, \bibinfo
  {author} {\bibfnamefont {Z.}~\bibnamefont {Zheng}}, \bibinfo {author}
  {\bibfnamefont {D.}~\bibnamefont {Barral}}, \bibinfo {author} {\bibfnamefont
  {Y.}~\bibnamefont {Zhang}}, \bibinfo {author} {\bibfnamefont
  {M.}~\bibnamefont {Xiao}}, \ and\ \bibinfo {author} {\bibfnamefont
  {K.}~\bibnamefont {Bencheikh}},\ }\href {http://arxiv.org/abs/2009.06348}
  {\bibfield  {journal} {\bibinfo  {journal} {arXiv:2009.06348}\ } (\bibinfo
  {year} {2020})}\BibitemShut {NoStop}%
\bibitem [{\citenamefont {Nickolls}\ \emph {et~al.}(2008)\citenamefont
  {Nickolls}, \citenamefont {Buck}, \citenamefont {Garland},\ and\
  \citenamefont {Skadron}}]{Nickolls2008}%
  \BibitemOpen
  \bibfield  {author} {\bibinfo {author} {\bibfnamefont {J.}~\bibnamefont
  {Nickolls}}, \bibinfo {author} {\bibfnamefont {I.}~\bibnamefont {Buck}},
  \bibinfo {author} {\bibfnamefont {M.}~\bibnamefont {Garland}}, \ and\
  \bibinfo {author} {\bibfnamefont {K.}~\bibnamefont {Skadron}},\ }\href
  {\doibase 10.1145/1365490.1365500} {\bibfield  {journal} {\bibinfo  {journal}
  {Queue}\ }\textbf {\bibinfo {volume} {6}},\ \bibinfo {pages} {40–53}
  (\bibinfo {year} {2008})}\BibitemShut {NoStop}%
\bibitem [{\citenamefont {{The GPyOpt authors}}(2016)}]{gpyopt2016}%
  \BibitemOpen
  \bibfield  {author} {\bibinfo {author} {\bibnamefont {{The GPyOpt
  authors}}},\ }\href@noop {} {\enquote {\bibinfo {title} {Gpyopt: A bayesian
  optimization framework in python},}\ }\bibinfo {howpublished}
  {\url{http://github.com/SheffieldML/GPyOpt}} (\bibinfo {year}
  {2016})\BibitemShut {NoStop}%
\bibitem [{\citenamefont {Kushner}(1964)}]{Kushner:1964}%
  \BibitemOpen
  \bibfield  {author} {\bibinfo {author} {\bibfnamefont {H.~J.}\ \bibnamefont
  {Kushner}},\ }\href {\doibase 10.1115/1.3653121} {\bibfield  {journal}
  {\bibinfo  {journal} {Journal of Basic Engineering}\ }\textbf {\bibinfo
  {volume} {86}},\ \bibinfo {pages} {97} (\bibinfo {year} {1964})}\BibitemShut
  {NoStop}%
\bibitem [{\citenamefont {Mockus}(1989)}]{Mockus:1989}%
  \BibitemOpen
  \bibfield  {author} {\bibinfo {author} {\bibfnamefont {J.}~\bibnamefont
  {Mockus}},\ }\href@noop {} {\bibfield  {journal} {\bibinfo  {journal}
  {Springer, Dordrecht}\ } (\bibinfo {year} {1989})}\BibitemShut {NoStop}%
\bibitem [{\citenamefont {Bell}\ and\ \citenamefont
  {Hoberock}(2012)}]{thrust1}%
  \BibitemOpen
  \bibfield  {author} {\bibinfo {author} {\bibfnamefont {N.}~\bibnamefont
  {Bell}}\ and\ \bibinfo {author} {\bibfnamefont {J.}~\bibnamefont
  {Hoberock}},\ }in\ \href {\doibase
  https://doi.org/10.1016/B978-0-12-385963-1.00026-5} {\emph {\bibinfo
  {booktitle} {GPU Computing Gems Jade Edition}}},\ \bibinfo {series and
  number} {Applications of GPU Computing Series},\ \bibinfo {editor} {edited
  by\ \bibinfo {editor} {\bibfnamefont {W.}~\bibnamefont {mei W.~Hwu}}}\
  (\bibinfo  {publisher} {Morgan Kaufmann},\ \bibinfo {address} {Boston},\
  \bibinfo {year} {2012})\ pp.\ \bibinfo {pages} {359 -- 371}\BibitemShut
  {NoStop}%
\bibitem [{\citenamefont {Kaczmarski}\ and\ \citenamefont
  {Rz{k{a}}{\.{z}}ewski}(2013)}]{thrust2}%
  \BibitemOpen
  \bibfield  {author} {\bibinfo {author} {\bibfnamefont {K.}~\bibnamefont
  {Kaczmarski}}\ and\ \bibinfo {author} {\bibfnamefont {P.}~\bibnamefont
  {Rz{k{a}}{\.{z}}ewski}},\ }in\ \href@noop {} {\emph {\bibinfo {booktitle}
  {New Trends in Databases and Information Systems}}},\ \bibinfo {editor}
  {edited by\ \bibinfo {editor} {\bibfnamefont {M.}~\bibnamefont
  {Pechenizkiy}}\ and\ \bibinfo {editor} {\bibfnamefont {M.}~\bibnamefont
  {Wojciechowski}}}\ (\bibinfo  {publisher} {Springer Berlin Heidelberg},\
  \bibinfo {address} {Berlin, Heidelberg},\ \bibinfo {year} {2013})\ pp.\
  \bibinfo {pages} {37--46}\BibitemShut {NoStop}%
\bibitem [{\citenamefont {Wynters}(2013)}]{thrust3}%
  \BibitemOpen
  \bibfield  {author} {\bibinfo {author} {\bibfnamefont {E.}~\bibnamefont
  {Wynters}},\ }\href {https://dl.acm.org/doi/pdf/10.5555/2460156.2460185}
  {\bibfield  {journal} {\bibinfo  {journal} {J. Comput. Sci. Coll.}\ }\textbf
  {\bibinfo {volume} {28}},\ \bibinfo {pages} {148–155} (\bibinfo {year}
  {2013})}\BibitemShut {NoStop}%
\bibitem [{\citenamefont {{Kennedy}}\ and\ \citenamefont
  {{Eberhart}}(1995)}]{kennedy:1995}%
  \BibitemOpen
  \bibfield  {author} {\bibinfo {author} {\bibfnamefont {J.}~\bibnamefont
  {{Kennedy}}}\ and\ \bibinfo {author} {\bibfnamefont {R.}~\bibnamefont
  {{Eberhart}}},\ }\href {\doibase 10.1109/ICNN.1995.488968} {\bibfield
  {journal} {\bibinfo  {journal} {Proceedings of ICNN'95 - International
  Conference on Neural Networks}\ }\textbf {\bibinfo {volume} {4}},\ \bibinfo
  {pages} {1942} (\bibinfo {year} {1995})}\BibitemShut {NoStop}%
\bibitem [{\citenamefont {Yang}\ \emph {et~al.}(2011)\citenamefont {Yang},
  \citenamefont {Deb},\ and\ \citenamefont {Fong}}]{yang:2011}%
  \BibitemOpen
  \bibfield  {author} {\bibinfo {author} {\bibfnamefont {X.-S.}\ \bibnamefont
  {Yang}}, \bibinfo {author} {\bibfnamefont {S.}~\bibnamefont {Deb}}, \ and\
  \bibinfo {author} {\bibfnamefont {S.}~\bibnamefont {Fong}},\ }\href@noop {}
  {\bibfield  {journal} {\bibinfo  {journal} {In: Fong S. (eds) Networked
  Digital Technologies. Communications in Computer and Information Science, vol
  136. Springer, Berlin, Heidelberg}\ } (\bibinfo {year} {2011})}\BibitemShut
  {NoStop}%
\bibitem [{\citenamefont {Johansson}\ \emph {et~al.}(2012)\citenamefont
  {Johansson}, \citenamefont {Nation},\ and\ \citenamefont {Nori}}]{qutip1}%
  \BibitemOpen
  \bibfield  {author} {\bibinfo {author} {\bibfnamefont {J.}~\bibnamefont
  {Johansson}}, \bibinfo {author} {\bibfnamefont {P.}~\bibnamefont {Nation}}, \
  and\ \bibinfo {author} {\bibfnamefont {F.}~\bibnamefont {Nori}},\ }\href
  {\doibase 10.1016/j.cpc.2012.02.021} {\bibfield  {journal} {\bibinfo
  {journal} {Computer Physics Communications}\ }\textbf {\bibinfo {volume}
  {183}},\ \bibinfo {pages} {1760} (\bibinfo {year} {2012})}\BibitemShut
  {NoStop}%
\bibitem [{\citenamefont {Johansson}\ \emph {et~al.}(2013)\citenamefont
  {Johansson}, \citenamefont {Nation},\ and\ \citenamefont {Nori}}]{qutip2}%
  \BibitemOpen
  \bibfield  {author} {\bibinfo {author} {\bibfnamefont {J.}~\bibnamefont
  {Johansson}}, \bibinfo {author} {\bibfnamefont {P.}~\bibnamefont {Nation}}, \
  and\ \bibinfo {author} {\bibfnamefont {F.}~\bibnamefont {Nori}},\ }\href
  {\doibase 10.1016/j.cpc.2012.11.019} {\bibfield  {journal} {\bibinfo
  {journal} {Computer Physics Communications}\ }\textbf {\bibinfo {volume}
  {184}},\ \bibinfo {pages} {1234} (\bibinfo {year} {2013})}\BibitemShut
  {NoStop}%
\bibitem [{\citenamefont {BROYDEN}(1970)}]{Broyden1970}%
  \BibitemOpen
  \bibfield  {author} {\bibinfo {author} {\bibfnamefont {C.~G.}\ \bibnamefont
  {BROYDEN}},\ }\href {\doibase 10.1093/imamat/6.1.76} {\bibfield  {journal}
  {\bibinfo  {journal} {IMA Journal of Applied Mathematics}\ }\textbf {\bibinfo
  {volume} {6}},\ \bibinfo {pages} {76} (\bibinfo {year} {1970})}\BibitemShut
  {NoStop}%
\bibitem [{\citenamefont {Fletcher}(1970)}]{Fletcher1970}%
  \BibitemOpen
  \bibfield  {author} {\bibinfo {author} {\bibfnamefont {R.}~\bibnamefont
  {Fletcher}},\ }\href {\doibase 10.1093/comjnl/13.3.317} {\bibfield  {journal}
  {\bibinfo  {journal} {The Computer Journal}\ }\textbf {\bibinfo {volume}
  {13}},\ \bibinfo {pages} {317} (\bibinfo {year} {1970})}\BibitemShut
  {NoStop}%
\bibitem [{\citenamefont {Shanno}(1970)}]{Shanno1970}%
  \BibitemOpen
  \bibfield  {author} {\bibinfo {author} {\bibfnamefont {D.~F.}\ \bibnamefont
  {Shanno}},\ }\href {http://www.jstor.org/stable/2004840} {\bibfield
  {journal} {\bibinfo  {journal} {Mathematics of Computation}\ }\textbf
  {\bibinfo {volume} {24}},\ \bibinfo {pages} {647} (\bibinfo {year}
  {1970})}\BibitemShut {NoStop}%
\bibitem [{\citenamefont {Goldfarb}(1970)}]{Goldfarb1970}%
  \BibitemOpen
  \bibfield  {author} {\bibinfo {author} {\bibfnamefont {D.}~\bibnamefont
  {Goldfarb}},\ }\href {http://www.jstor.org/stable/2004873} {\bibfield
  {journal} {\bibinfo  {journal} {Mathematics of Computation}\ }\textbf
  {\bibinfo {volume} {24}},\ \bibinfo {pages} {23} (\bibinfo {year}
  {1970})}\BibitemShut {NoStop}%
\bibitem [{\citenamefont {{Sanderson}}\ and\ \citenamefont
  {{Curtin}}(2016)}]{Sanderson2016}%
  \BibitemOpen
  \bibfield  {author} {\bibinfo {author} {\bibfnamefont {C.}~\bibnamefont
  {{Sanderson}}}\ and\ \bibinfo {author} {\bibfnamefont {R.}~\bibnamefont
  {{Curtin}}},\ }\href {\doibase 10.21105/joss.00026} {\bibfield  {journal}
  {\bibinfo  {journal} {The Journal of Open Source Software}\ }\textbf
  {\bibinfo {volume} {1}},\ \bibinfo {eid} {26} (\bibinfo {year}
  {2016})}\BibitemShut {NoStop}%
\bibitem [{\citenamefont {Sanderson}\ and\ \citenamefont
  {Curtin}(2018)}]{Sanderson2018}%
  \BibitemOpen
  \bibfield  {author} {\bibinfo {author} {\bibfnamefont {C.}~\bibnamefont
  {Sanderson}}\ and\ \bibinfo {author} {\bibfnamefont {R.}~\bibnamefont
  {Curtin}},\ }in\ \href
  {https://link.springer.com/chapter/10.1007%2F978-3-319-96418-8_50} {\emph
  {\bibinfo {booktitle} {A User-Friendly Hybrid Sparse Matrix Class in C++}}}\
  (\bibinfo {organization} {Springer},\ \bibinfo {year} {2018})\ pp.\ \bibinfo
  {pages} {422--430}\BibitemShut {NoStop}%
\bibitem [{\citenamefont {{O'Hara}}(2020)}]{OptimLib}%
  \BibitemOpen
  \bibfield  {author} {\bibinfo {author} {\bibfnamefont {K.}~\bibnamefont
  {{O'Hara}}},\ }\href@noop {} {\enquote {\bibinfo {title} {Optimlib},}\
  }\bibinfo {howpublished} {\url{https://github.com/kthohr/optim}} (\bibinfo
  {year} {2020})\BibitemShut {NoStop}%
\bibitem [{\citenamefont {Storn}\ and\ \citenamefont
  {Price}(1997)}]{Storn1997}%
  \BibitemOpen
  \bibfield  {author} {\bibinfo {author} {\bibfnamefont {R.}~\bibnamefont
  {Storn}}\ and\ \bibinfo {author} {\bibfnamefont {K.}~\bibnamefont {Price}},\
  }\href {\doibase 10.1023/A:1008202821328} {\bibfield  {journal} {\bibinfo
  {journal} {J. of Global Optimization}\ }\textbf {\bibinfo {volume} {11}},\
  \bibinfo {pages} {341–359} (\bibinfo {year} {1997})}\BibitemShut {NoStop}%
\bibitem [{\citenamefont {Ekström}\ \emph {et~al.}(2019)\citenamefont
  {Ekström}, \citenamefont {Forssén}, \citenamefont {Dimitrakakis},
  \citenamefont {Dubhashi}, \citenamefont {Johansson}, \citenamefont
  {Muhammad}, \citenamefont {Salomonsson},\ and\ \citenamefont
  {Schliep}}]{ekstrom:2019}%
  \BibitemOpen
  \bibfield  {author} {\bibinfo {author} {\bibfnamefont {A.}~\bibnamefont
  {Ekström}}, \bibinfo {author} {\bibfnamefont {C.}~\bibnamefont {Forssén}},
  \bibinfo {author} {\bibfnamefont {C.}~\bibnamefont {Dimitrakakis}}, \bibinfo
  {author} {\bibfnamefont {D.}~\bibnamefont {Dubhashi}}, \bibinfo {author}
  {\bibfnamefont {H.~T.}\ \bibnamefont {Johansson}}, \bibinfo {author}
  {\bibfnamefont {A.~S.}\ \bibnamefont {Muhammad}}, \bibinfo {author}
  {\bibfnamefont {H.}~\bibnamefont {Salomonsson}}, \ and\ \bibinfo {author}
  {\bibfnamefont {A.}~\bibnamefont {Schliep}},\ }\href
  {https://iopscience.iop.org/article/10.1088/1361-6471/ab2b14/meta} {\bibfield
   {journal} {\bibinfo  {journal} {J. Phys. G: Nucl. Part. Phys.}\ }\textbf
  {\bibinfo {volume} {46}},\ \bibinfo {pages} {095101} (\bibinfo {year}
  {2019})}\BibitemShut {NoStop}%
\bibitem [{\citenamefont {Cahill}\ and\ \citenamefont
  {Glauber}(1969{\natexlab{b}})}]{PhysRev.177.1857}%
  \BibitemOpen
  \bibfield  {author} {\bibinfo {author} {\bibfnamefont {K.~E.}\ \bibnamefont
  {Cahill}}\ and\ \bibinfo {author} {\bibfnamefont {R.~J.}\ \bibnamefont
  {Glauber}},\ }\href {\doibase 10.1103/PhysRev.177.1857} {\bibfield  {journal}
  {\bibinfo  {journal} {Phys. Rev.}\ }\textbf {\bibinfo {volume} {177}},\
  \bibinfo {pages} {1857} (\bibinfo {year} {1969}{\natexlab{b}})}\BibitemShut
  {NoStop}%
\bibitem [{\citenamefont {Donodov}\ and\ \citenamefont
  {Mank'o}(2003)}]{donodov2003}%
  \BibitemOpen
  \bibfield  {author} {\bibinfo {author} {\bibfnamefont {V.~V.}\ \bibnamefont
  {Donodov}}\ and\ \bibinfo {author} {\bibfnamefont {V.~I.}\ \bibnamefont
  {Mank'o}},\ }\href@noop {} {\emph {\bibinfo {title} {Theory of Nonclassical
  states of light}}}\ (\bibinfo  {publisher} {CRC Press},\ \bibinfo {year}
  {2003})\BibitemShut {NoStop}%
\bibitem [{\citenamefont {Leonhardt}(2010)}]{leonhardt2010essential}%
  \BibitemOpen
  \bibfield  {author} {\bibinfo {author} {\bibfnamefont {U.}~\bibnamefont
  {Leonhardt}},\ }\href@noop {} {\emph {\bibinfo {title} {Essential Quantum
  Optics: from Quantum Measurements to Black Holes}}}\ (\bibinfo  {publisher}
  {Cambridge University Press},\ \bibinfo {year} {2010})\BibitemShut {NoStop}%
\bibitem [{\citenamefont {Paris}\ \emph {et~al.}(2003)\citenamefont {Paris},
  \citenamefont {Cola},\ and\ \citenamefont {Bonifacio}}]{paris2003}%
  \BibitemOpen
  \bibfield  {author} {\bibinfo {author} {\bibfnamefont {M.~G.~A.}\
  \bibnamefont {Paris}}, \bibinfo {author} {\bibfnamefont {M.}~\bibnamefont
  {Cola}}, \ and\ \bibinfo {author} {\bibfnamefont {R.}~\bibnamefont
  {Bonifacio}},\ }\href {\doibase 10.1103/PhysRevA.67.042104} {\bibfield
  {journal} {\bibinfo  {journal} {Physical Review A}\ }\textbf {\bibinfo
  {volume} {67}},\ \bibinfo {pages} {042104} (\bibinfo {year}
  {2003})}\BibitemShut {NoStop}%
\bibitem [{\citenamefont {Douce}\ \emph {et~al.}(2017)\citenamefont {Douce},
  \citenamefont {Markham}, \citenamefont {Kashefi}, \citenamefont {Diamanti},
  \citenamefont {Coudreau}, \citenamefont {Milman}, \citenamefont {van Loock},\
  and\ \citenamefont {Ferrini}}]{douce2017}%
  \BibitemOpen
  \bibfield  {author} {\bibinfo {author} {\bibfnamefont {T.}~\bibnamefont
  {Douce}}, \bibinfo {author} {\bibfnamefont {D.}~\bibnamefont {Markham}},
  \bibinfo {author} {\bibfnamefont {E.}~\bibnamefont {Kashefi}}, \bibinfo
  {author} {\bibfnamefont {E.}~\bibnamefont {Diamanti}}, \bibinfo {author}
  {\bibfnamefont {T.}~\bibnamefont {Coudreau}}, \bibinfo {author}
  {\bibfnamefont {P.}~\bibnamefont {Milman}}, \bibinfo {author} {\bibfnamefont
  {P.}~\bibnamefont {van Loock}}, \ and\ \bibinfo {author} {\bibfnamefont
  {G.}~\bibnamefont {Ferrini}},\ }\href {\doibase
  10.1103/PhysRevLett.118.070503} {\bibfield  {journal} {\bibinfo  {journal}
  {Physical Review Letters}\ }\textbf {\bibinfo {volume} {118}},\ \bibinfo
  {pages} {070503} (\bibinfo {year} {2017})}\BibitemShut {NoStop}%
\bibitem [{\citenamefont {Grimsmo}\ \emph {et~al.}(2020)\citenamefont
  {Grimsmo}, \citenamefont {Combes},\ and\ \citenamefont
  {Baragiola}}]{grimsmo2019}%
  \BibitemOpen
  \bibfield  {author} {\bibinfo {author} {\bibfnamefont {A.~L.}\ \bibnamefont
  {Grimsmo}}, \bibinfo {author} {\bibfnamefont {J.}~\bibnamefont {Combes}}, \
  and\ \bibinfo {author} {\bibfnamefont {B.~Q.}\ \bibnamefont {Baragiola}},\
  }\href {\doibase 10.1103/PhysRevX.10.011058} {\bibfield  {journal} {\bibinfo
  {journal} {Physical Review X}\ }\textbf {\bibinfo {volume} {10}},\ \bibinfo
  {pages} {011058} (\bibinfo {year} {2020})}\BibitemShut {NoStop}%
\bibitem [{\citenamefont {Sabapathy}\ and\ \citenamefont
  {Weedbrook}(2018)}]{Sabapathy2018}%
  \BibitemOpen
  \bibfield  {author} {\bibinfo {author} {\bibfnamefont {K.~K.}\ \bibnamefont
  {Sabapathy}}\ and\ \bibinfo {author} {\bibfnamefont {C.}~\bibnamefont
  {Weedbrook}},\ }\href {\doibase 10.1103/PhysRevA.97.062315} {\bibfield
  {journal} {\bibinfo  {journal} {Physical Review A}\ }\textbf {\bibinfo
  {volume} {97}},\ \bibinfo {pages} {062315} (\bibinfo {year} {2018})},\
  \Eprint {http://arxiv.org/abs/1802.05220} {1802.05220} \BibitemShut {NoStop}%
\bibitem [{\citenamefont {Dall'Arno}\ \emph {et~al.}(2010)\citenamefont
  {Dall'Arno}, \citenamefont {D'Ariano},\ and\ \citenamefont
  {Sacchi}}]{dall2010purification}%
  \BibitemOpen
  \bibfield  {author} {\bibinfo {author} {\bibfnamefont {M.}~\bibnamefont
  {Dall'Arno}}, \bibinfo {author} {\bibfnamefont {G.~M.}\ \bibnamefont
  {D'Ariano}}, \ and\ \bibinfo {author} {\bibfnamefont {M.~F.}\ \bibnamefont
  {Sacchi}},\ }\href {\doibase 10.1103/PhysRevA.82.042315} {\bibfield
  {journal} {\bibinfo  {journal} {Physical Review A}\ }\textbf {\bibinfo
  {volume} {82}},\ \bibinfo {pages} {042315} (\bibinfo {year}
  {2010})}\BibitemShut {NoStop}%
\end{thebibliography}%


\providecommand{\noopsort}[1]{}\providecommand{\singleletter}[1]{#1}%
%

\end{document}